\documentclass[a4paper,fleqn,usenatbib]{mnras}

\usepackage{mathptmx}
\usepackage[utf8]{inputenc}
\usepackage[T1]{fontenc}
\usepackage{ae,aecompl}
\usepackage{supertabular}

\usepackage{graphicx}	% Including figure files
\usepackage{amsmath}	% Advanced maths commands
\usepackage{amssymb}	% Extra maths symbols
\usepackage{pdflscape}
\title[Stellar activity with LAMOST]{Stellar activity with LAMOST - II. Chromospheric activity in open clusters}

\author[X.-S. Fang et al.]{
Xiang-Song Fang,\thanks{E-mail: xsfang@bao.ac.cn}
Gang Zhao,
Jing-Kun Zhao
and
Yerra Bharat Kumar
\\
Key Laboratory of Optical Astronomy, National Astronomical Observatories, Chinese Academy of Sciences, Beijing 100012, China\\
}

\date{Accepted 2018 January 22. Received 2018 January 18; in original form 2017 August 9}

\pubyear{2018}

\begin{document}
\label{firstpage}
\pagerange{\pageref{firstpage}--\pageref{lastpage}}
\maketitle

\begin{abstract}
We use the LAMOST spectra of member stars in Pleiades, M34, Praesepe, and Hyades to study how chromospheric activity vary as a function of mass and rotation 
at different age. 
We measured excess equivalent widths of H$\alpha$, H$\beta$, and Ca~{\sc ii} K based on estimated chromospheric contributions from old and inactive field dwarfs, 
and excess luminosities are obtained by normalizing bolometric luminosity, for more than 700 late-type stars in these open clusters. 
Results indicate two activity sequences in cool spot coverage and H$\alpha$ excess emission among GK dwarfs in Pleiades and M dwarfs 
in Praesepe and Hyades, paralleling with well known rotation sequences. 
A weak dependence of chromospheric emission on rotation exists among ultra fast rotators in saturated regime with Rossby number Ro$\lesssim0.1$. 
In the unsaturated regime, chromospheric and coronal emission show similar dependence on Ro, but with a shift toward larger Ro, 
indicating chromospheric emission gets easily saturated than coronal emission, and/or convective turnover time-scales based on X-ray data do not work well with chromospheric emission. More interestingly, our analysis show fully convective slow 
rotators obey the rotation-chromospheric activity relation similar to hotter stars, confirming the previous finding. 
We found correlations among H$\alpha$, H$\beta$, and Ca~{\sc ii} K emissions, in which H$\alpha$ losses are more important 
than Ca~{\sc ii} K for cooler and more active stars.
In addition, a weak correlation is seen between chromospheric emission and photospheric activity that shows dependency on stellar 
spectral type and activity level, which provides some clues on how spot configuration vary as a function of mass and activity level.
\end{abstract}


%%%%%%%%%%%%%%%%%%%%%%%%%%%%%%%%%%%%%%%%%%%%%%%%%%

%%%%%%%%%%%%%%%%% BODY OF PAPER %%%%%%%%%%%%%%%%%%

\section{Introduction}

Stellar magnetic activity, triggered by magnetic fields and maintained by a magnetic dynamo, is an ubiquitous phenomenon in late-type stars. Such activity 
includes a variety of phenomena, e.g., appearance of cool star-spots that are thought to be the fingerprints of emergent magnetic lines on the photosphere; 
the chromospheric and coronal heating that are responsible for excess chromospheric emissions and thermal X-rays, respectively; impulsive events like 
flares; and other solar phenomena. 

Stellar mass and rotation are thought to be two main basic parameters that affect the dynamo configuration and thus the properties of stellar magnetic 
activity. On the other hand, the loss of stellar angular momentum through the magnetized stellar wind depends on the magnetic field configuration on stellar 
surface \citep{gall2013}. As the star brakes due to angular momentum loss, its magnetic field weakens (or the efficiency of the dynamo decreases), 
and thus activity decays. The correlations between rotation, activity and age are evident from observations of various tracers of activity, e.g., the 
rotation-activity connection is found based on stellar X-ray emissions \citep[e.g.][]{pizz2003,mama2008,wrig2011,rein2014}, chromospheric Ca {\sc ii} HK emissions 
\citep[e.g.][]{noye1984,mama2008} and H$\alpha$ emissions \citep[e.g.][]{stau1997,doug2014,newt2017}, and photospheric activity indicator such as cool spot fractional 
coverage \citep[][hereafter Paper I]{fang2016}. The connection between rotation and activity among cool stars directly allows us to understand the nature of 
the magnetic dynamo in these stars, however, such connection appears to be sensitive to tracers of activity, e.g., chromospheric emission show different dependency on 
rotation compared to coronal X-ray emission \citep{doug2014}. 

Chromospheric activity (hereafter CA) compasses a wide range of phenomena that produce excess emission (e.g., Ca {\sc ii} and Balmer 
lines) with respect to a radiative equilibrium atmosphere, stemming largely from changes in their magnetic field \citep[see a comprehensive review on this 
topic by][]{hall2008}. Thanks to classic work of Mount Wilson Observatory Ca~{\sc ii} HK long-term observations that are operated from 1966 through 
2003, for the initial purpose of searching stellar analogues of the solar cycle \citep{wils1968}, wherein variety of CA properties including chromospheric variations are explored \citep{wils1978,bali1995,bali1998}, e.g., the well-known CA long-term variation patterns 
(cyclic variation, irregular variation, flat or little variation). What followed was a series of attempts trying to gain the general understanding 
(statistical characters) of CA among nearby late-type stars, e.g., statistical studies based on chromospheric emissions at the core of Ca {\sc ii} H and 
K lines among nearby FGK field stars \citep[][]{henr1996,gray2003,gray2006}, confirmed the bi-modal distribution in stellar activity \citep{vaug1980}; 
\citet{west2004,west+2008} investigate the fraction of active stars as a function of spectral type and the vertical distance from the Galactic plane, based on 
a large sample of nearby M-type stars using H$\alpha$ emission line as an activity indicator.  

Compared to field stars, open cluster members have an advantage for studying CA properties in light of rotation-activity-age connection due to their 
homogeneity in age and chemical composition. Therefore, CA in several open clusters got special attention, e.g., the young open cluster Pleiades 
\citep{sode1993} and NGC 2516 \citep{jack2010}, the median-age open cluster Hyades \citep{stau1991,stau1997} and Praesepe \citep{doug2014}. Based on these 
investigations a general picture is drawn that CA also varies as a function of both stellar mass and rotation. However, due to the limitation of sample stars 
used in previous studies in terms of mass coverage and available rotation periods (e.g., Pleiades), the behaviours of CA with respect to stellar mass 
changes and rotation is still not well understood, which demands further investigation on this issue. 

LAMOST \citep{cui+2012} observed millions of stellar spectra for cool stars in our Milky Way \citep{liu+2015}, and more and more stars' 
rotation periods become available thanks to recent and on-going photometric surveys, e.g., for $\sim$1000 Pleiades members \citep{hart2010,rebu2016}, both of 
which allow us to initiate systematic study of CA among low mass stars in different population. In Paper I, we found that almost all Pleiades 
members show excess emission in various indicators (e.g., Ca~{\sc ii} HK, H$\alpha$, H$\beta$, Ca~{\sc i} $\lambda 4226$, Mg~{\sc i} b triplet, Na~{\sc i} D 
lines, and Ca~{\sc ii} IRT lines, see Figure 7 and 8 in Paper I) relative to old and inactive field stars, indicating they are chromospherically active. 
In this paper, we continue to investigate their CA properties in detail among members of nearby open clusters including Pleiades, M34, Praesepe and Hyades by 
focusing on three chromospheric indicators: H$\alpha$, H$\beta$ and Ca~{\sc ii} K lines. We examined their incidence and strength of chromospheric activity 
as a function of effective temperature, and checked the correlation between activity and rotation. 
We also investigated the relationship amongst these three 
chromospheric indicators, and the relations between chromospheric emission and several manifestations of magnetic 
activity in different stellar atmosphere layers (cool spot coverage, light variation amplitude, and coronal emission), which holds key clues not only 
to the configuration of the magnetically active region on the stellar surface but also to the magnetically heating processes in stellar outer atmosphere.

%%%%%%%%%%%%%%%%%% 
\section{Data and sample}
\subsection{LAMOST DR3}
The LAMOST, characterized by both wide field of view (5 deg in diameter) and large effective aperture of $\sim$4 m, is a reflecting Schmidt 
telescope located at the Xinglong Observatory, China. LAMOST spectra are obtained in low resolution $R\approx1800$ (e.g., $\sim$3.6~\AA~around 6500~\AA) 
\citep[see][for more details]{zhao2006,zhao2012,luo+2015}.

The LAMOST Date Release~3 (DR3) was made available to the Chinese astronomical community and international partners during December 2015, containing nearly 5.
3 million stellar spectra that were collected from pilot survey through three years regular survey (autumn 2011 to summer 2015). LAMOST stellar 
parameter pipeline (LASP) provides the stellar parameters like effective temperature ($T_{\text{eff}}$), surface gravity ($\log g$), metallicity ([Fe/H]) and 
radial velocity (RV) for $\sim$3.2 million stellar spectra of AFGK stars whose spectra meet the signal-to-noise ratio criterion 
(e.g., $g$-band SNR of SNR$g \geq 15$ and SNR$g \geq 6$ for the bright and dark nights, respectively) \citep{wu++2011,luo+2015}. In addition, DR3 also include a catalogue of $\sim$0.3 million M-type stars.  

\subsection{Sample stars} 
We have selected four nearby open clusters with a range of ages from 100 to 700 Myr (Pleiades, M34, Praesepe and Hyades), 
and retrieved the available spectra of good quality from LAMOST DR3 archive.

The Pleiades (Seven Sisters) is a young open cluster with an age of $\sim$125 Myr \citep{stau1998} and metallicity of [Fe/H]$\sim$0.03 \citep{sode2009}, 
at a distance of around 130 pc \citep[e.g., $\sim$136 pc;][]{meli2014}. In addition to LAMOST DR2 archive used in Paper I, DR3 provide spectra for more than 
70 targets from the Pleiades potential member catalogues provided by \citet{stau2007} and \citet{bouy2015} (whose probability more than 75 
percent). In total, we selected 378 probable Pleiades members with 458 LAMOST spectra having signal-to-noise ratio in $r$-band SNR$r>10$. 

M34 (NGC 1039) has an age of about 220 Myrs \citep[see][and references therein]{meib2011}, locates at a 
distance of $\sim$470 pc toward the galactic anti-centre \citep{jone1996}. M34 has metallicity close to solar, [Fe/H]$\sim$0.07 \citep{schu2003}. 
\citet{jone1996} provided proper motion membership probability analysis for more than 600 stars with brightness down to V$\sim$16.2 in the M34 field, 
wherein, the cooler members tend to have lower membership probability due to the faintness, 
e.g., the typical probability of faint stars with V$>$15 ($\sim$K2 and cooler) is $\sim$40 percent, 
while the brighter members have probabilities close to 100 percent. 
We selected candidate M34 members with membership probability greater than 40 percent. However, real cool members might be excluded in this criterion. 
To collect more member stars, in particular at cooler end, we chose periodic variables detected in M34  
\citep{irwi2006,jame2010,meib2011}, considering them as younger and more active. 
Unfortunately, only a small number of cool stars in the vicinity of M34 field have been 
observed by LAMOST (DR3), partly due to their faintness. In total, we collected 45 candidates that have LAMOST spectra with SNR$r>10$, 
most of which are FG-type stars.

The Hyades is the nearest open cluster at a distance of about 46~pc \citep{perr1998}, and has solar metallicity [Fe/H]$\sim0.1$ \citep{tayl2005,carr2011}. 
The Praesepe (NGC 2632; the Beehive Cluster) has a distance of 180-190 pc \citep[e.g.][]{an++2007,vanl2009} and a metallicity of [Fe/H]$\sim$0.16 
\citep{carr2011,yang2016}. The Hyades and Praesepe are both part of the Hyades supercluster, and seem to share an intermediate-age around 600-800 Myr 
\citep{perr1998,foss2008,bran2015}. We adopt the membership catalogue provided by \citet{doug2014} for this two clusters, and retrieved 197 spectra for 142 
Hyades candidates, and 227 spectra for 176 candidates of Praesepe from LAMOST DR3 archive with SNR$r>10$.

\begin{figure}
\centering
\includegraphics[width=\columnwidth]{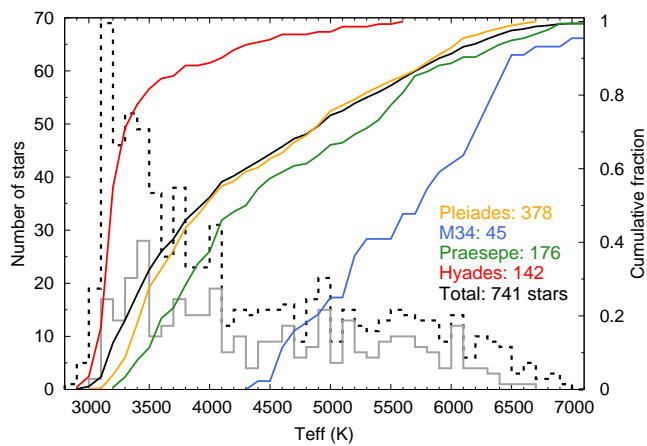}
\caption{Distribution of the sample stars as a function of temperature. The dashed black line shows the distribution for total sample stars, while the 
grey line represents these stars for which rotation period is now known. The solid lines with different colours correspond to cumulative fraction 
profiles (right y-axis). }
\label{fig:sample_data}
\end{figure}

Fig.~\ref{fig:sample_data} and Table~\ref{tab:sample_data} show the summary of total sample (741 stars), 
and easy to note that more than half ($\sim60$ percent) of them are late K- and M-type stars.

\begin{table}
\centering
\caption{Basic information about the sample stars in current work}
\label{tab:sample_data}
\begin{tabular}{lcccccc}
   \hline
Cluster    &  d     &    Age    &   [Fe/H]    &   E(B-V)   & Sample$^{a}$    \\
           & (pc)   &    (Myr)  &             &   (mag)    &                 \\
  \hline
Pleiades   &  136   & $\sim$125 &    +0.03    &  0.03$^{b}$&  378(458), 231 \\
M34        &  470   & $\sim$220 &    +0.07    &  0.07      &  45(49), 21    \\
Praesepe   &  184   & $\sim$650 &    +0.16    &  0.027     &  176(227), 76  \\   
Hyades     &   46   & $\sim$650 &    +0.11    &  0.001     &  142(197), 24  \\
  \hline
\multicolumn{6}{l}{$^a$ total number of stars (spectra), and those with rotation period.}\\
\multicolumn{6}{l}{$^b$ there is significant differential extinction in the Pleiades, }\\
\multicolumn{6}{l}{see \citet{sode1993}. }\\
\end{tabular}
\end{table}

%%%%%%%%%%%%%%%%%%%
\section{Quantifying chromospheric activity}
It is well known that enhanced emission of H$\alpha$, produced by collisional excitation in the relatively dense chromospheres of low-mass dwarf stars, 
is the strongest and widely-used activity indicator \citep[e.g.][]{stau1986,sode1993,west2004,west2015,newt2017}. 
In this work, H$\alpha$ line was used as main chromospheric activity indicator, considering that more than 50 percent of our sample stars 
are late K- and M-type stars (see Fig.~\ref{fig:sample_data}). However, we also investigated the behaviours of other two activity indicators, 
Ca~{\sc ii} K and H$\beta$ lines. 

\subsection{Excess equivalent width}
The equivalent widths were measured from RV-corrected spectra of all sample stars. For FGK dwarfs, we adopted RV values derived by LASP, 
while we measured RVs for M dwarfs following methodology described in Paper~I. The equivalent widths of H$\alpha$ line (hereafter EW$_{\text{H}\alpha}$) 
were measured using the formula (equation 4) in Paper I, where the line was centred at 6563~\AA~over 12~\AA~bandpasses and continuum 
flux was taken to be the average flux between 6547-6557~\AA~and 6570-6580~\AA. 
The measurements of EW$_{\text{H}\alpha}$ for entire sample are plotted against effective (quiescent) temperatures (see Appendix~\ref{sec:teffq} for details) 
in Fig.~\ref{fig:ewha_teff}, wherein the mean values of EW$_{\text{H}\alpha}$ for inactive reference stars (a large sample of very less active field FGKM dwarfs with solar metallicities over the temperature range $3000-6500$ K, see Paper I for more details) are treated as the minimum or basal value of chromospheric emissions, 
and shown as black solid lines. It is clear from the figure that the members of Pleiades show higher EW$_{\text{H}\alpha}$ compared to inactive counterparts. 
In fact, the correlation between EW$_{\text{H}\alpha}$ and chromospheric activity level is complex, due to presence of non-monotonic behaviour \citep{cram1979}, 
e.g., firstly the H$\alpha$ absorption increases with increase in activity, then fill-in, and eventually appear as emission line. 
Therefore, among the less active stars at the same temperature, the stronger H$\alpha$ absorption line indicates its higher activity. 
In this work, we adapted simple approach by ignoring such non-monotonic effect, and consider excess EW$_{\text{H}\alpha}$ above the basal value 
as the tracer of pure chromospheric emission. 
To determine the excess H$\alpha$ emission, we calculated the excess equivalent widths (the 
$\Delta\text{EW}_{\text{H}\alpha}$ in Paper I, hereafter denote as EW$^{'}_{\text{H}\alpha}$) by subtracting the basal value at the same temperature, 
$\text{EW}_{\text{H}\alpha,\text{basal}}$, i.e., 
\begin{equation}
\text{EW}^{'}_{\text{H}\alpha} = \text{EW}_{\text{H}\alpha} - \text{EW}_{\text{H}\alpha,\text{basal}} .
\label{equ:exew}
\end{equation} 
Similarly, we obtained the excess equivalent widths, EW$^{'}_{\text{Ca K}}$ and EW$^{'}_{\text{H}\beta}$, 
for Ca~{\sc ii} K and H$\beta$ lines, respectively (see Appendix~\ref{sec:otherlines} for details).
\subsection{Excess fractional luminosity}
\begin{figure*}
\centering
\includegraphics[width=\columnwidth]{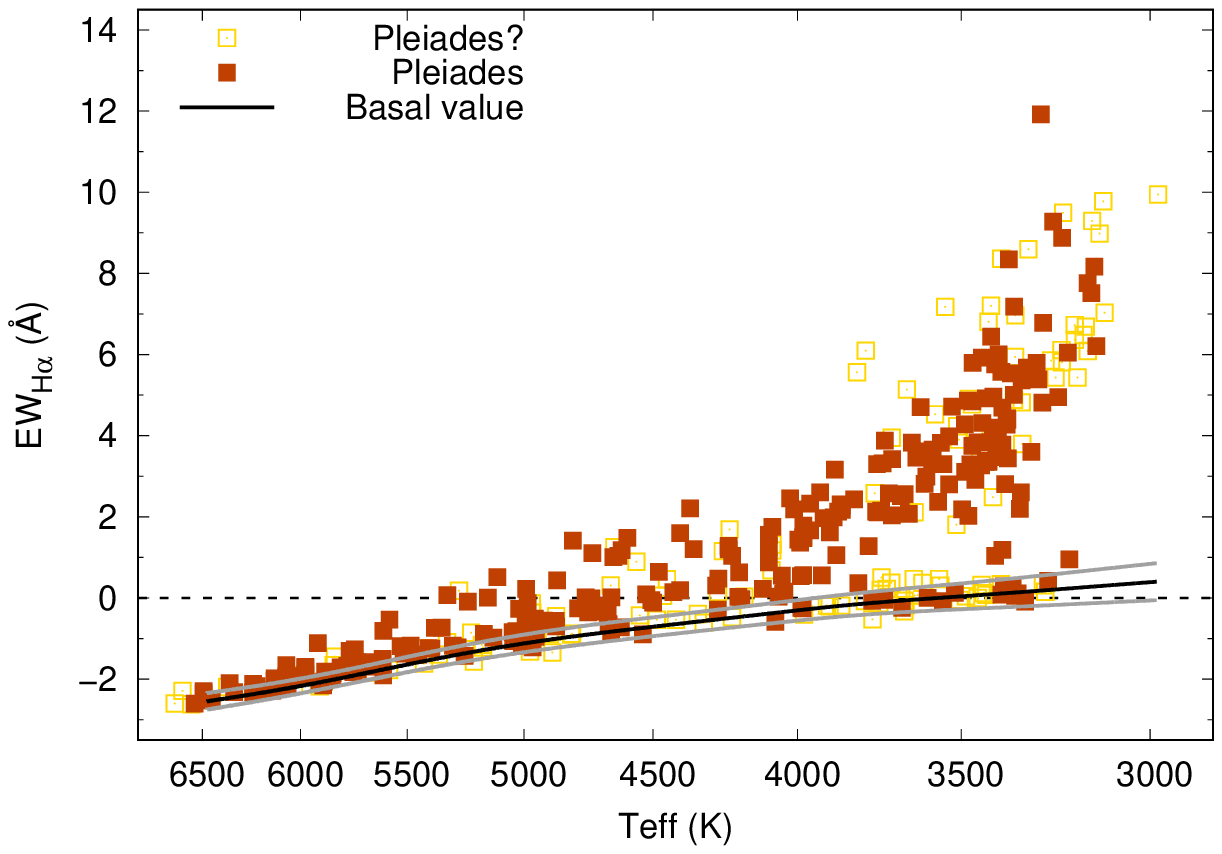}
\includegraphics[width=\columnwidth]{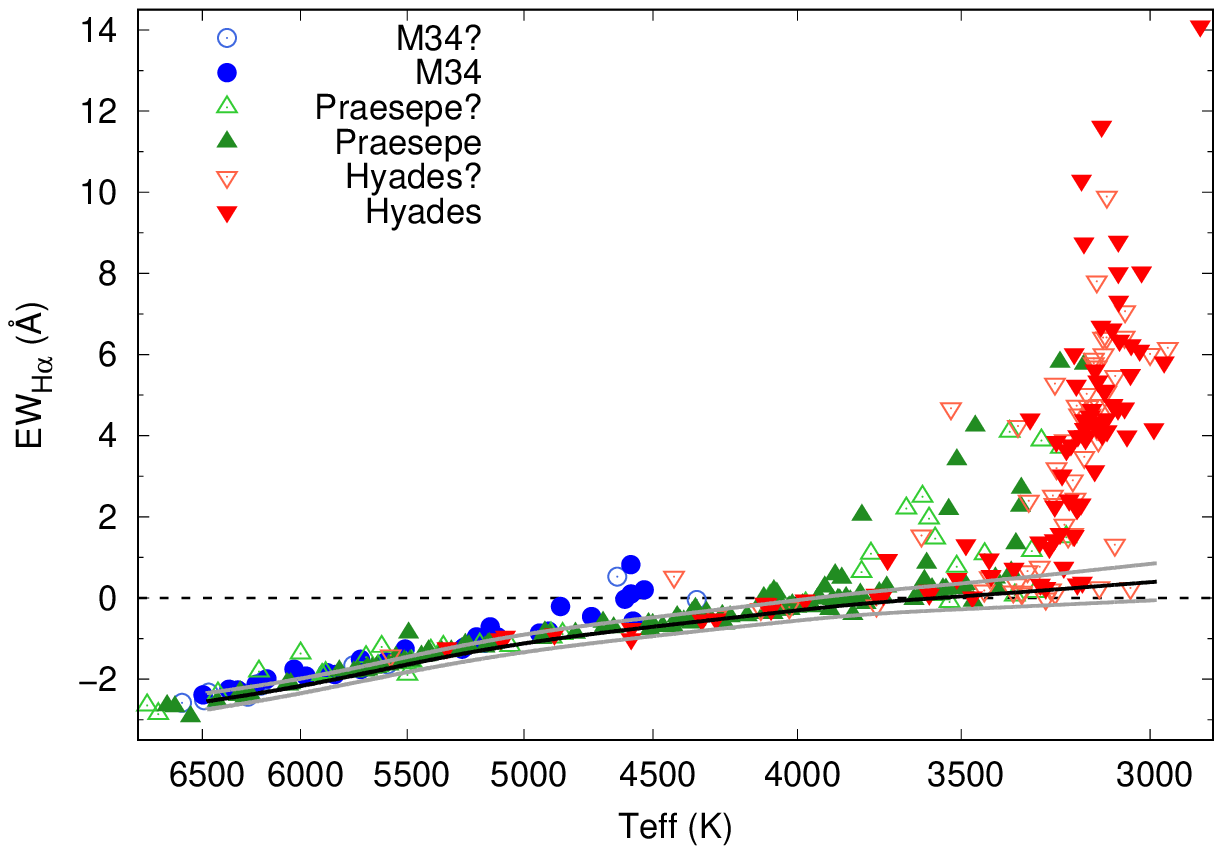}
\includegraphics[width=\columnwidth]{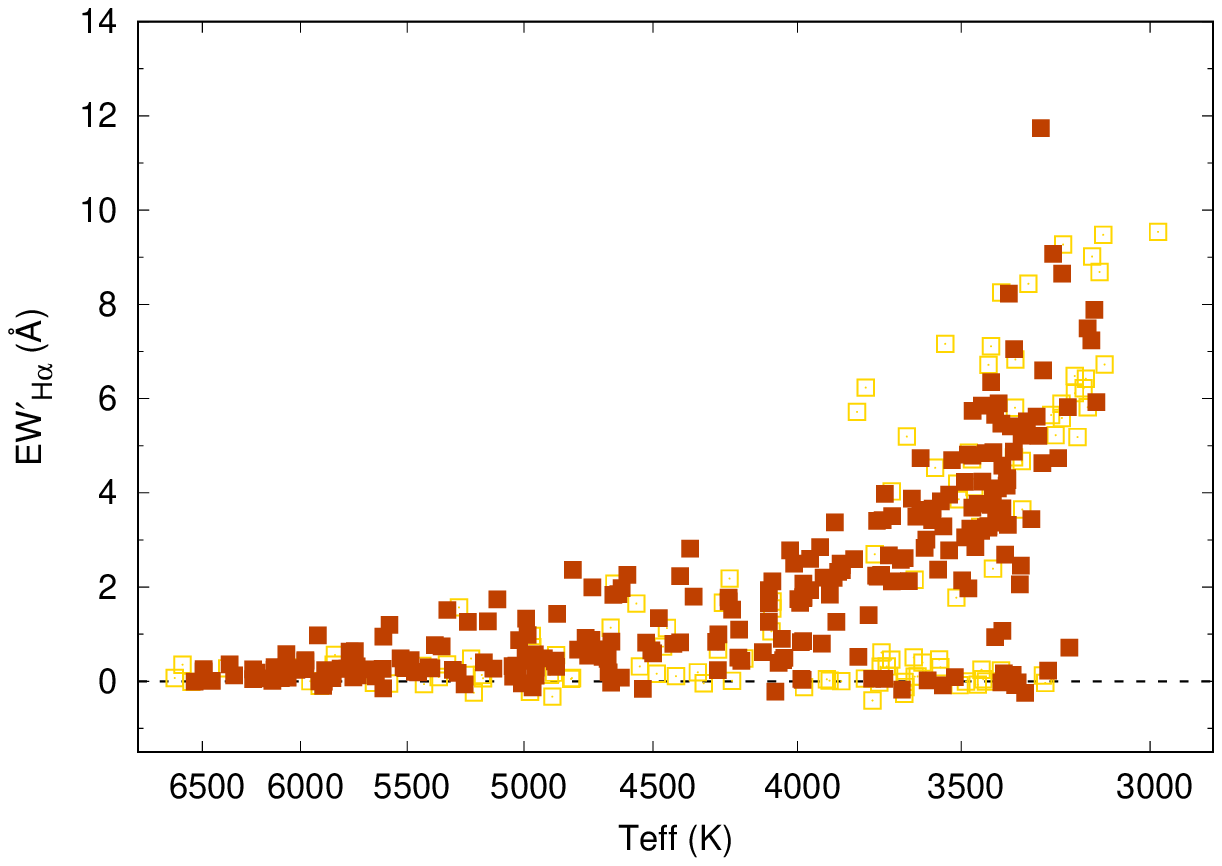}
\includegraphics[width=\columnwidth]{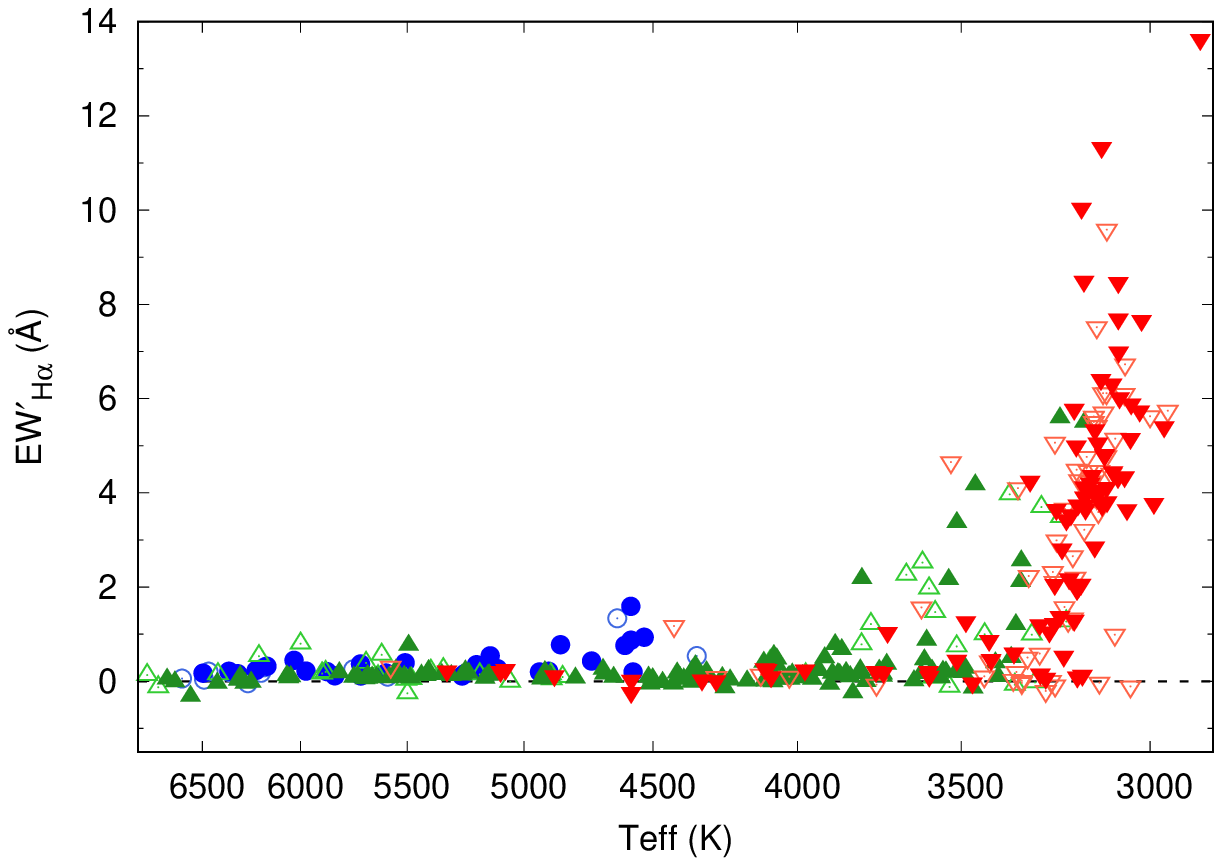}
\includegraphics[width=\columnwidth]{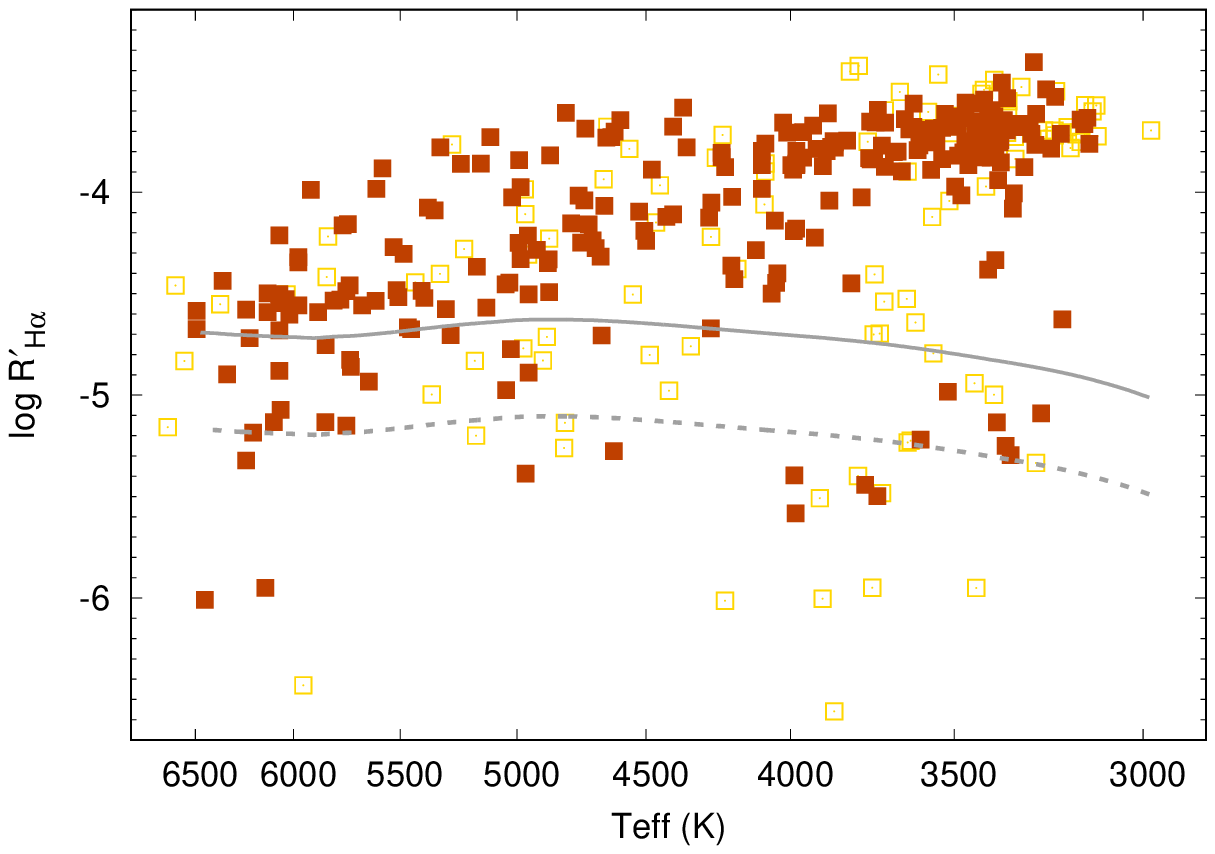}
\includegraphics[width=\columnwidth]{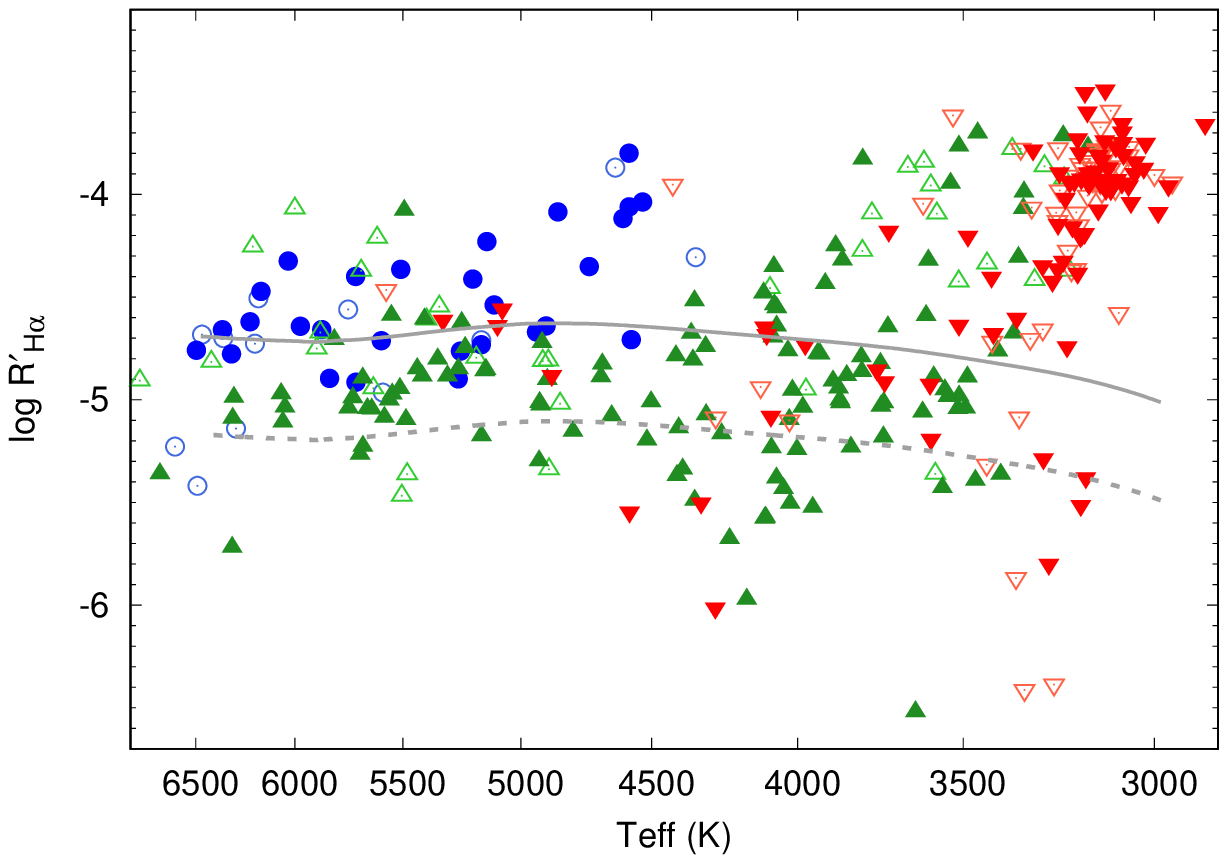}
\caption{Top: The equivalent widths of H$\alpha$ line for stars in Pleiades (left), and M 34, Praesepe and Hyades (right). Black solid lines show the mean 
relations for inactive reference stars, upper and lower 3$\sigma$ to the mean value 
are shown as grey lines. Filled symbols represent probable single members and open symbols denote probable binary members or non-members. For 
stars having LAMOST multi-epoch observations, we displayed their average values rather than individual observational values. Middle: The excess 
equivalent width of H$\alpha$, EW$^{'}_{\text{H}\alpha}$. 
Bottom: The excess fractional luminosity of H$\alpha$, $R^{'}_{\text{H}\alpha}$, where the grey solid (dashed) lines denote the stars locating in the upper 3$\sigma$ ($\sigma$) regime.}
\label{fig:ewha_teff}
\end{figure*}

As shown in Fig.~\ref{fig:ewha_teff}, the strong change of EW$_{\text{H}\alpha}$ (\& EW$^{'}_{\text{H}\alpha}$) with temperature does not only 
reflect the intrinsic change of chromospheric activity, but also the effects of the drop in the nearby continuum level. To remove the dependency of the measured 
EW$_{\text{H}\alpha}$ on the surrounding continuum level, and to compare the strength of chromospheric activity in different spectral types, we used the fractional luminosity of H$\alpha$ (normalized with the bolometric luminosity) \citep[e.g.][]{reid1995,west2004}. However, quantifying fractional luminosity for a given line requires a flux-calibrated spectrum and the distance. For all activity indicators in this work, 
the excess fractional luminosities were derived from the excess equivalent widths using a distance-independent value, $\chi$, 
the ratio between the continuum flux near the line of interest and the bolometric flux, following commonly used $\chi$-methods \citep[e.g.][]{west2004}. 
Using the $\chi$ ratios of H$\alpha$ line (${\chi}_{\text{H}\alpha}$) obtained from the model spectra (see Appendix~\ref{sec:chi}), 
we obtained the H$\alpha$ excess fractional luminosity (denote as $R^{'}_{\text{H}\alpha}$), 
the ratio of excess H$\alpha$ emission luminosity ($L^{'}_{\text{H}\alpha}$) to the bolometric luminosity ($L_{\text{bol}}$), as follows, 
\begin{equation}
R^{'}_{\text{H}\alpha} \equiv \frac{L^{'}_{\text{H}\alpha}}{L_{\text{bol}}}= {\chi}_{\text{H}\alpha} \times \text{EW}^{'}_{\text{H}\alpha}.
\label{equ:exlum}
\end{equation}
In similar way, we estimated Ca~{\sc ii} K and H$\beta$ excess fractional luminosities, $R^{'}_{\text{Ca K}}$ and $R^{'}_{\text{H}\beta}$, 
respectively (see Appendix~\ref{sec:otherlines}). These values provide the fraction of respective bolometric 
luminosity emitted from the chromosphere in the lines, and are analogous to the chromospheric Ca {\sc ii} emission ratio $R^{'}_{\text{HK}}$ 
\citep[e.g.][]{noye1984} and $R^{'}_{\text{IRT}}$ \citep[e.g.][] {mars2009,jack2010}.

\subsection{Comparison with literature measurements}
We compared our EW$_{\text{H}\alpha}$ measurements with \citet{doug2014}, wherein they measured EW$_{\text{H}\alpha}$ from medium resolution 
(about two times higher than LAMOST) spectra, for common stars in Hyades and Praesepe (see Fig.~\ref{fig:ewha_com}), 
and found to be in good agreement. Furthermore, \citet{sode1993} and \citet{sode2001} obtained excess H$\alpha$ emission from high-resolution echelle spectra 
for about 100 FGK dwarfs in the Pleiades and about 50 FGK dwarfs in the M34, respectively, using a synthetic subtraction technique, i.e., 
the H$\alpha$ profile of each target star is subtracted by the rotationally broadened H$\alpha$ of inactive counterpart.  
In Fig.~\ref{fig:rha_com} we compared their measurements of excess H$\alpha$ emission with our measurements for common stars in the Pleiades and M34, 
and found they are in close agreement. However, it seems there still exist systematic offset in fractional luminosity for less active hotter stars with $\log R^{'}_{\text{H}\alpha} \lesssim-4.5$, e.g., our measured $R^{'}_{\text{H}\alpha}$ are $\sim$0.2 dex larger. In fact, as discussed in Appendix~\ref{sec:errors}, 
a small error in excess equivalent width would result in larger error in fractional luminosity (e.g., see Table~\ref{tab:mc}), 
particularly when the excess emission is very weak. In a careful check on the excess equivalent width, 
we indeed found there is a small offset ($\sim$0.1~\AA, see the upper panel of Fig.~\ref{fig:rha_com}) that is comparable 
with the uncertainties due to 100 K uncertainty in temperature (see Table~\ref{tab:mc}). 

\subsection{Variability in chromospheric emission}
\begin{figure}
\centering
\includegraphics[width=\columnwidth]{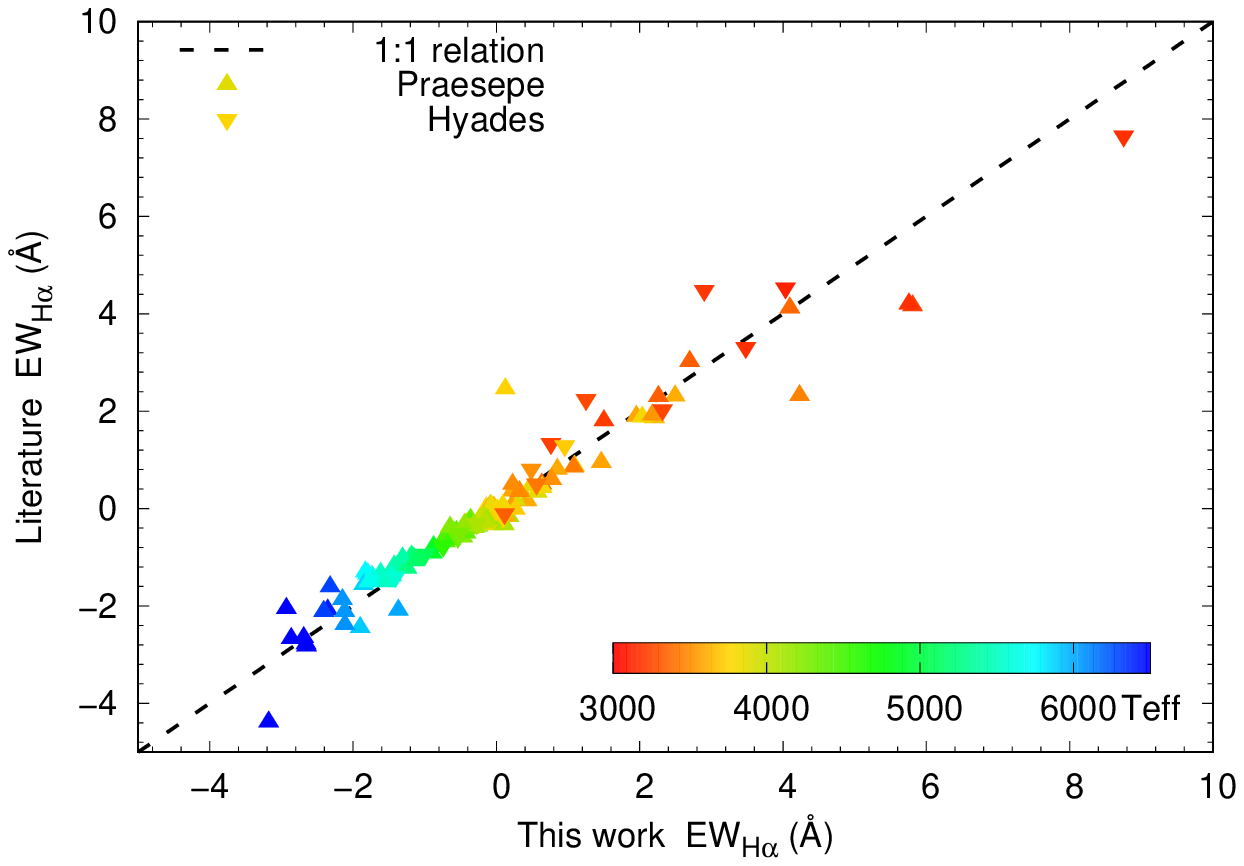}
\caption{Comparison of our measured EW$_{\text{H}\alpha}$ with those of \citet{doug2014} for common stars in Hyades and Praesepe.}
\label{fig:ewha_com}
\end{figure}

\begin{figure}
\centering
\includegraphics[width=\columnwidth]{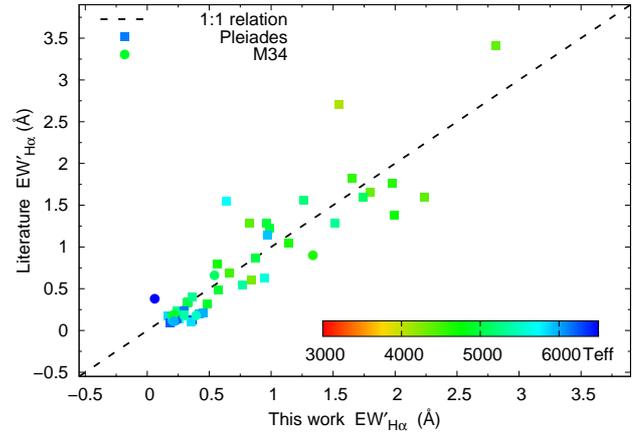}
\includegraphics[width=\columnwidth]{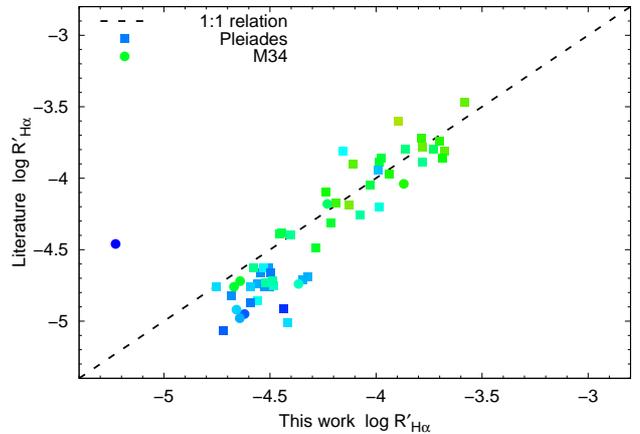}
\caption{Comparison of H$\alpha$ excess emissions with those of \citet{sode1993} (Pleiades members) and \citet{sode2001} (M34 members) for common stars.}
\label{fig:rha_com}
\end{figure}

\begin{figure}
\centering
\includegraphics[width=\columnwidth]{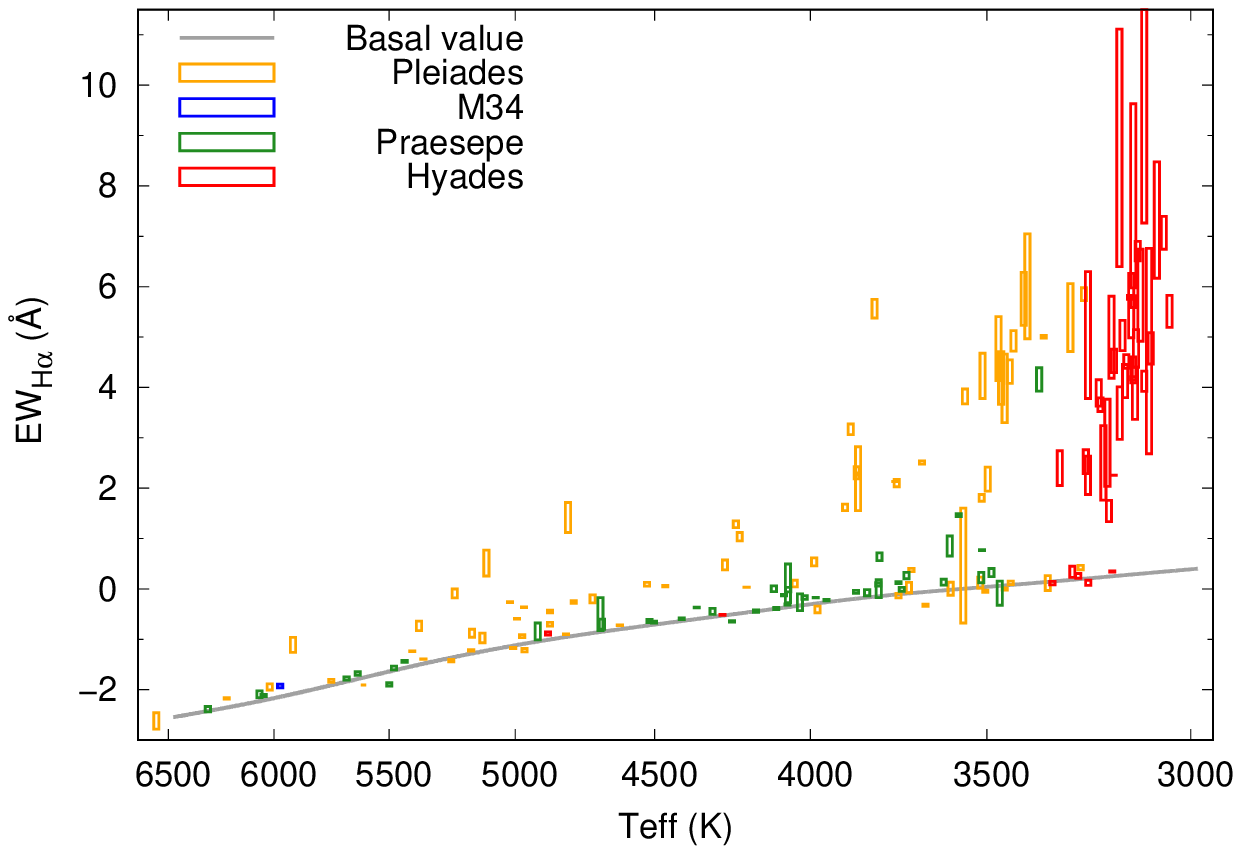}
\caption{Variation of EW$_{\text{H}\alpha}$ for stars having multi-epoch observations. Also shown by grey-colour solid line are the basal values. }
\label{fig:vdewha}
\end{figure}

The larger scatter in Fig.~\ref{fig:ewha_com} and Fig.~\ref{fig:rha_com} are ofcourse partly from the uncertainties of our measurements. 
However, we believe that they are mainly related to intrinsic variation in activity, which is well-known for Pleiades members \citep[e.g.][]{sode1993}. 
To further check the variation in H$\alpha$ emission, we have plotted EW$_{\text{H}\alpha}$ against the temperature in Fig.~\ref{fig:vdewha}, in which we 
show minimum and maximum values of EW$_{\text{H}\alpha}$ for stars having multi-epoch spectra. From the figure it is clear that very cool members show high 
variation in H$\alpha$ emission, which is evidently larger than H$\alpha$ equivalent width measurement errors, indicating 
intrinsic variability of chromospheric activity. Also, it shows that more active stars have large equivalent width variations, indicating stars 
with high level activity suffer more often large variations, which maybe due to short-time scale evolution of active regions or flare-like events.   
%%
%%%%%%%%%%%%%%%%%%%%%%%%%%%%
\section{Results and Discussion} 
\subsection{Chromospheric activity as a function of mass} 
As shown in top-left panel of Fig.~\ref{fig:ewha_teff}, the EW$_{\text{H}\alpha}$ is generally a strong function of stellar effective temperature, 
i.e., EW$_{\text{H}\alpha}$ increases with decreasing temperature, from H$\alpha$ absorption at $T_{\text{eff}}\sim6500$ K (EW$_{\text{H}\alpha}\sim-2$~\AA) 
to H$\alpha$ emission at $T_{\text{eff}}\sim3000$ K (EW$_{\text{H}\alpha}\sim10$~\AA). Such trend in the Pleiades was also observed by \citet{stau1987}. 
The upper envelop of EW$_{\text{H}\alpha}$ vs. $T_{\text{eff}}$ profile indicates the activity levels of most active stars at a given temperature, 
provides the clues on chromospherically cooling effects for stars with different mass. Also, it shows the dependency of the upper envelop on spectral 
types among Pleiades members, e.g., the EW$_{\text{H}\alpha}$ increases moderately as temperature decreases for late-F and G-type stars ($T_{\text{eff}}\gtrsim5000$ K), and for K-type stars it goes into a plateau where exists a slight increment of EW$_{\text{H}\alpha}$, and then it increases sharply when 
$T_{\text{eff}}\lesssim4000$ K. However, at the same time, the spread in EW$_{\text{H}\alpha}$ also increases. 
Such feature is more overt in the middle-left panel of Fig.~\ref{fig:ewha_teff}. The lower EW$_{\text{H}\alpha}$ of Pleiades members 
follow the mean relations of inactive reference stars, which vary with temperature, i.e., from EW$_{\text{H}\alpha}\sim-2.5$~\AA~at 
$T_{\text{eff}}\sim6500$ K to EW$_{\text{H}\alpha}\sim0$~\AA~at $T_{\text{eff}}\sim3800$ K.

\begin{figure}
\centering
\includegraphics[width=\columnwidth]{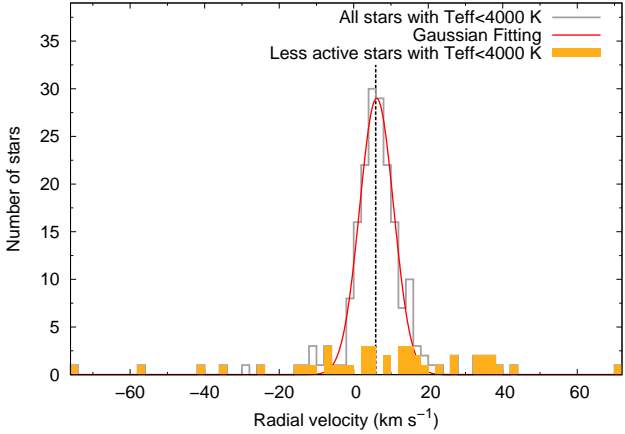}
\caption{RV distributions of Pleiades candidate members with $T_{\text{eff}}<4000$ K in inactive regime, shown by the grey line. 
The red line is a Gaussian fitting, which gives $\mu\approx6.2~km\,s^{-1}$ and $\sigma\approx4.6~km\,s^{-1}$. 
The expected Pleiades RV at 5.9 $km\,s^{-1}$ provided by \citet{merm2009} is marked as a black dashed line.}
\label{fig:rv_lessactive}
\end{figure}

We noticed that some Pleiades G-type members have larger EW$_{\text{H}\alpha}$ values compared to inactive counterparts, 
indicating they have shallower H$\alpha$ absorption line profile (i.e., the core of the line is filled in). 
The excess H$\alpha$ emission for all clusters are shown in middle panels of Fig.~\ref{fig:ewha_teff}. 
It is clear from the figure that most of Pleiades candidates show excess emission 
compared to inactive counterparts, i.e., EW$^{'}_{\text{H}\alpha}>0$, 
indicating they are in chromospherically active phase, which is consistent with its young age. However, a small fraction of Pleiades stars have weaker excess 
H$\alpha$ emission compared to other stars in the same spectral type, e.g., about 40 very low-mass stars with $T_{\text{eff}}$<4000 K have 
EW$^{'}_{\text{H}\alpha}<0.5$~\AA, while other members in this temperature regime have evident larger 
EWs (e.g. EW$^{'}_{\text{H}\alpha}$>2~\AA), making a gap around $T_{\text{eff}}\sim3500$ K in the diagram. 
One general interpretation is that these stars may be field dwarfs rather than Pleiades members. 
We found that about one third of these stars are photometrically single candidate members according to their location on CMDs. 
To further check their membership, we plotted their RVs distribution in Fig.~\ref{fig:rv_lessactive}, and found about half of them have 
RVs from -8 to 20 $km\,s^{-1}$  suggesting they are likely Pleiades members within 3$\sigma$-RV regime by adopting the mean RV value of $\mu\sim$6 $km\,s^{-1}$ 
and a scatter of $\sigma\sim$5 $km\,s^{-1}$ for Pleiades (see Fig.~\ref{fig:rv_lessactive}). Several stars (e.g., about one sixth of them) are 
probably Pleiades members which pass both the CMD and RV criterion, such as BPL167, DH 668 and DH 908. 
On the other hand, we noticed that all these less active stars have no rotation periods available till the date. 
In other words, stars having detected rotation periods show larger activity in the cooler end, which makes the issue complex. 
Therefore, their memberships and potential light variation through monitoring indeed need to be further investigated.  

The large contrast of continuum flux at H$\alpha$ line between M-type and hotter stars dilutes their larger differences in equivalent widths, 
which makes the upper envelops of $R^{'}_{\text{H}\alpha}$ among these stars have similar values, e.g., $\log R^{'}_{\text{H}\alpha}\sim-3.7$ 
(see bottom panels of Fig.~\ref{fig:ewha_teff}). Still, the upper envelop of $R^{'}_{\text{H}\alpha}$ in Pleiades M-type stars (e.g., $T_{\text{eff}}\lesssim3900$ 
K) show a slight increment with decreasing effective temperature, confirms the previous finding that the low-mass stars in young clusters with age less 
than 150 Myr show an increase in activity strength ($L_{\text{H}\alpha}/L_{\text{bol}}$) toward lower mass \citep{hawl1999}. 
Also, previous studies show that later spectral type stars have lower activity level among late M-type stars \citep[e.g.][]{west2004}. 
Though such trends are evident from $R^{'}_{\text{H}\alpha}$ when $T_{\text{eff}}\lesssim$3300 K, and $R^{'}_{\text{H}\beta}$ (see Fig.~\ref{fig:ewxx}), 
we can not make a solid conclusion about this issue based on current data.

Another remarkable feature in Fig.~\ref{fig:ewha_teff} is a large scatter of $R^{'}_{\text{H}\alpha}$ among hotter Pleiades members. 
In particular, there evidently are two sequences among GK-type Pleiades candidates, i.e., one sequence of stars have similar activity strengths, 
and other sequence being with lower activity strength showing an increment towards lower temperature. The largest difference of activity level between 
these two sequences, e.g., 0.5-0.6 dex in $\log R^{'}_{\text{H}\alpha}$ (a factor of about four in $R^{'}_{\text{H}\alpha}$), occurs at about 5000 K. 
Such a difference in activity levels is unlikely due to measurement uncertainties, indicating bi-sequence behaviour among these stars. 
Therefore, apart from the stellar mass, other factors may play a role in producing the observed dispersion pattern. In fact, we have noticed these two 
sequences correspond to rotational sequences (see Section~\ref{sec:rot-act}).

It is known that the chromospheric activity depends not only on stellar mass but also on age, e.g., the $\text{H}\alpha$ emission lasts longer in cooler stars 
\citep{stau1991,hawl1999}. Indeed, Fig.~\ref{fig:ewha_teff} shows a clear age effect on chromospheric activity, i.e., compared to Pleiades, Hyades and Praesepe are 
typically less active, and M34 with intermediate age between Pleiades and Praesepe/Hyades shows activity levels in-between them. Additionally, the 
onset of H$\alpha$ direct emission occurs at about $T_{\text{eff}}\sim$5000-5500, 4500-5000 K, 4000-4500 K for Pleiades, 
M34 and Hyades/Praesepe members, respectively. Furthermore, two activity sequences appears among 
cooler members (e.g., late-K and M-type stars) in Praesepe and Hyades, similar to the feature observed for GK-type members in Pleiades.
Interestingly, the members of Praesepe and Hyades with $T_{\text{eff}}$ less then 3600 K (spectral type later than M1-M2) have activity levels 
similar to those of Pleiades active members, indicates the activity lifetime for M1-M2 type stars is around 600-700 Myr, a value agree with that reported 
by \citet{west+2008}. It is possible to get such kind of activity-timescale for earlier spectral type stars by comparing M34 with Pleiades, 
however, we could not able to provide any solid conclusion due to limited sample of M34 candidates in our data.  

\subsection{Rotation-activity connection}
\subsubsection{Rotation and activity sequences}
\label{sec:rot-act}
\begin{figure*}
\centering
\includegraphics[width=\columnwidth]{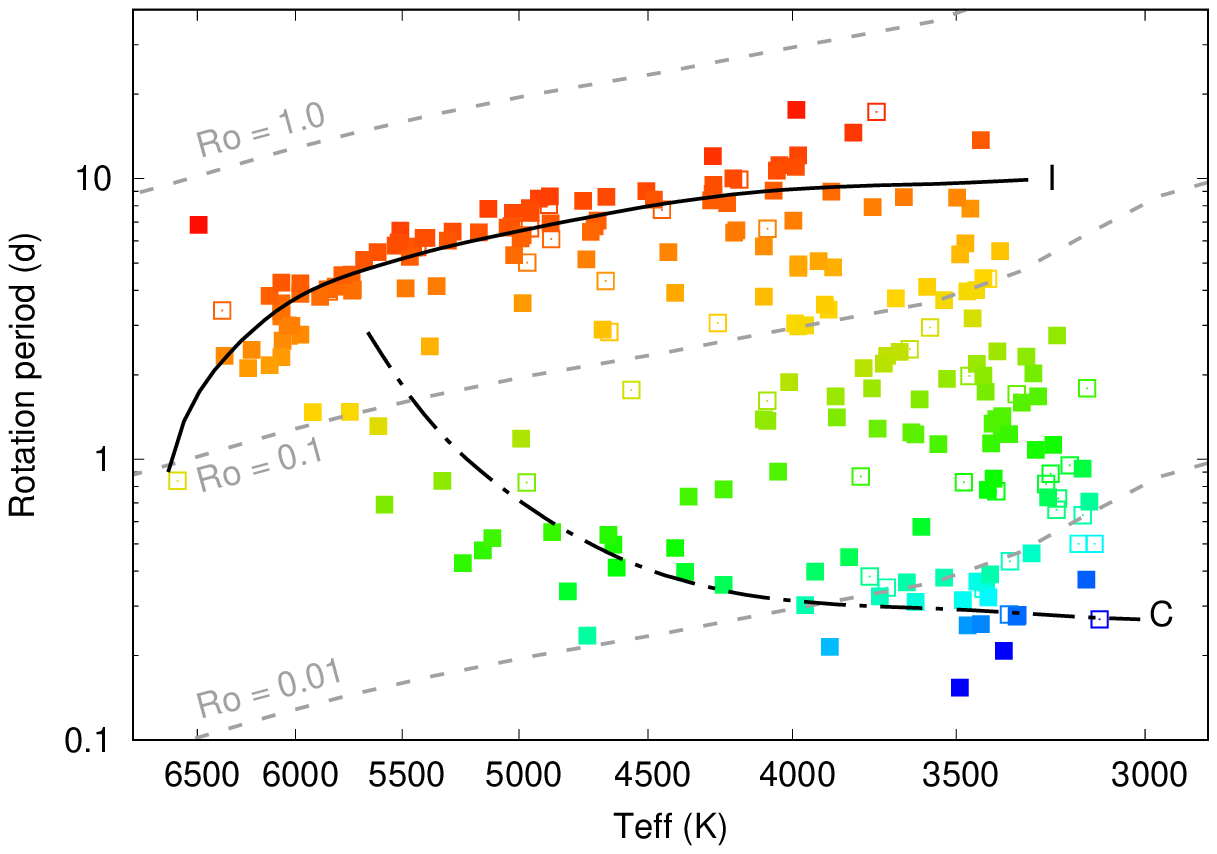}
\includegraphics[width=\columnwidth]{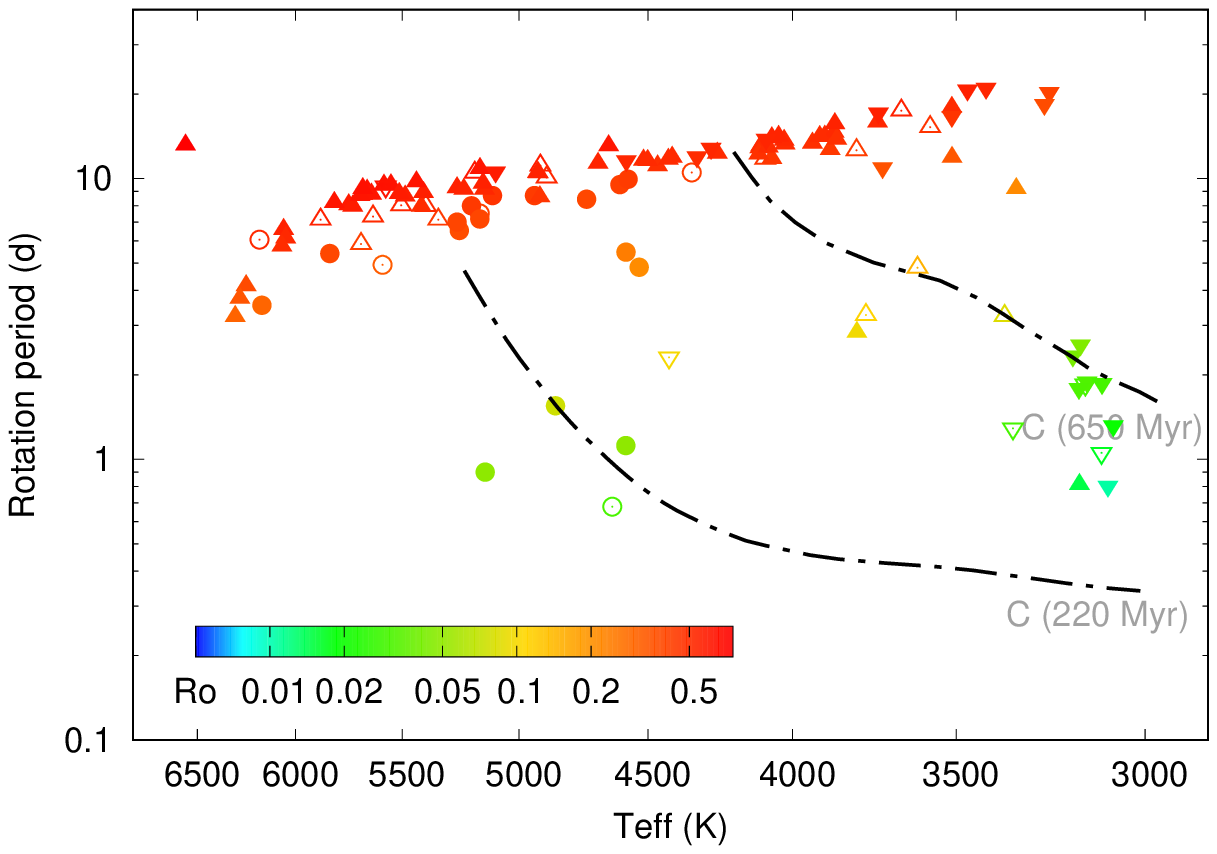}
\includegraphics[width=\columnwidth]{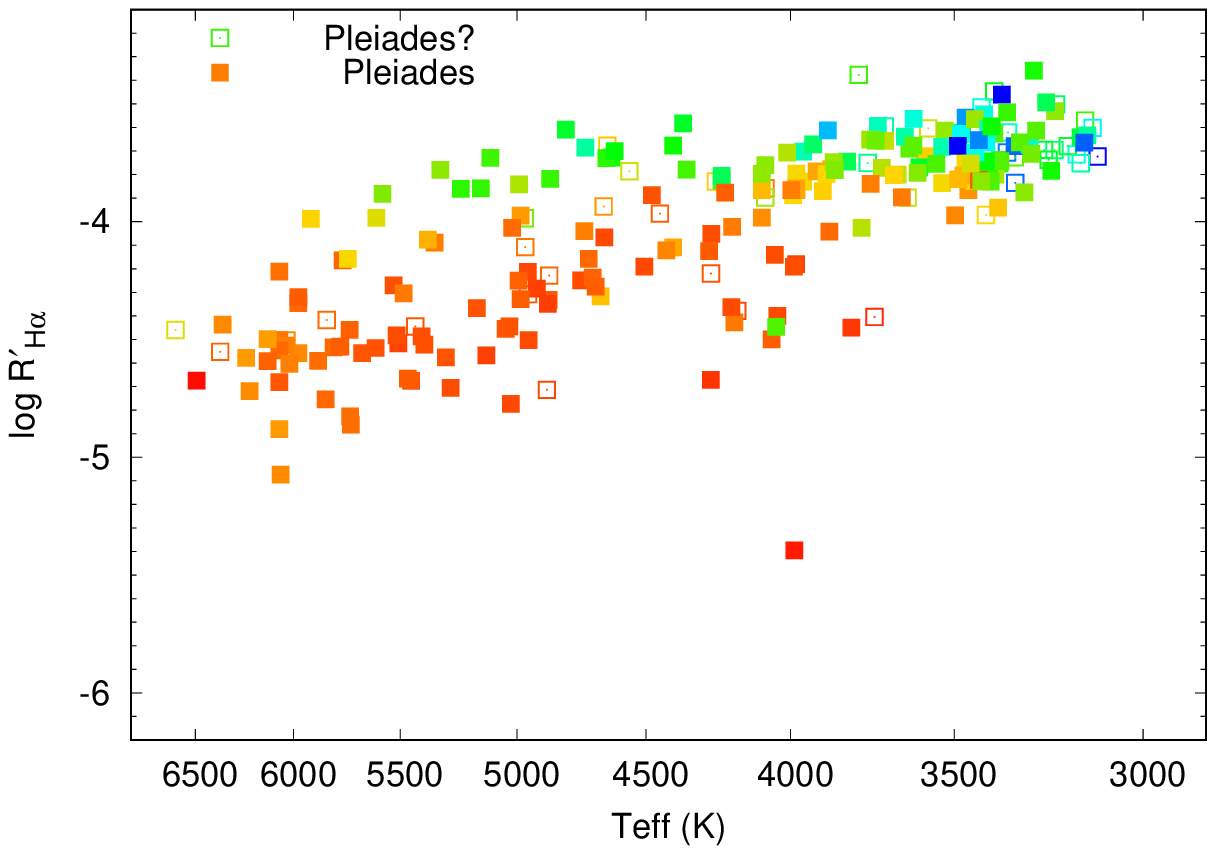}
\includegraphics[width=\columnwidth]{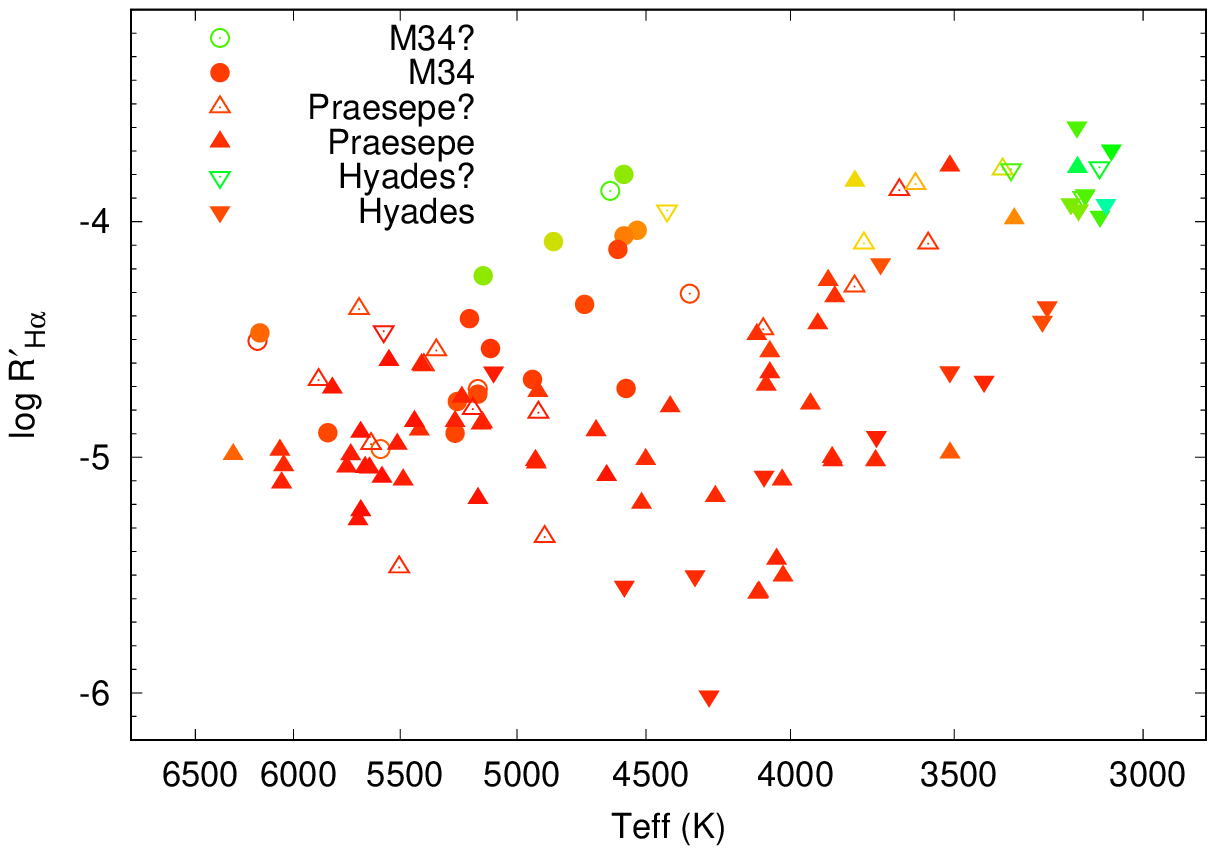}
\includegraphics[width=\columnwidth]{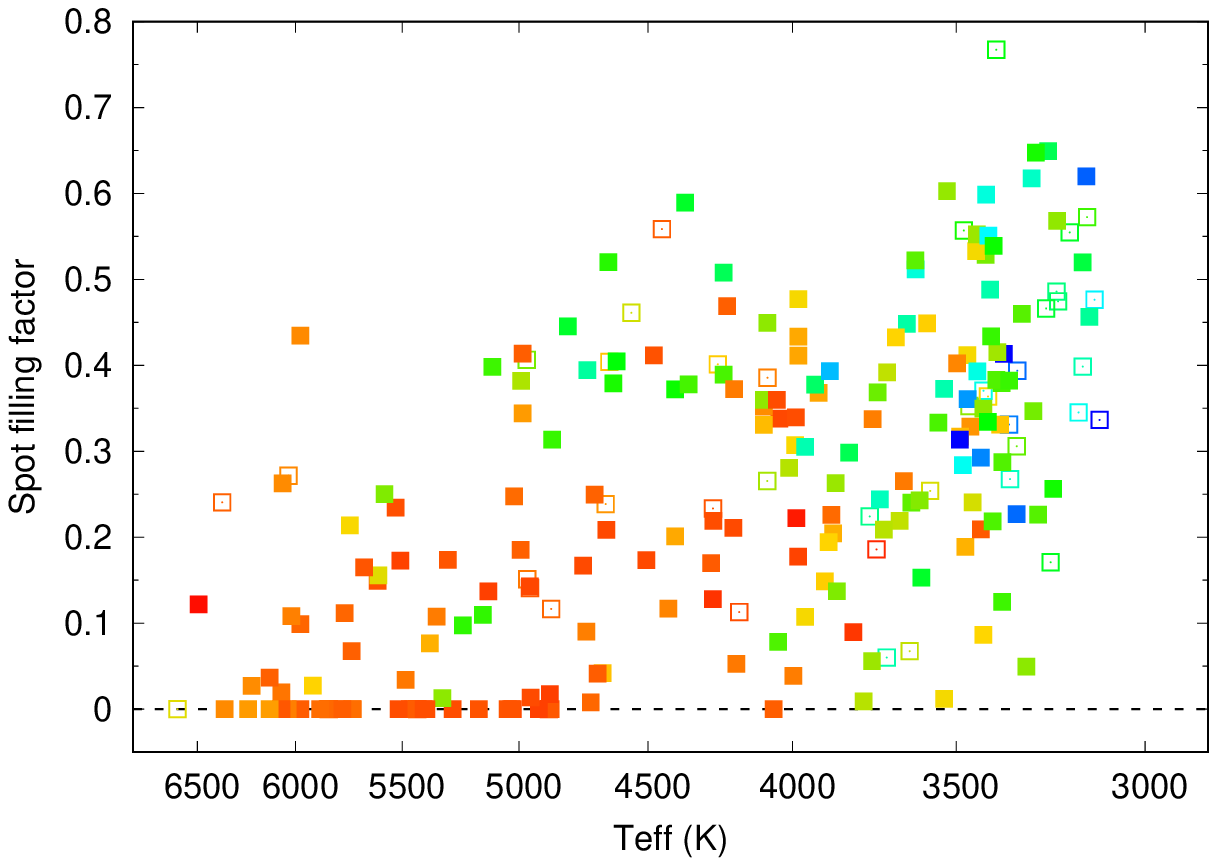}
\includegraphics[width=\columnwidth]{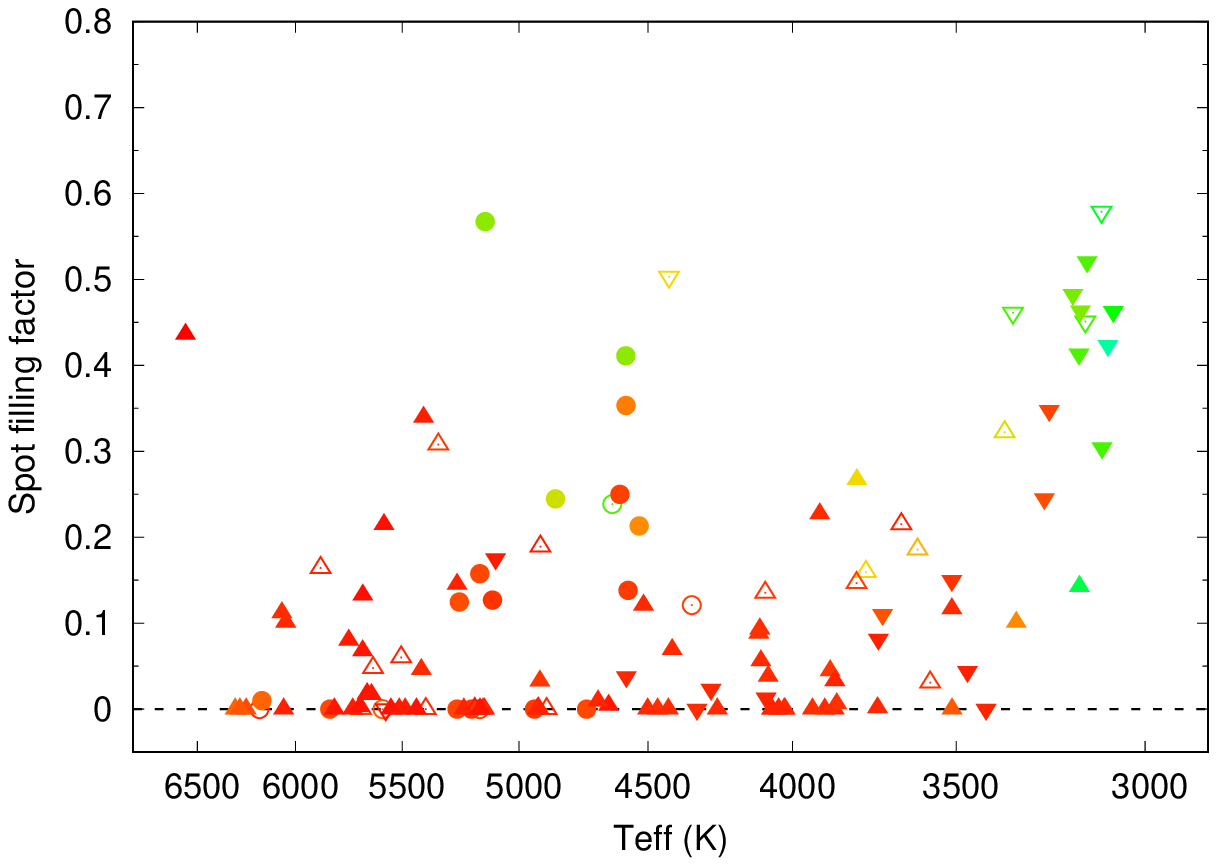}
\caption{Rotation period (top), $R^{'}_{\text{H}\alpha}$ (middle), and spot filling factor (bottom) as a function of $T_{\text{eff}}$ for stars with 
rotation period available. Left: Pleiades. Right: M34, Praesepe and Hyades. The colour gradient indicates Rossby numbers. Over-plotted in top-right panel 
are contours of constant Rossby numbers (Ro=0.01, 0.1 and 1.0; grey dashed lines) based on the empirically calibrated mass-convective overturn time 
relation by \citet{wrig2011}. The canonical rotational I/C sequences are illustrated by black lines in top panels.}
\label{fig:ha_rot}
\end{figure*}

As discussed above, there exist two sequences among GK-type stars in Pleiades, and M-type stars in Praesepe and Hyades 
in $R^{'}_{\text{H}\alpha}$-$T_{\text{eff}}$ diagram, i.e., H$\alpha$ emissions in Pleiades begin to bifurcate at $\sim$6000 K and merge again at 
$\sim$4000 K; similar sequences in Praesepe and Hyades appear at $\sim$4000 K and disappear at about 3100 K. In order to understand further on this 
scenario, we provide $R^{'}_{\text{H}\alpha}$ as a function of temperature again in Fig.~\ref{fig:ha_rot}, wherein shown only members with known rotation 
periods. For comparison purpose, we also plot the corresponding rotation period ($P_{\text{rot}}$) distribution in the upper panels, where the 
$P_{\text{rot}}$ values were collected from the literature (see Appendix~\ref{sec:rotation}). The colour contrast in Fig.~\ref{fig:ha_rot} denotes the Rossby 
Number Ro (Ro$=P_{\text{rot}}/\tau$, normalized rotation period by the convective turnover time $\tau$). The $\tau$ was estimated from stellar mass (mass 
were converted from the quiescent photosphere temperature using PARSEC models, see Appendix~\ref{sec:teffq}) using the correlation between convective 
turnover time and mass, derived by \citet{wrig2011}. It is clear that stars in the upper sequence rotate more faster than those in the lower activity 
sequence.  

\begin{figure*}
\centering
\includegraphics[width=\columnwidth]{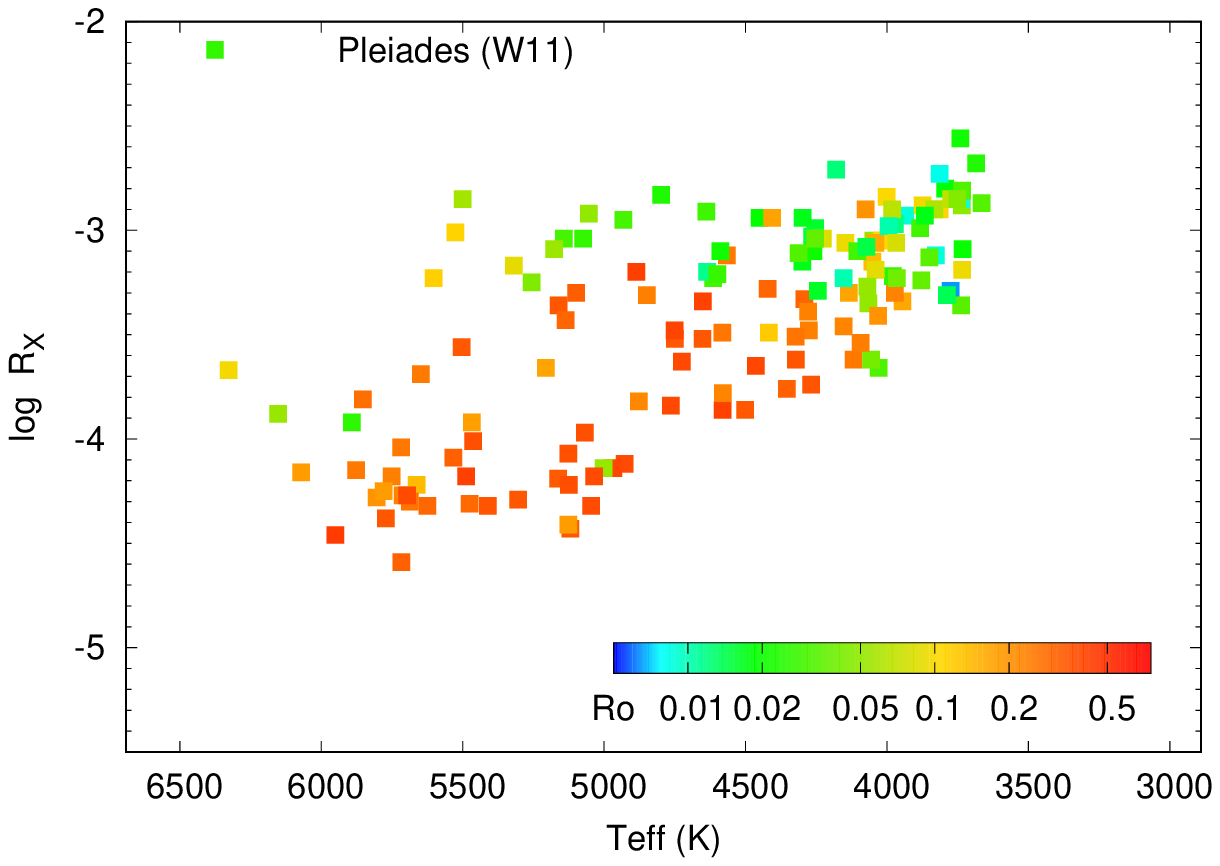}
\includegraphics[width=\columnwidth]{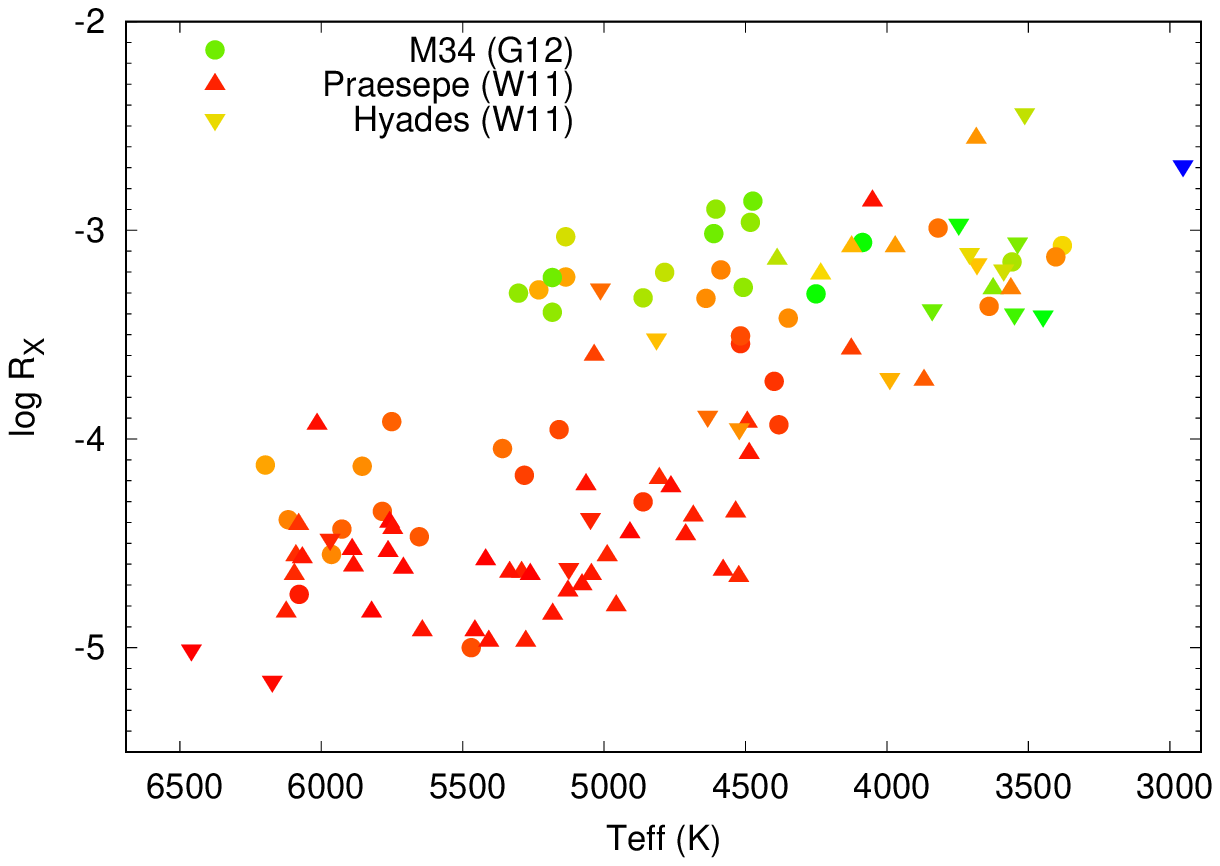}
\caption{Fractional X-ray emission luminosities $R_{\text{X}}$ as a function of $T_{\text{eff}}$ for stars in Pleiades (left), 
and M34, Praesepe and Hyades (right). These data were collected from literature, as labelled in the panels, where W11: \citet{wrig2011}, G12: \citet{gond2012}.}
\label{fig:lx_rot}
\end{figure*}

We noticed that there exist similar features in cool starspot coverages (spot filling factors), as shown in the lower panels of Fig.~\ref{fig:ha_rot}, 
where the spot filling factors were estimated by modelling their TiO band near 7050~\AA~(TiO2) using the standard values obtained based on references stars with 
solar metallicity (see Paper I for more details). 
It is clear that there exists a trend for fast rotators among K-type stars in Pleiades having larger spot coverages, 
and the spot coverages belong to two branches among GK-type Pleiades members. 
Further, we collected X-ray emission data from literature \citep{wrig2011,gond2012} for stars in these open clusters, 
and found the existence of similar feature with dual sequences, as shown in Fig.~\ref{fig:lx_rot}. 

It is known that the open cluster members locate in three well known regions in the rotation-color diagram \citep{barn2003a}, i.e., I (Interface) and C 
(Convective) sequence stars and gap stars between I and C sequence (e.g., see the top-left panel of Fig.~\ref{fig:ha_rot}). From Fig.~\ref{fig:ha_rot}, 
we can get a general conclusion that the activity level sequences correspond to the rotation sequences, namely, the more active sequence stars correspond to 
the C sequence stars, and the stars in less active sequence map to I sequence stars. The dynamo configuration in I sequences may be different with that in C 
sequences, e.g., a interface dynamo dominates in I sequence stars, and a turbulent or convective dynamo works in C sequence stars, correspondingly causing a 
large-scale interface magnetic field and a small-scale turbulent field \citep{barn2003a,barn2003b}. Therefore, such a bifurcation of chromospheric activity of 
Pleiades GK-type members is probably due to a change in the dynamo configuration. For older open clusters such as Hyades with an age of 600-700 Myr their 
GK-type member stars rotate slowly and should be in I sequence in the period-colour diagram, 
but the M members display clearly both I and C sequence (see the top-right panel of Fig.~\ref{fig:ha_rot}). As mentioned above, indeed there is 
a presence of two H$\alpha$ emission sequences for early and Mid-M stars in Praesepe and Hyades. Moreover, it is clear that the two activity sequences 
correspond to I- and C-rotation sequences, as shown in Fig.~\ref{fig:ha_rot}. For M34, an open cluster with an age of about 220 Myrs, the two activity 
sequences should be observed among K-type stars. However, we could not notice such feature due to lack of observational data of the M34 K-type stars. 

\subsubsection{Rotation-activity correlation}
\begin{figure*}
\centering
\includegraphics[width=\columnwidth]{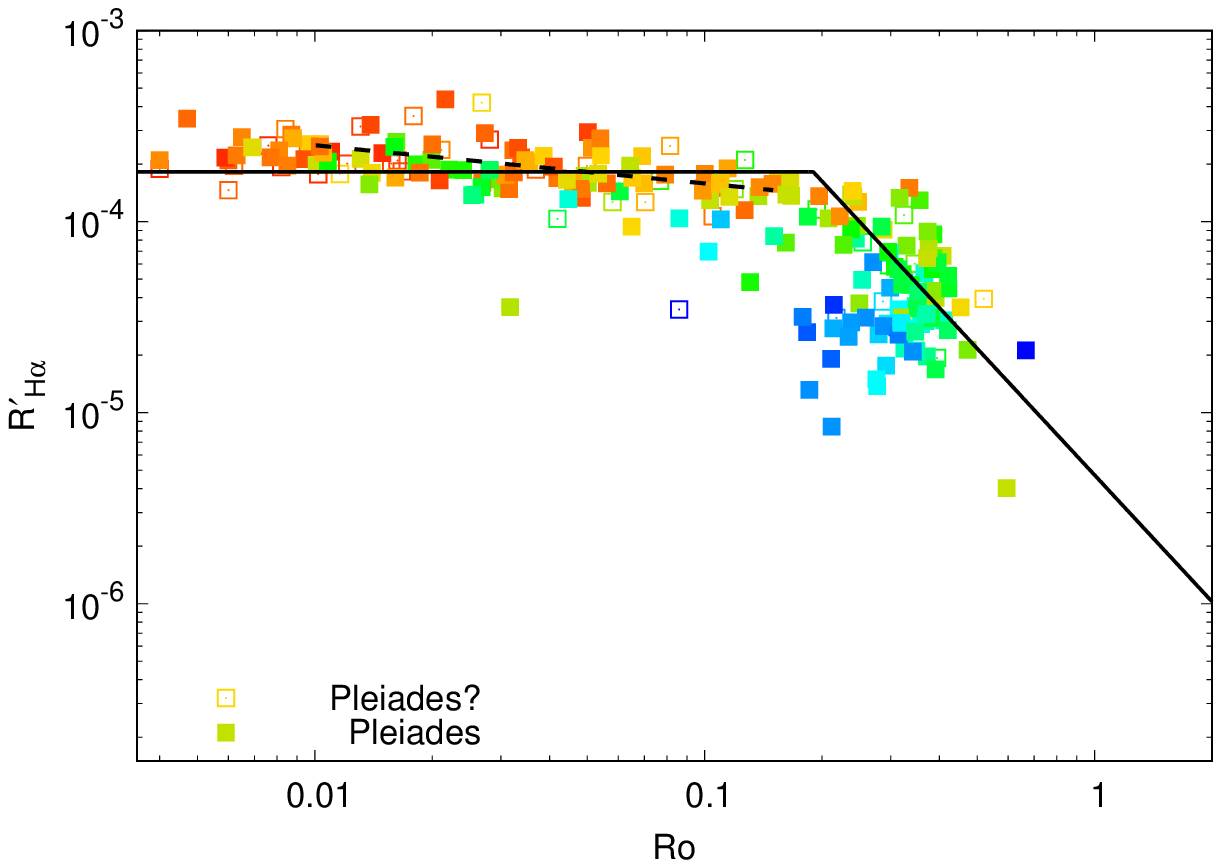}
\includegraphics[width=\columnwidth]{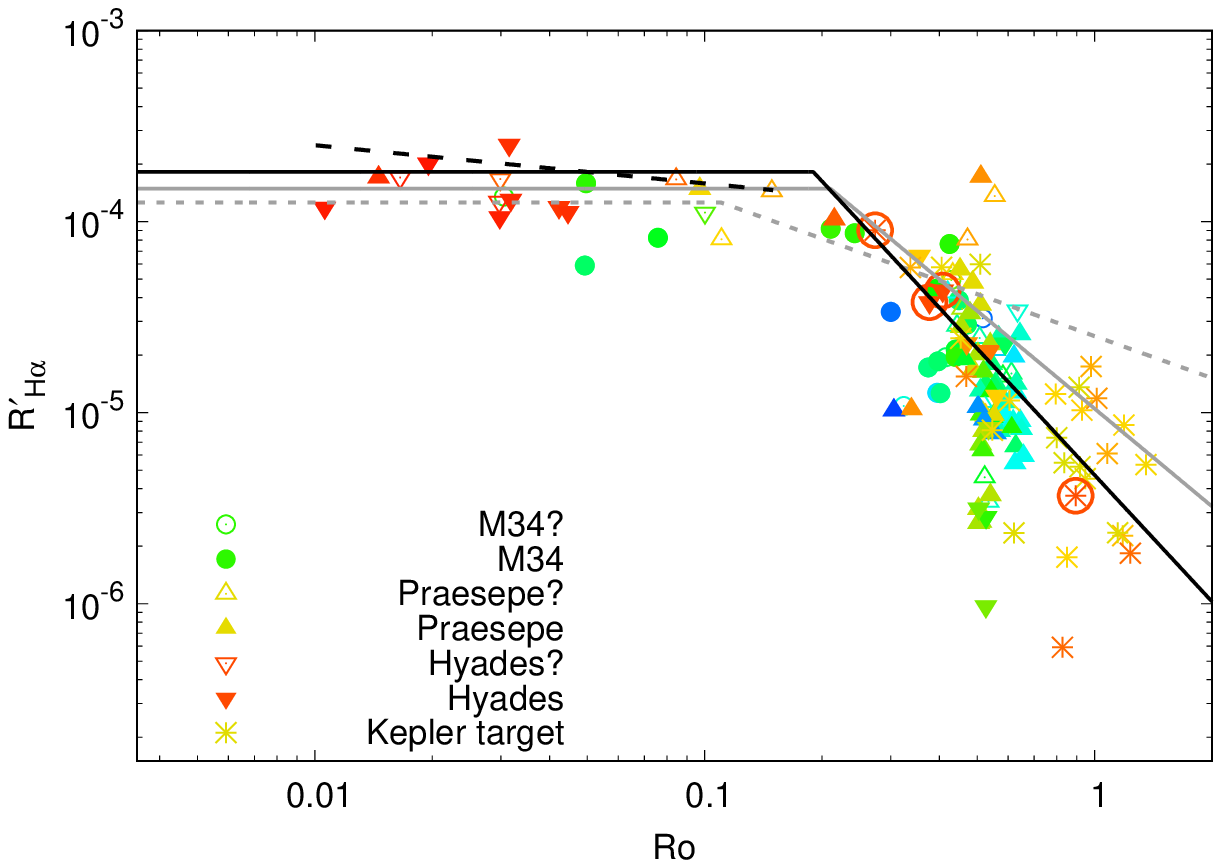}
\includegraphics[width=\columnwidth]{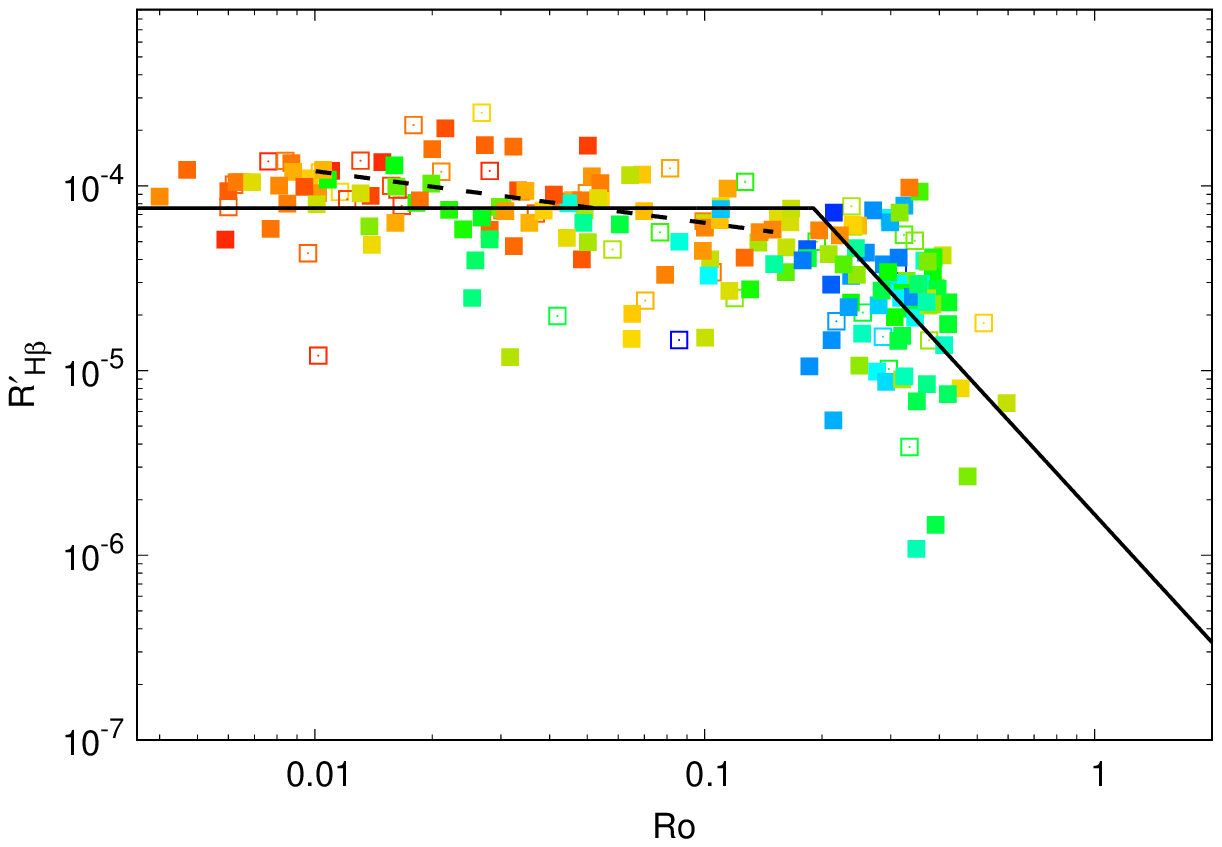}
\includegraphics[width=\columnwidth]{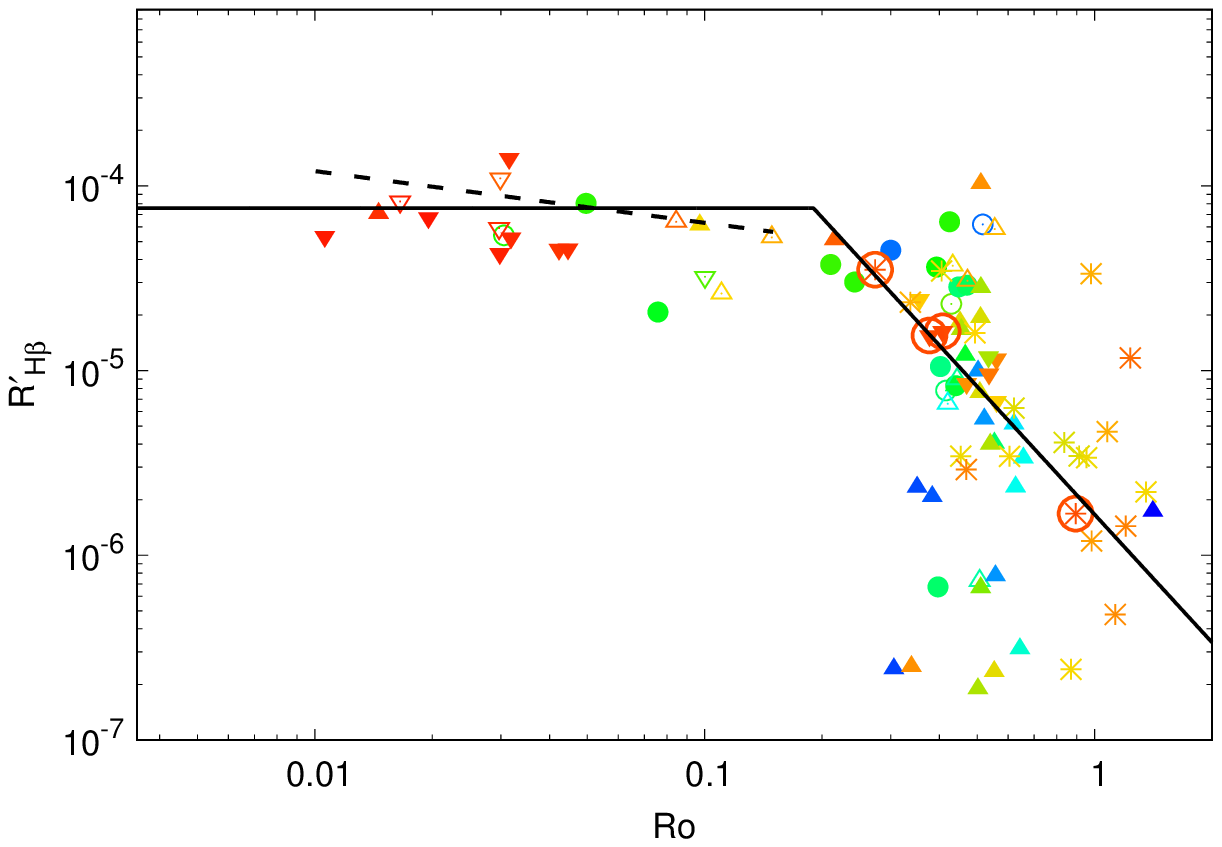}
\includegraphics[width=\columnwidth]{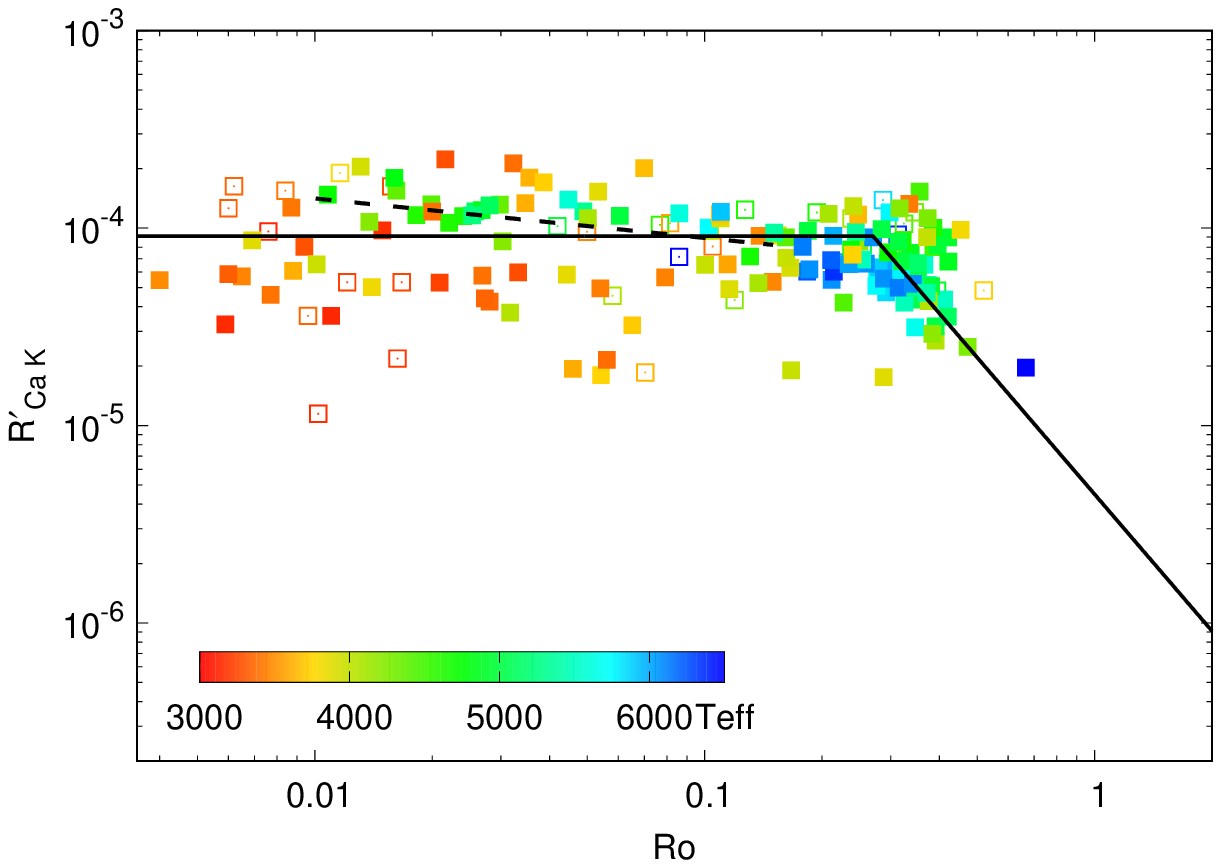}
\includegraphics[width=\columnwidth]{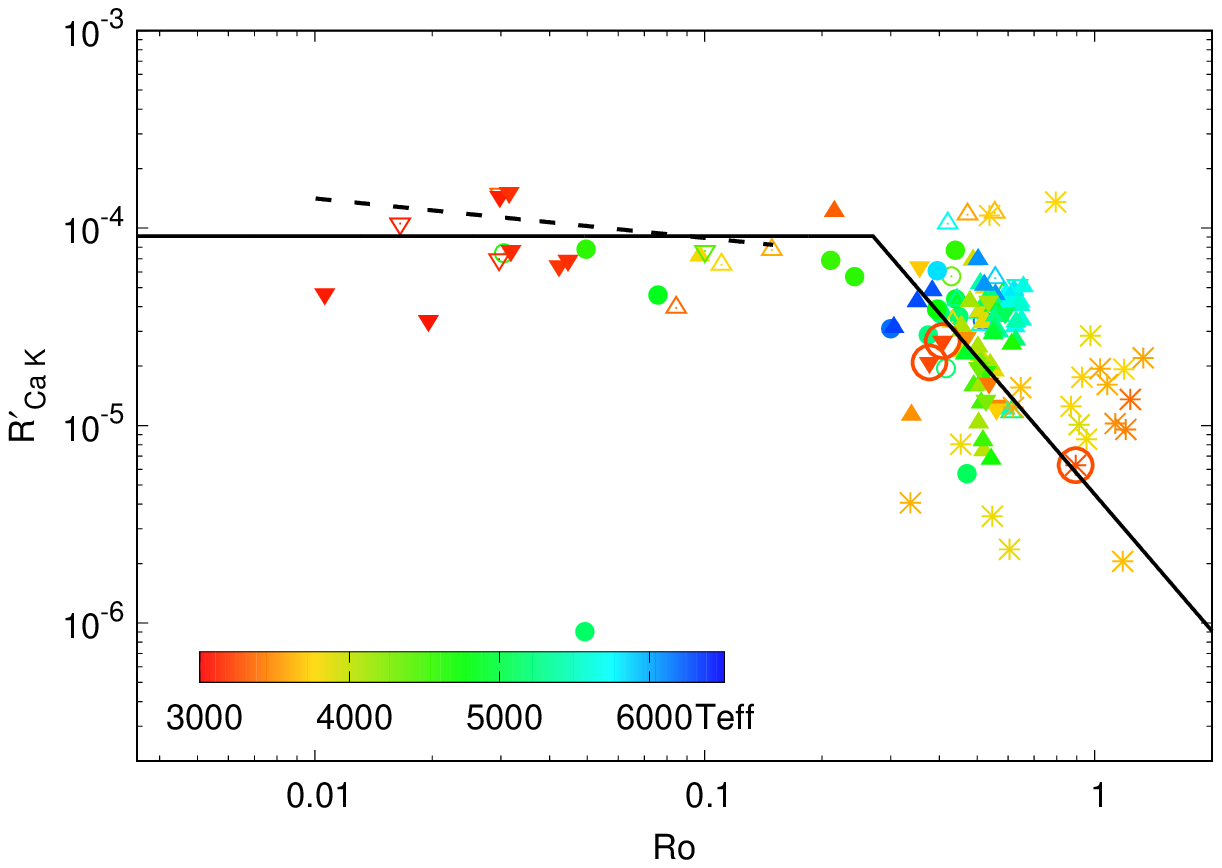}
\caption{$R^{'}_{\text{H}\alpha}$, $R^{'}_{\text{H}\beta}$, and $R^{'}_{\text{Ca K}}$ vs. Ro for stars in Pleiades (left) and M 34, Praesepe and Hyades (right). 
The colour gradient represents different effective temperatures. 
The cross symbols show slowly rotating $Kepler$ M dwarfs ($P_{\text{rot}}>10$ days, Mass$<0.6~M_{\odot}$). 
The fully convective candidates are marked by large circles. 
The grey solid and dashed line in top-right panel denote the relation of \citet{newt2017} and \citet{doug2014}, respectively.}
\label{fig:lha_rossby}
\end{figure*}

We have investigated the rotation-activity connection in terms of Rossby number Ro, 
however, it is difficult to judge whether Rossby number is superior to rotation period when modernizing this rotation-activity relation 
\citep[e.g., see a comprehensive study on this topic by][]{rein2014}. In Fig.~\ref{fig:lha_rossby} the $R^{'}_{\text{H}\alpha}$ against  
Ro is displayed for our sample stars whose periods are available. As expected, it shows two different regimes, saturated and unsaturated, 
with different dependency of activity on rotation around a turnoff point (e.g., Ro$\sim0.1$), 
usually shown by X-ray emissions \citep[e.g.][]{wrig2011}, chromospheric Ca {\sc ii} and H$\alpha$ emissions \citep[e.g.][]{noye1984,jack2010,doug2014,newt2017}, 
and cool spot coverages \citep{fang2016}. 

The rotation-activity relationship is often parameterized by a flat region connected to a power law decay, i.e., 
the activity has constant (saturated) value below a turnoff of Ro (namely $\text{Ro}_{\text{sat}}$), 
and above $\text{Ro}_{\text{sat}}$ it declines as a power law of Ro, indicates activity is proportional to $\text{Ro}^{\beta}$. 
\citet{wrig2011} found that the decline in coronal emission follow a power-law with $\beta=-2.18$ for 
Ro$>0.13$ based on large sample of 824 late-type stars. Furthermore, they derived $\beta=-2.7$ for a set of solar-type stars. However, the 
chromospheric emissions show a much shallower power law decay for slow rotators, e.g., the first two Ca {\sc ii} IRT excess emissions 
in early M-type dwarfs in NGC 2516 decreases as a power-law of $\beta\sim-1$ for Ro$\gtrsim0.1$ \citep{jack2010}. 
Recently, \citet{doug2014} derived much shallower power-law with 
$\beta=-0.73$ for Ro$>0.11$ in H$\alpha$ emissions of late K and M dwarfs of Praesepe and Hyades. 
More recently, \citet{newt2017} detected a power law index $\beta=-1.7$ with $\text{Ro}_{\text{sat}}=0.21$ in excess H$\alpha$  
emission based on large sample of nearby M dwarfs.

The rapid rotators with Ro$\lesssim$0.1 in Fig.~\ref{fig:lha_rossby} indicates that the chromospheric emissions have similar values shaping a plateau feature 
\citep[e.g. ][]{delf1998,rein2012,doug2014,newt2017}. Approximately, H$\alpha$ emission saturates at 
$\log R^{'}_{\text{H}\alpha}\approx-3.7$ dex, while H$\beta$ and Ca{\sc ii} K emission saturate respectively at $\log R^{'}_{\text{H}\beta}\approx-4.1$ dex and 
$\log R^{'}_{\text{Ca K}}\approx-4.0$ dex. The activity level is usually thought to be a constant value in the saturated regime. 
However, our data shows that the chromospheric emissions still increase slightly as Ro decreases, a phenomenon exists in other activity indicators, e.g., 
Ca {\sc ii} IR emissions of early M-type members of NGC 2516 \citep{jack2010}. We found that it can be roughly fitted by some power of Ro, e.g., 
$R^{'}_{\text{H}\alpha}\propto \text{Ro}^{-0.2}$ for Pleiades stars with Ro$\lesssim0.1$, as shown by black dashed line in the top-left panel 
of Fig.~\ref{fig:lha_rossby}, a relationship consistent with that ($L_{\text{X}}/L_{\text{bol}}\propto \text{Ro}^{-0.16}$) derived by \citet{rein2014} based 
on X-ray emission. A similar increment could be seen in the H$\beta$ emission, e.g., $R^{'}_{\text{H}\beta}\propto \text{Ro}^{-0.28}$ (the black 
dashed line in middle-left panel of Fig.~\ref{fig:lha_rossby}). We could not see such trend in Ca {\sc ii} K emission in this regime for total sample 
stars, which maybe due to large uncertainties in $R^{'}_{\text{Ca K}}$ for cooler stars. However, one can see that there also exist a slight increase 
relation in GK-type stars, e.g., $R^{'}_{\text{Ca K}}\propto \text{Ro}^{-0.2}$ (the dashed line in bottom-left panel of Fig.~\ref{fig:lha_rossby}). 
Such a slight slope of the activity-rotation relationship in the saturated regime indicates some unclear dependency of the activity on rotation 
is remaining even when saturation is reached, as pointed out by \citet{rein2014}.

The activity levels of ultra fast rotating G/K stars are decreasing below the saturation level, 
a phenomenon known as supersaturation which has been detected in the coronal emission \citep[e.g.][]{rand1996,wrig2011}. 
we found no clear evidence in our sample for any systematic fall in chromospheric activity levels when 
the Rossby numbers become very small. However, we noticed that the average CA of M-type stars follows a decreasing trend as rotation 
rate increases at Ro$\lesssim$0.01 (particularly in terms of the excess Ca {\sc ii} K emission), 
showing a weak evidence of damped cooling efficiency in ultra fast rotators. 

The regime with Ro$\gtrsim$0.1 are usually thought to be unsaturated regime. At first sight, the activity 
decreases with increasing Ro, which is expected from qualitative arguments based on the $\alpha\Omega$-type shell 
dynamo theory \citep[e.g.][and references therein]{noye1984,char2014}. 
Moreover, the stars with Ro in the regime of 0.1-0.4 have large scatter in activity with upper envelop near to saturated activity level, 
which makes it difficult to explain the location of turnoff point. Even worse is the strong dependency of power law with onset of 
saturation, Ro$_{\text{sat}}$, i.e., $\beta$ anti-correlates with Ro$_{\text{sat}}$, as seen in \citet{doug2014} and \citet{newt2017}. Anyway, we sought to 
represent the potential trends using the canonical rotation-activity relation (a saturated regime and a power law decay) following the previous studies, 
being fitted by eye rather than using any $\chi^{2}$-methods. For Pleiades candidate members (see left panels), we found that the 
saturation may occurs at Ro near to 0.2 rather than the canonical value of 0.1, and a power law is likely in the range of $-2.0\sim-2.5$, 
e.g., it can be represented by a relation of $R^{'}_{\text{H}\alpha}\propto \text{Ro}^{-2.2}$ for Ro$\gtrsim0.19$, 
as illustrated by the black solid line in top-left panel of Fig.~\ref{fig:lha_rossby}. 
We found a similar power law decay with Ro$_{\text{sat}}$ near to 0.2 in H$\beta$ emissions for the Pleiades sample stars, 
as illustrated by the black solid line ($R^{'}_{\text{H}\beta}\propto \text{Ro}^{-2.3}$ for Ro$\gtrsim0.19$) in 
middle-left panel of Fig.~\ref{fig:lha_rossby}. Compared to these two Balmer emissions, the Ca {\sc ii} K emissions show a slight different trend, 
i.e., the unsaturated regime start at a larger Ro (e.g., Ro$_{\text{sat}}\gtrsim0.25$), but appears to be a reasonably fit by a similar power 
law (an example relation with $\beta=-2.3$ and Ro$_{\text{sat}}=0.27$ is shown by the black solid line in bottom-left panel).

The results for M34, Hyades and Praesepe are shown in the right panels of Fig.~\ref{fig:lha_rossby}. For comparison purpose, in each right panel, 
we re-plotted the black lines in corresponding left panel. Among these older stars, most hotter stars are slow rotators, in particular members of Praesepe and 
Hyades, which are in unsaturated regime as expected, and 
only some M dwarfs in the Praesepe and Hyades and K-type members in M34 are fast rotators (see top panels of Fig.~\ref{fig:ha_rot}) that are occupied in the 
saturated regime. In the saturated regime, their excess chromospheric emissions increase as Ro decreases similar to Pleiades stars, 
but systematically having lower level. In the unsaturated regime, they appear to still follow the power law in a similar way as the Pleiades members, 
but with larger dispersion. However, it seems that Ca {\sc ii} K emission saturates at a larger Ro$_{\text{sat}}$ value, 
e.g., Ro$_{\text{sat}}\gtrsim0.3$ (alternatively, these stars follow a shallower power laws at the same onset saturation in Pleiades). 
In top-right panel of Fig.~\ref{fig:lha_rossby}, also shown is the relation of $\beta=-0.73$ with Ro$_{\text{sat}}=0.11$ in dashed grey line, 
which was derived by \citet{doug2014} for M dwarfs with H$\alpha$ in emission in the Praesepe and Hyades. 
However, our data are inconsistent with their results, our H$\alpha$ data can be fitted by a clearly deeper power law 
with a larger Ro$_{\text{sat}}$ value. The strong degeneracy between $\beta$ and Ro$_{\text{sat}}$ may be a potential contributor. 
Moreover, \citet{doug2014} sample consists of stars with $r'-K\gtrsim3$ (roughly later than K7) and definite H$\alpha$ emission, 
neglected potential less active members with H$\alpha$ in absorption (but in fact having excess emissions such as these GK-type members), 
which may also accounted for the notable difference between their results and ours. 
The relation for $\beta=-1.7$ with Ro$_{\text{sat}}=0.21$ derived by \citet{newt2017} for nearby M dwarfs also shown in solid grey line in the top-right panel. The Ro$_{\text{sat}}$ shown by our H$\alpha$ data is consistent with their value, however, 
the power law index is slightly deeper than their value. 

The transformation between the solar-type stars (with radiative cores and convective envelopes) 
and fully convective stars takes place roughly in stars of spectral type M3-M4. 
The interface layer between radiative cores and convective envelopes (known as the tachocline) 
is believed to play an important role in the generation of the magnetic field in solar-type stars \citep{spie1992}. 
Therefore, the dynamo mechanism in the fully convective stars such as middle/late M-type stars 
is expected to be very different because of the absence of a tachocline. 
However, it is evident that slowly rotating, 
fully convective stars still show rotation-activity (X-ray emission) connection in the same way similar to solar-type stars \citep{wrig2016}. 
Recently, \citet{newt2017} found a single relation between excess H$\alpha$ emission and Ro for stars including both early- and late-type M dwarfs, 
supporting the previous finding. Unfortunately, most M-type stars in our sample are fast rotators (see Fig.~\ref{fig:ha_rot}), 
and only two slow rotators in Hyades are fully convective candidate stars (marked as large open circles, and see Table~\ref{tab:fustars}). 
To take a simple look at this issue, we picked up LAMOST spectra for 47 slowly rotating 
M dwarfs ($P_{\text{rot}}>$10 days, Mass$<$0.6 M$_{\odot}$) from the main-sequence $Kepler$ targets with known rotation periods provided by \citet{mcqu2014}, 
and measured their excess CA emissions, as shown by crosses in right panels of Fig.~\ref{fig:lha_rossby} (only 29 of them show excess H$\alpha$ emission). 
Note that two of them are the fully convective 
candidates, which were marked with large open circles (but also see Table~\ref{tab:fustars}). 
It is clear that these slowly rotating M dwarfs, including the fully convective candidates, 
still follow the power law decay that show no evident difference from hotter stars, 
which is consistent with the finding of \cite{newt2017}, though our results show a slight different power law decay. 
If this scenario is real, it then suggests that the magnetic dynamo may be similar to that in solar-type stars. 
\citet{wrig2016} suggest a common magnetic dynamo in both partially and fully convective stars, 
supporting those models in which the dynamo originates throughout the stellar convection zone. 

\begin{table*}
\caption{Slow rotating, potential fully convective stars with detected excess emissions in both H$\alpha$ and H$\beta$ lines}
\label{tab:fustars}
\begin{tabular}{lccccccccccccccccccc}
   \hline
Object name & SpT.&$T_{\text{eff}}$&Mass& $P_{\text{rot}}$ & Ro &$\log R^{'}_{\text{H}\alpha}$& $\log R^{'}_{\text{H}\beta}$&$\log R^{'}_{\text{Ca K}}$& 
Note\\
            &     & (K)         &(M$_{\odot}$)&  (days)    &    &                        &                        &                     &      \\
   \hline
04202761+1853499  & M3.6  & 3244  &  0.375  &  20.3105 & 0.4079 & -4.3609  & -4.7871  & -4.5722  &  Hyades \\
04303385+1444532  & M3.5  & 3257  &  0.382  &  18.4100 & 0.3771 & -4.4220  & -4.8098  & -4.6805  &  Hyades  \\
KIC 10731839      & M3.3  & 3272  &  0.362  &  46.1150 & 0.8946 & -5.4351  & -5.7745  & -5.2008  &  $Kepler$ target\\
KIC 11495571      & M3.4  & 3292  &  0.374  &  13.6400 & 0.2737 & -4.0450  & -4.4538  &  --      &  $Kepler$ target\\    
\hline 
\end{tabular}
\end{table*}

%%%%%%%%%%%%%%%%%%%%%%%%%%%%%%%%%%%%%%%%%aaaaaaaaaaaaa
\subsection{Correlations between activity indicators}
The correlations among activity indicators of different temperature layers contain information of overall heating process and systematic deficiencies in 
these layers \citep{gude2004}. Therefore, the correlations between different chromospheric activity indicators and the relation of excess 
chromospheric emission with other activity indicators had gained lot of attention \citep[e.g.][]{schr1992,mont1995,mont1996,mart2011}. 
In fact, this topic need further investigation, e.g., need complex physical models to interpret the correlations, which is beyond the scope of this 
paper. Our intent here is to merely investigate the basic observational correlations between different tracers of activity.

\subsubsection{\text{H}$\alpha$, H$\beta$ and Ca~{\sc ii} K}
\begin{figure*}
\centering
\includegraphics[width=\columnwidth]{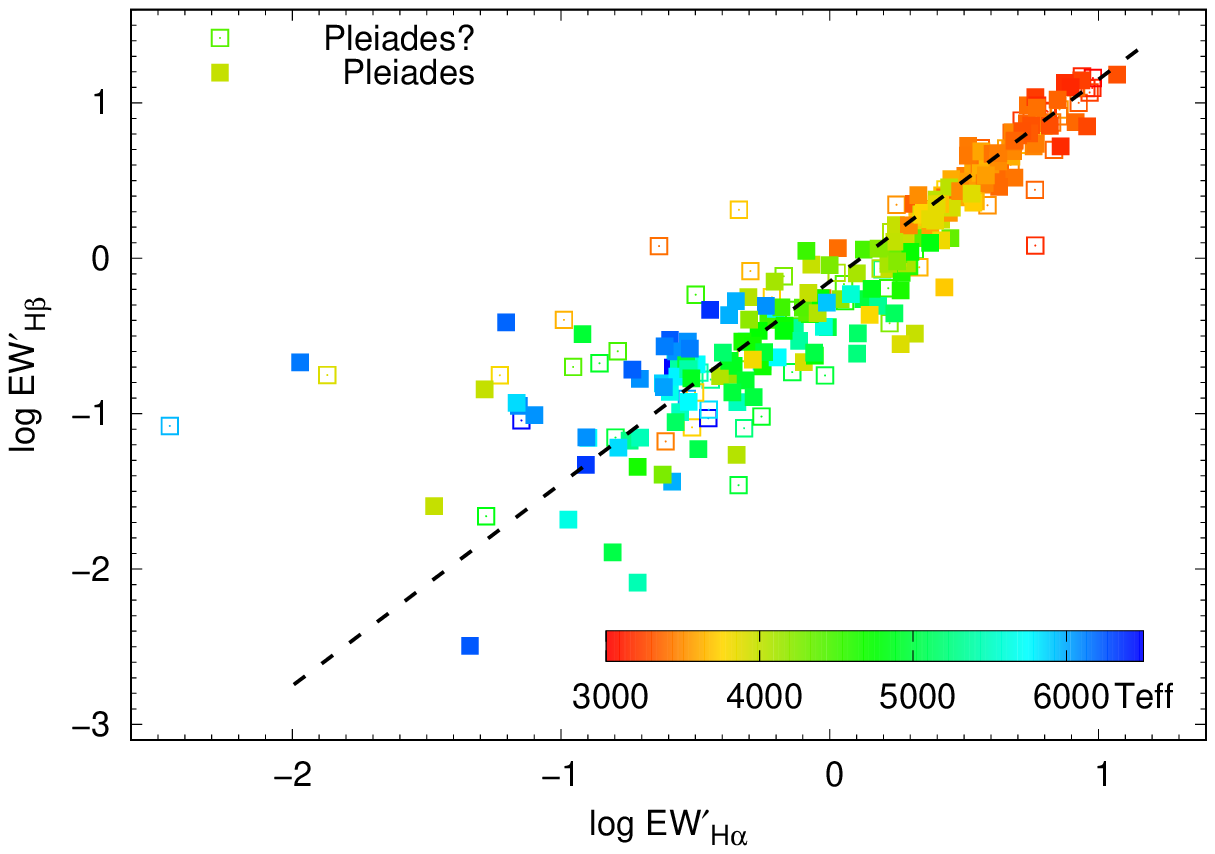}
\includegraphics[width=\columnwidth]{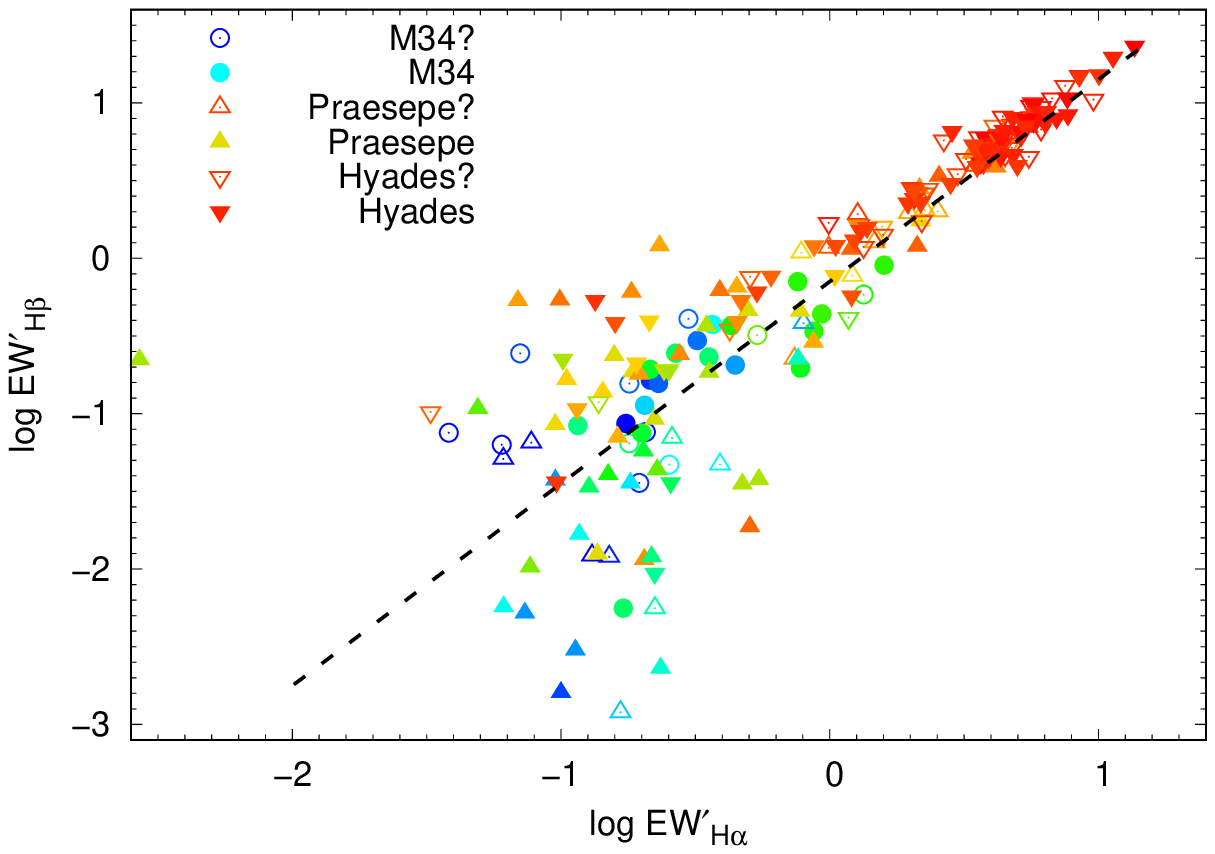}
\includegraphics[width=\columnwidth]{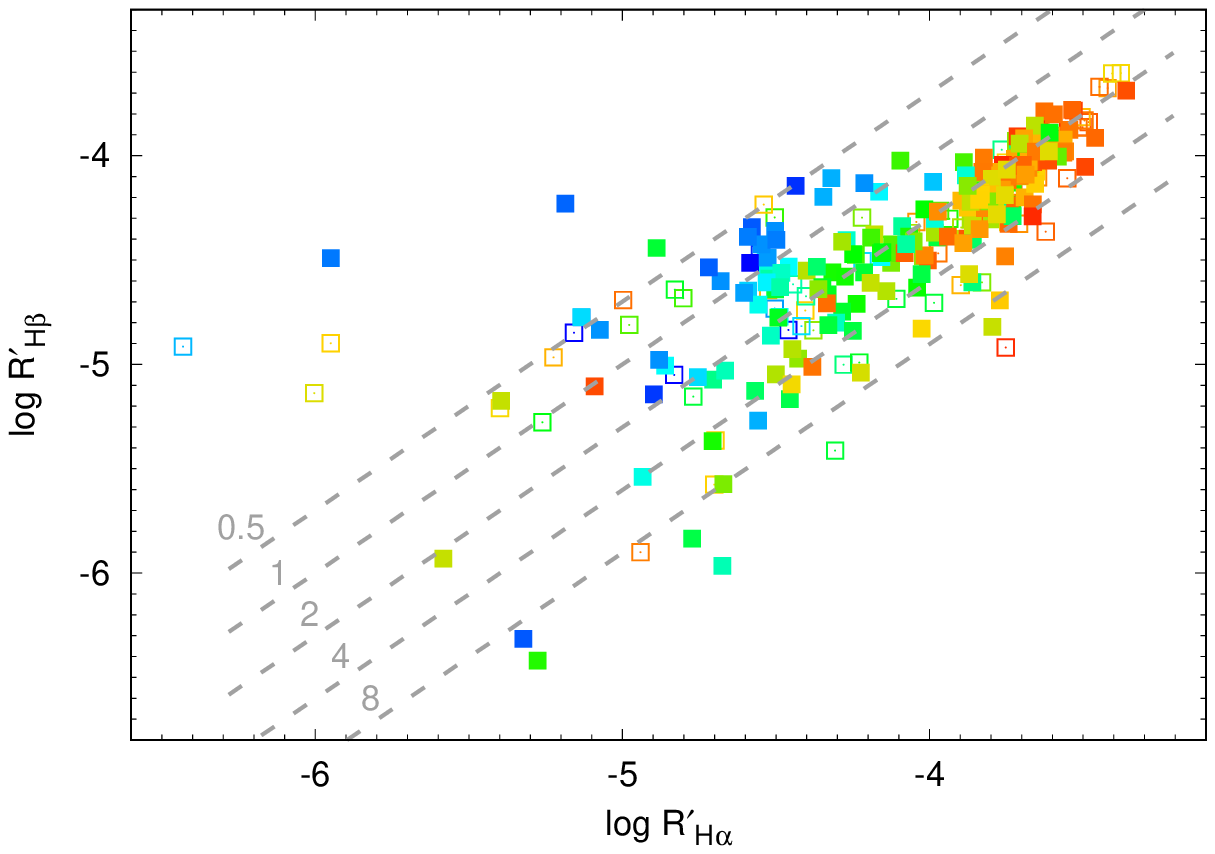}
\includegraphics[width=\columnwidth]{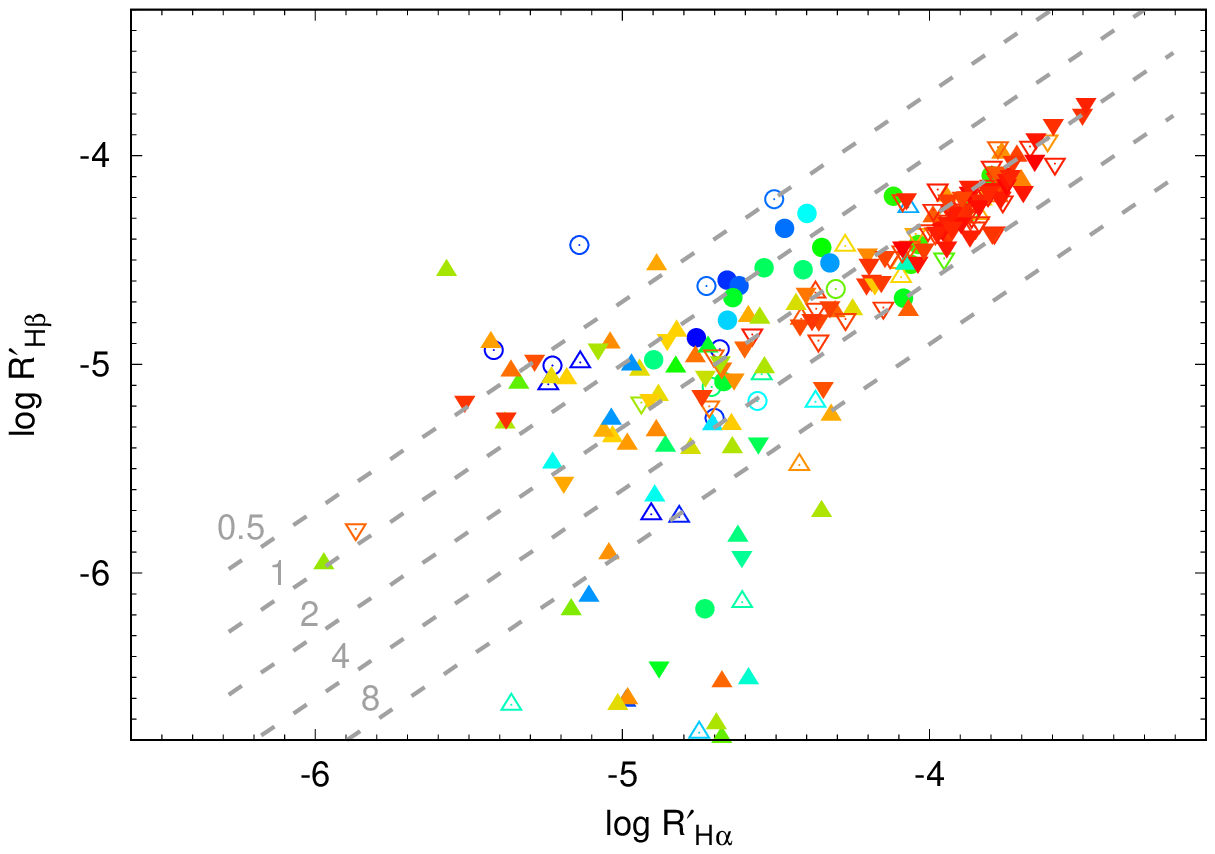}
\caption{The correlation of excess equivalent widths (upper panels) and fractional excess luminosities (lower panels) of H$\beta$ and H$\alpha$ lines. 
Dashed lines in upper panels: power-law approximation (see the text). The dashed lines with grey colours in the lower panels denote constant ratios 
of $R^{'}_{\text{H}\alpha}/R^{'}_{\text{H}\beta} = 0.5,~1,~2,~4~\text{and}~8$, respectively. }
\label{fig:hb_ha}
\end{figure*}

\begin{figure*}
\centering
\includegraphics[width=\columnwidth]{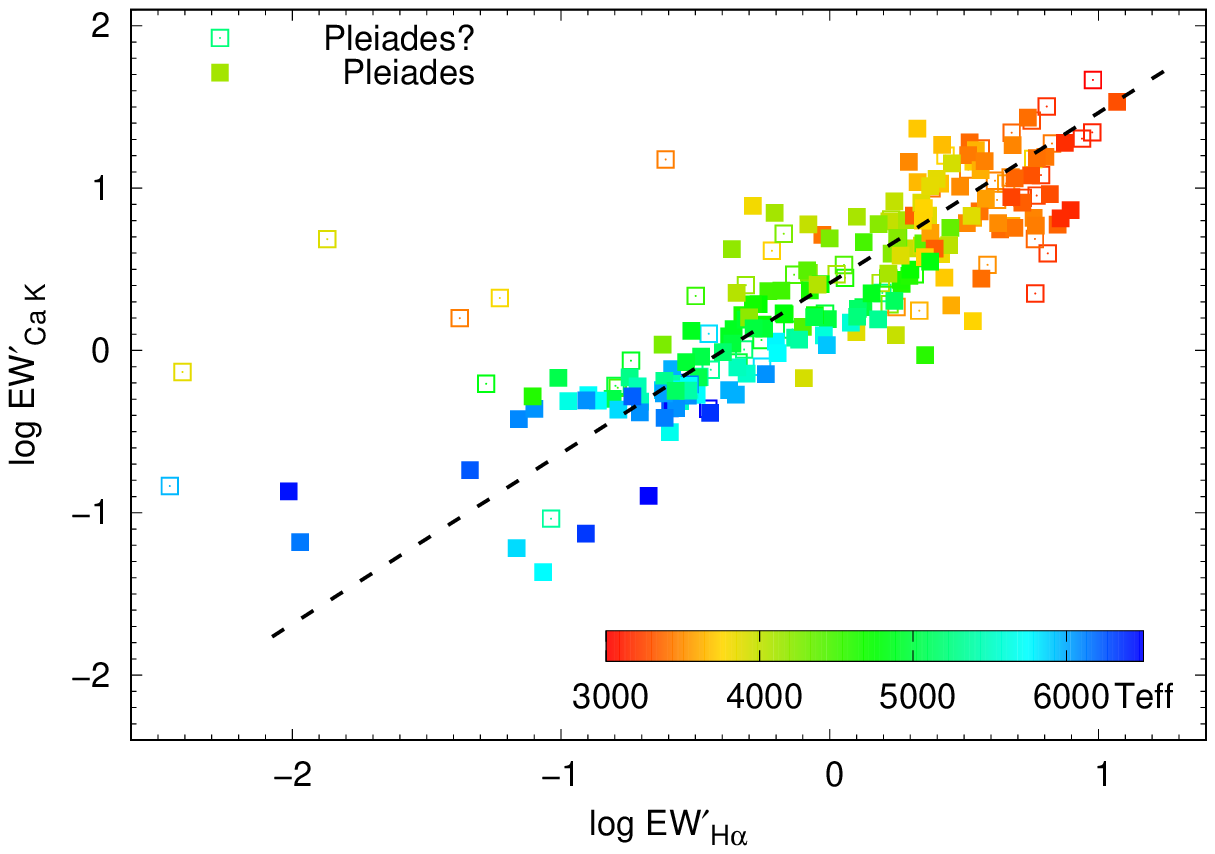}
\includegraphics[width=\columnwidth]{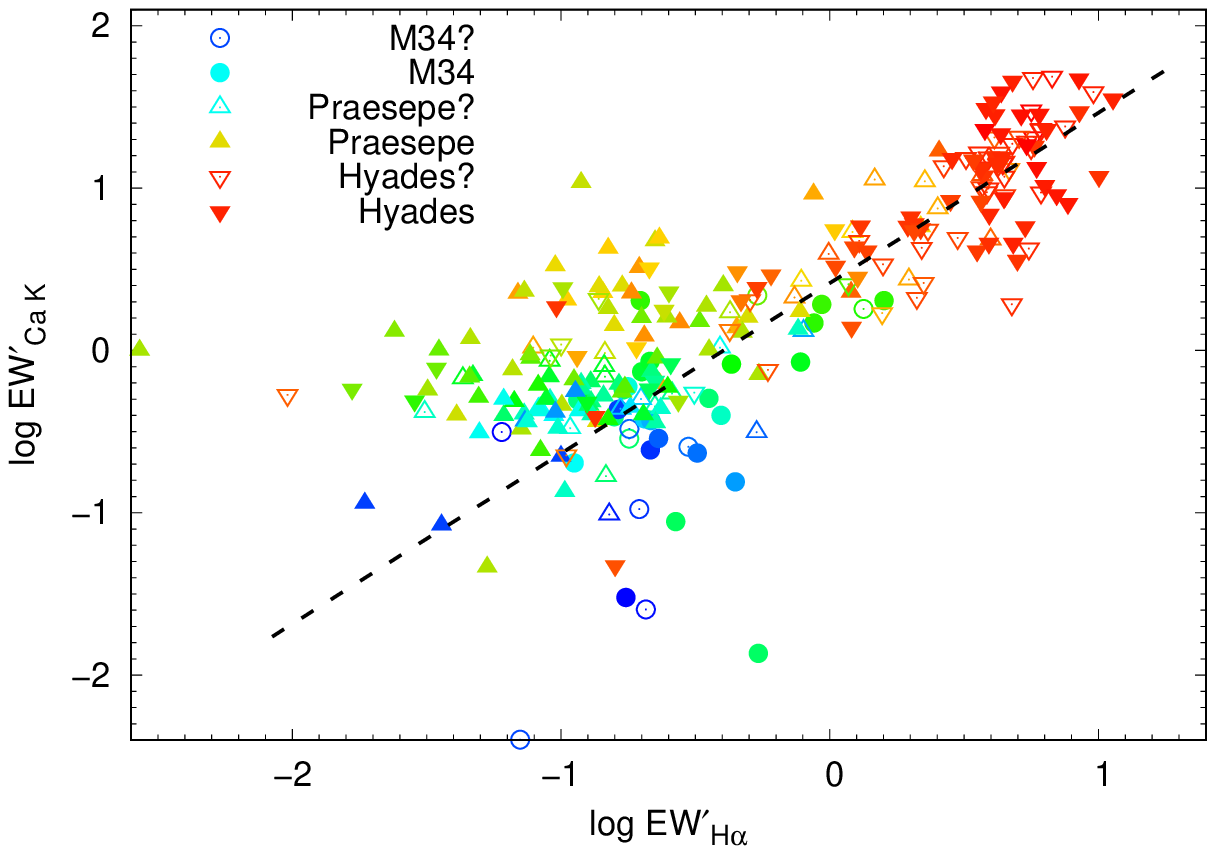}
\includegraphics[width=\columnwidth]{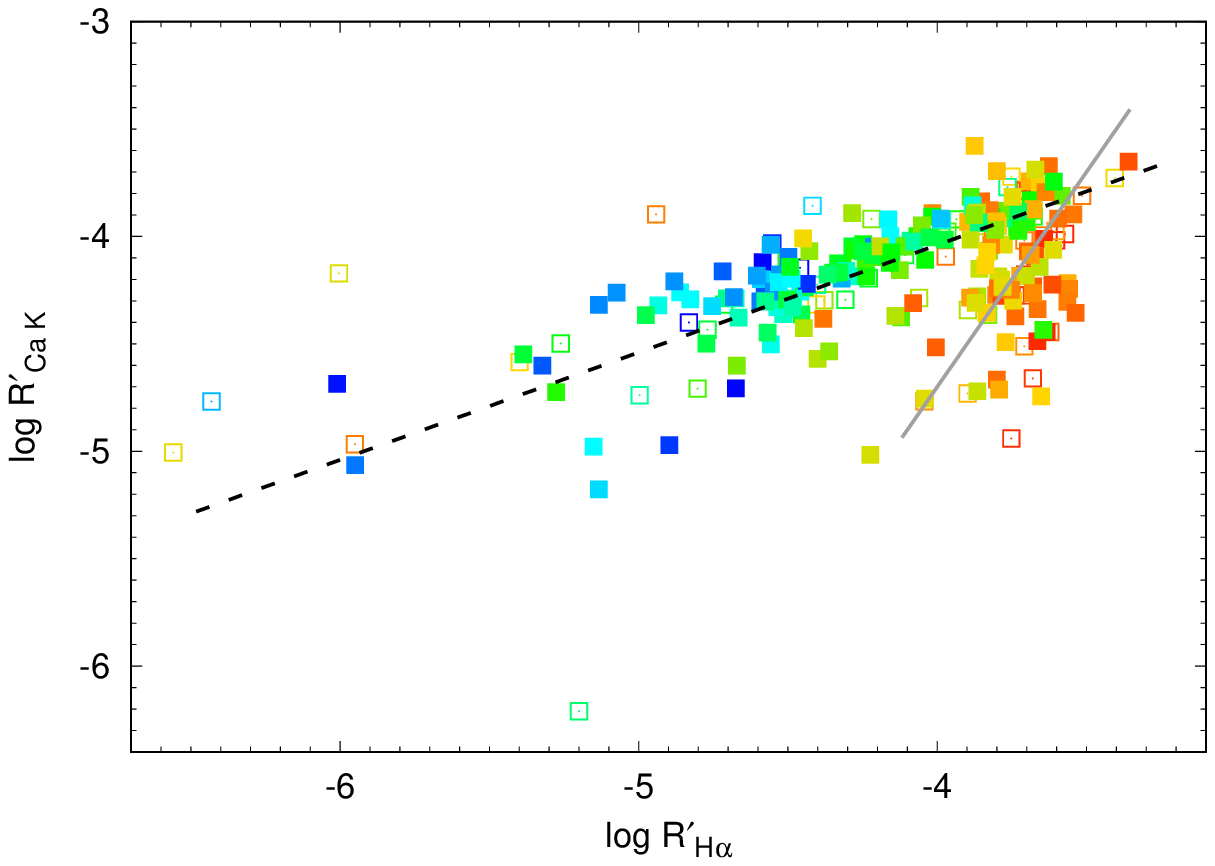}
\includegraphics[width=\columnwidth]{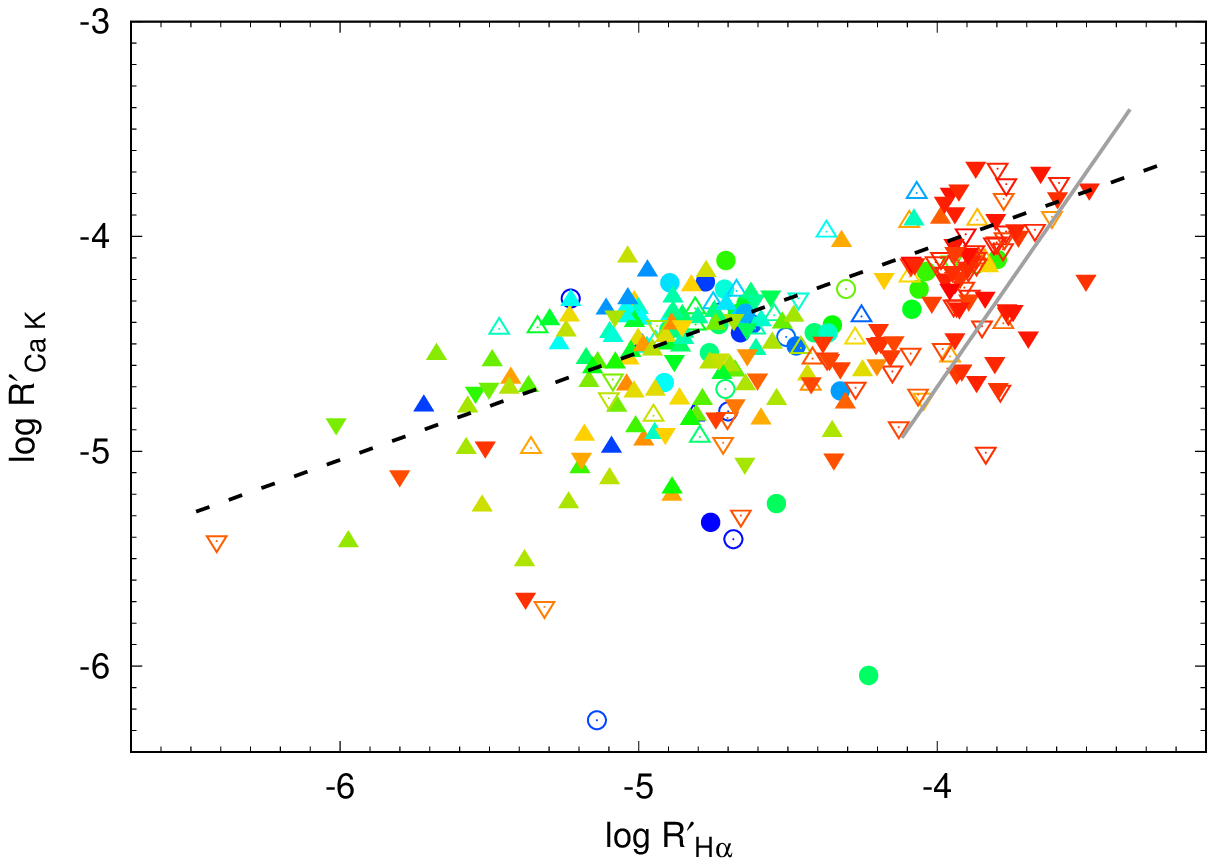}
\caption{Relationships between excess emissions of H$\alpha$ and Ca {\sc ii} K lines (see the text).}
\label{fig:ck_ha}
\end{figure*}

Ca~{\sc ii} H\&K lines are the most widely used chromospheric activity indicators as their source function are collisionally dominated. H$\alpha$ and 
H$\beta$ are the two strongest lines among Balmer series, and their excess emissions are formed at the middle chromosphere \citep{mont2004}. Our data 
shows that there exist good correlations of excess emissions of H$\alpha$ to that of H$\beta$ and Ca~{\sc ii} K lines, as shown in Fig.~\ref{fig:hb_ha} and 
Fig.~\ref{fig:ck_ha}. 

Fig.~\ref{fig:hb_ha} shows a power law correlation (linear relation in log-log scale) between excess equivalent widths of the two Balmer 
lines, EW$^{'}_{\text{H}\alpha}$ and EW$^{'}_{\text{H}\beta}$, among the Pleiades stars. We found that it can be represented well by a power 
law index in the range of $1.2-1.4$, 
e.g., EW$^{'}_{\text{H}\beta} \propto ({\text{EW}}^{'}_{\text{H}\alpha})^{1.3}$, as illustrated by the black dashed line in top-left panel of 
Fig.~\ref{fig:hb_ha}. For members in M34, Praesepe and Hyades, it also can be represented roughly by the same power law relation, as shown by the black 
dashed line in top-right panel. It is noticeable that the main trend between $R^{'}_{\text{H}\beta}$ and $R^{'}_{\text{H}\alpha}$ also 
roughly linear in log-log scale. Moreover, we found that the power law index is around 1.0, which indicates a linear relation roughly even in linear scale, 
e.g., $R^{'}_{\text{H}\beta} \propto R^{'}_{\text{H}\alpha}$, as demonstrated by dashed lines with constant ratios 
($R^{'}_{\text{H}\alpha}/R^{'}_{\text{H}\beta} = 0.5,~1,~2,~4~\text{and}~8$, from top to bottom, respectively) in lower panels of Fig.~\ref{fig:hb_ha}. 
A constant H$\alpha$ to H$\beta$ ratio has been detected in previous studies, 
e.g., \citet{reid1995} found a good linear correlation between the flux emitted at H$\alpha$ and H$\beta$ for 
low mass stars in the Hyades ($F_{\text{H}\alpha}\sim4.6 \times F_{\text{H}\beta}$). 
However, some late-F or early G-type stars show departures from the general trend, e.g., 
their H$\beta$ losses seem to be more important than H$\alpha$ losses. 
In fact, compared to H$\alpha$, H$\beta$ is preferred the higher chromospheric density, 
e.g., while H$\beta$ enhanced during a flare-like event, the H$\alpha$ is not proportionally increased in strength \citep{huen1987}. 
Thus the ratio of energy emitted in the H$\alpha$ to H$\beta$ ($\text{E}_{\text{H}\alpha}/\text{E}_{\text{H}\beta}$) 
is widely used as indicator of the presence of flare-like events \citep{huen1987}, 
and also a diagnostic indicator for discriminating between stellar plages and prominences \citep{hall1992}. 
The typical value of $\text{E}_{\text{H}\alpha}/\text{E}_{\text{H}\beta}$ is around 1 in flare-like events, 
it then become 4 or larger for RS CVn stars \citep{huen1987}. Therefore, the departure that some earlier type stars with lower H$\alpha$ to H$\beta$ ratios 
($R^{'}_{\text{H}\alpha}/R^{'}_{\text{H}\beta}<1$) were likely underlying flare/plage-like activities. 
If this is the real case, then it suggests that these stars preferred to suffer flare events than cooler stars.

It is clear from Fig.~\ref{fig:ck_ha} that there exist power-law relationship (linear relation in log-log scale) between excess equivalent widths of H$\alpha$ 
and Ca~{\sc ii} K lines. Here we did not attempt any $\chi^{2}$ fitting but merely sought to reproduce the correlation by eye as before. 
It can be represented approximately by a power-law index of $1.0-1.1$, e.g., 
EW$^{'}_{\text{Ca K}} \propto ({\text{EW}}^{'}_{\text{H}\alpha})^{1.05}$ (see the black dashed lines in upper panels), 
for which the power-law exponent is slight larger than the value (0.97) derived by 
\citet{mont1995} based on a sample of 51 chromospherically active binaries. The lower panels of Fig.~\ref{fig:ck_ha} correspond to relations between 
excess fractional luminosities of Ca~{\sc ii} K and H$\alpha$, which could be represented approximately by a relation of $R^{'}_{\text{Ca K}}\propto 
({R^{'}_{\text{H}\alpha}})^{0.5}$ for GK-type Pleiades stars (see the black dashed lines in lower panels). 
However, cooler and more chromospherically active stars such as M-type stars do not obey the 
general relations, showing departure feature of much sharper slope compare to hotter stars. It is illustrated as an example relation 
$R^{'}_{\text{Ca K}}\propto ({R^{'}_{\text{H}\alpha}})^{2.0}$ by grey solid lines in the figure, which indicates that H$\alpha$ losses tend to be more 
important than Ca~{\sc ii} K losses for cooler stars that are more active. In fact, previous studies show such departures from the general trend in surface 
flux-flux relations between H$\alpha$ and other chromospheric indicators among some late-K and M dwarfs, and they are found to be young stars or flare stars being in the X-ray saturation regime \citep[e.g.][]{mart2011}. One can see that the departures hosted in M stars with non-universality 
trends appear both in young open cluster, Pleiades, and median-age open clusters, Hyades, and most of these stars are indeed in H$\alpha$ emission 
saturation regime. It has been suggested that the stars deviating from the general relation probably have a different magnetic structure 
\citep[][]{mart2011}, e.g., nanoflare heating.
 
\subsubsection{H$\alpha$ and photospheric activity}

\begin{figure}
\centering
\includegraphics[width=\columnwidth]{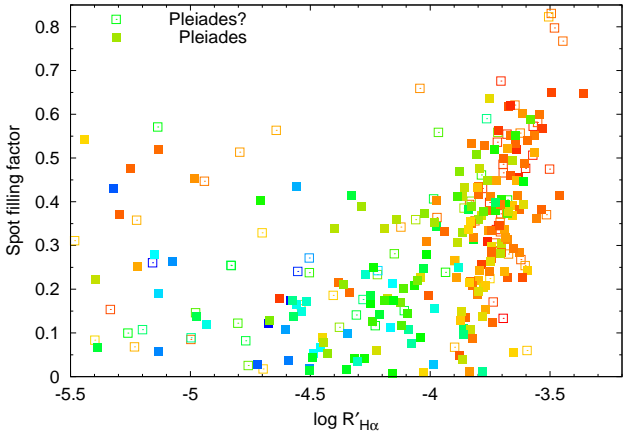}
\caption{Relations between $R^{'}_{\text{H}\alpha}$ and spot filling factor for the Pleiades sample stars with detected cool spots coverage. Colours as in Fig.~\ref{fig:lha_rossby}.}
\label{fig:lha_spot}
\end{figure}

In Fig.~\ref{fig:lha_spot}, we showed the spot filling factor (or fractional spot coverage, $f_{\text{s}}$) 
of stars with detected cool spots ($f_{\text{s}}>0$) in Pleiades as a function of $R^{'}_{\text{H}\alpha}$. 
The spot coverage of stars have large scatter, which makes harder to say about existence of any believable connections between these two activity indicators. 
However, the overall trend, in particular among more chromospherically active members 
(e.g., if we consider stars with $\log R^{'}_{\text{H}\alpha}>-5.0$ and keep aside the outliers at the end of weak H$\alpha$ emission), 
 shows that more active stars tend to have larger spot coverages. 
We use a Pearson's product-moment correlation analysis to test the statistical significance of the potential 
connections between $f_{\text{s}}$ and $\log R^{'}_{\text{H}\alpha}$. The Pearson's correlation coefficient $r\approx0.55$ among stars with $\log 
R^{'}_{\text{H}\alpha}>-5.0$ indicates that there exist some correlation between spot coverage and chromospheric emission. 
Compared to hotter stars such as GK-type members, most M-type Pleiades members (see red symbols in Fig.~\ref{fig:lha_spot}) 
are in H$\alpha$ emission saturation regime, showing a different and slightly tighter trend in the plot, 
which also evidenced by a stronger positive correlation of $r\approx0.63$ (for stars with $T_{\text{eff}}<3900$ K and 
$\log R^{'}_{\text{H}\alpha}>-4.0$). It is clear that such non-universality feature is very similar to the 
relation between chromospheric emissions of H$\alpha$ and Ca~{\sc ii} K, as discussed above. 
Note that the uncertainties are larger in determination of spot coverages for M-type star as discussed in Paper I, 
and the correlation showed in the plot need to be further investigated, which puts a caution for further discussion based on these features. 

\begin{figure*}
\centering
\includegraphics[width=\columnwidth]{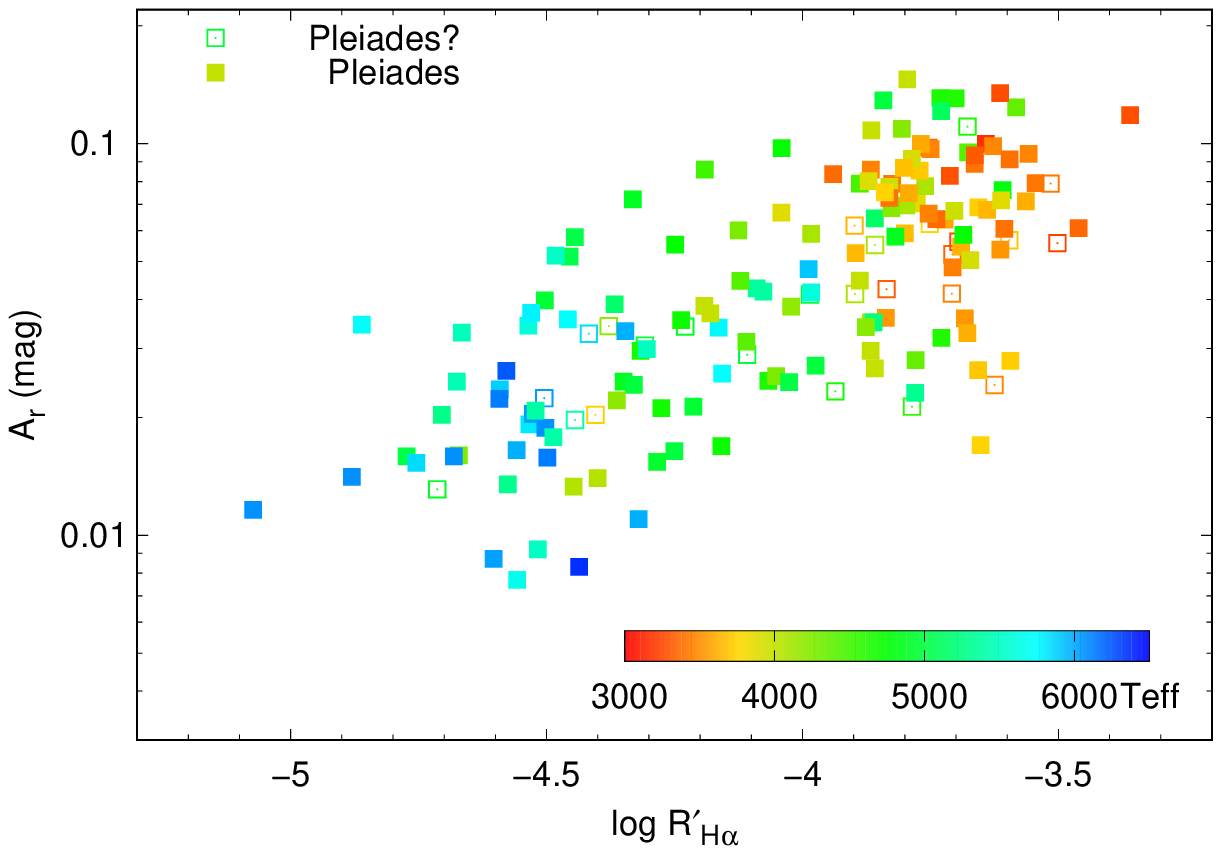}
\includegraphics[width=\columnwidth]{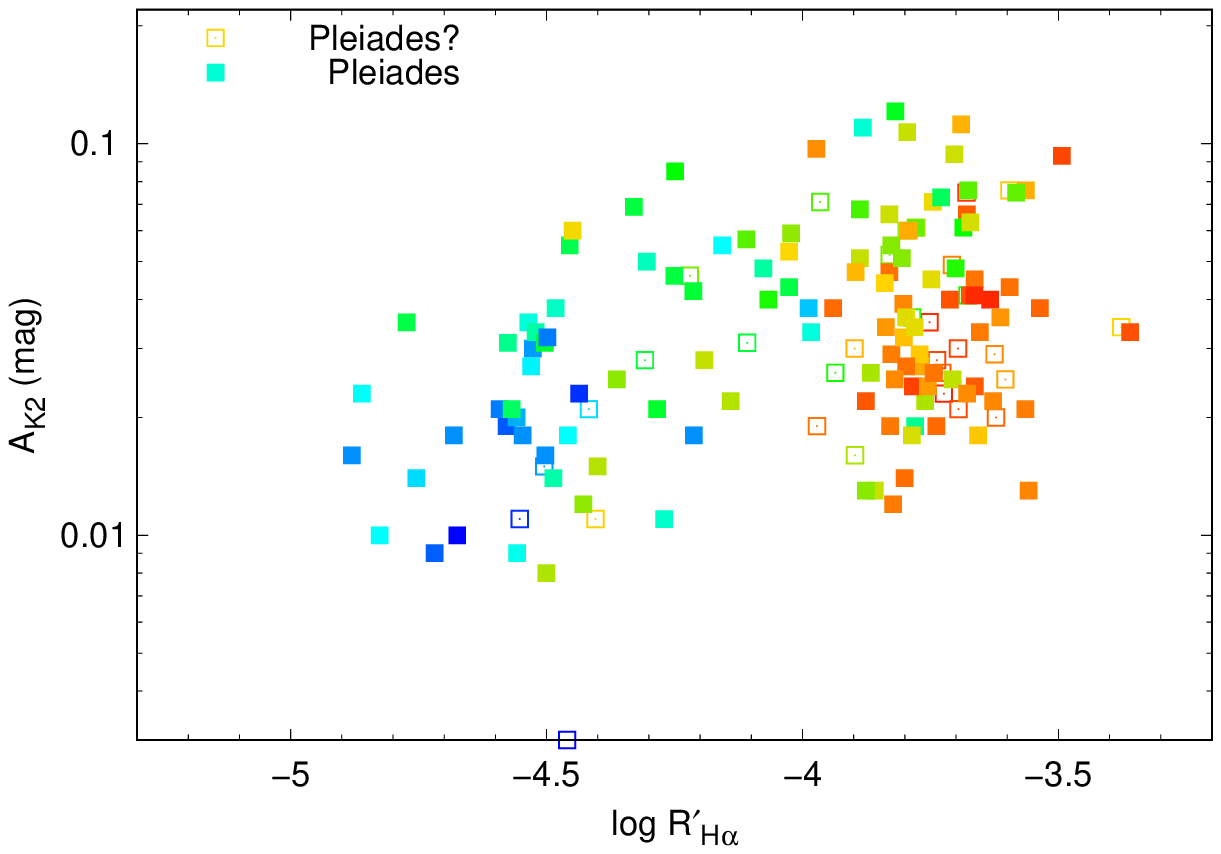}
\caption{Amplitudes of light variation (left: HATNet, right: $K2$) as a function of $\log R^{'}_{H\alpha}$. Note that the amplitude is in log scale. }
\label{fig:lha_amp}
\end{figure*}

The amplitude of light variation due to spot rotational modulation carries information of inhomogeneity in the stellar surface, 
thus acts as another indicator of photospheric activity, e.g, the light variation become larger with increase in chromospheric activity  
among Sun-like stars \citep[e.g.][]{lock1997,radi1998,lock2007} and M dwarfs \citep[e.g.][]{newt2017}.
The amplitudes of periodic light variation versus excess H$\alpha$ emissions for Pleiades candidates with 
detected rotation period are shown in Fig.~\ref{fig:lha_amp}, 
where the $r$-band peak-to-peak photometric amplitudes $A_{r}$ were derived by \citet{hart2010} based on the HATNet light curves. 
On the whole, there is a trend showing larger chromospheric emission corresponds to larger light variation, 
however, the dispersion is large as shown in the left panel. 
The Pearson's correlation coefficient in log-log scale is $r\sim0.71$ (with a $p$-value of $<2.2\times10^{-16}$). 
The right panel of Fig.~\ref{fig:lha_amp} is analogous to the left panel, wherein the amplitude of variation (between 10 percent and 90 
percent of the distribution of points) $A_{K2}$ is derived by \citet{rebu2016} based on $K2$ observations. 
We noticed some additional features from this panel. The correlation is weaker that is evidenced by a smaller value of $r\sim0.47$ ($p\sim1.4\times10^{-9}$). 
And there exist systematic offset between the $A_{K2}$ and $A_{r}$ for very cool stars (see red symbols). 
By neglecting the very cool stars with $T_{\text{eff}}<3900$ K, the correlation become tighter ($r\sim0.62$, $p\sim1.9\times10^{-10}$).
Previous studies show that the short-term photometric variability (mainly due to rotational spot modulation) 
in less active Sun-like stars can be related to their average level of chromospheric activity by a power law \citep[e.g.][]{radi1998}. 
The Pleiades GK-type members in the unsaturated regime are roughly linear on the log-log scale plot, indicating a power law relation. 
On the other hand, M dwarfs still follow the general trend of more active stars showing higher levels of photometric variability 
\citep[e.g.][]{newt2017}. As most of the Pleiades M-type stars are in saturated chromospheric activity regime, 
there is no believable trend among these very cool stars with $T_{\text{eff}}<3900$ 
K in both panels (left: $r\sim0.21$, $p\sim0.12$; right: $r\sim0.08$, $p\sim0.50$), which probably due to their very high activity levels. 
In fact, the chromospheric activity levels vary in short term (as shown in Fig.~\ref{fig:vdewha}), 
and long term such as cyclic variation \citep[e.g.][]{bali1995}, thus the panels of Fig.~\ref{fig:lha_amp} actually represents a snapshot of stellar variability. 
Moreover, the pattern and evolution of spots, and stellar inclination (the orientation angle of the axis of rotation relative to our line of sight) 
could effect the light variation pattern and contaminate the potential relation between chromospheric emission and light variation, 
thus the feature shown in this figure is only instructive rather than conclusive.

\subsubsection{H$\alpha$ and coronal X-ray emission}
\begin{figure}
\centering
\includegraphics[width=\columnwidth]{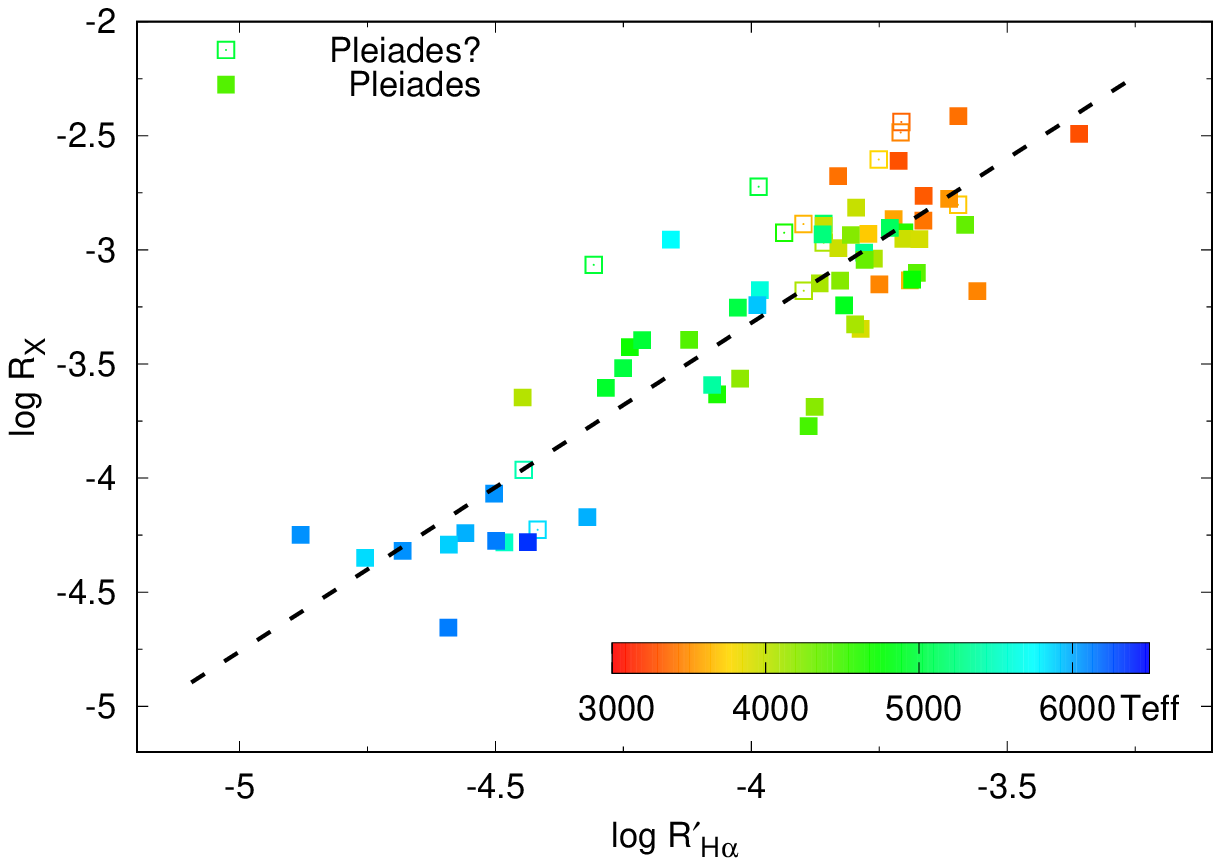}
\caption{The relations between H$\alpha$ fractional luminosity $R^{'}_{\text{H}\alpha}$ and X-ray fractional luminosity $R_{\text{X}}$.}
\label{fig:lha_lxr}
\end{figure}

To investigate the relation between chromospheric heating and coronal emission, 
we have taken $R_{\text{X}}$ ($=L_{\text{X}}/L_{\text{bol}}$) from \citet{wrig2011} for common Pleiades candidate members, 
and plotted against $R^{'}_{\text{H}\alpha}$ in Fig.~\ref{fig:lha_lxr}. 
A simple linear fit in log scale indicated that the power law index is around 1.44, e.g., 
$R^{'}_{\text{H}\alpha}\propto(R^{'}_{\text{H}\alpha})^{1.44}$ (see black dashed line in Fig.~\ref{fig:lha_lxr}). 
A slight steeper correlation of coronal and chromospheric emission flux with a power of 1.5-1.7 based on magnetic induced excess 
flux in the Ca~{\sc ii} H\&K lines was reported \citep[][and further references cited therein]{schr1992,gude2004}. 
Such a non-linearity in the chromospheric-coronal flux-flux correlation may be a manifestation of the chromospheric radiative losses \citep{gude2004}. 
The correlation between H$\alpha$ emission and X-ray emission indicates that the formation process may be same or at least correlated.
%%%%%%%%%%%
%%%%%%%%%%%
\subsection{What can we learn from the results}
The different onsets of saturation in various chromospheric activity indicators suggests the existence of transition regime at Ro range of 0.1-0.4, 
which connects the unsaturated and saturated regime with steep and shallower power law, respectively. In fact, 
\citet{gond2012,gond2013} showed the evidence for transition from saturated to non-saturated X-ray emission that occurs 
at Ro between $\sim$0.13 and 0.4 in open clusters M34 and M35. Indeed, this transition scenario could describe the scatter in Ro region of 0.1-0.4. 
However, such scenario fails to explain the whole observational features in this Ro regime, 
e.g. the Ca K emission that shows a clear rotation-activity relationship, 
suggests the investigation towards other unknown parameters or the activity behaviours in this regime. 

In addition, the chromospheric emission follow the similar (slight shallower) power laws in the unsaturated regime compared to coronal X-ray emission, 
but appears to be shifted to a larger Ro$_{\text{sat}}$ value, 
which indicates that chromospheric emission gets easily saturated at a given rotation rate than coronal emission. 
Alternatively, it may suggest that the convective overturn time scale, empirically derived based on X-ray emission \citep{wrig2011}, 
is not suitable for chromospheric emission.

A corresponding relation between the saturated/non-saturated regimes of X-ray emission and the C/I rotation sequences among 
late-type stars has been reported in literature \citep[e.g.][]{barn2003b,gond2012,gond2013}. 
In our work, we detected a similar correspondence in H$\alpha$ emission among cool stars in Pleiades, Praesepe and Hyades, 
as discussed in previous sections. \citet{barn2003b} and \citet{gond2012} suggested such transition from the saturated 
to the non-saturated regime of activity is the manifestation of a dynamo transition, e.g., from a turbulent dynamo to an interface-type dynamo. 
\citet{durn1993} showed a candidate turbulent dynamo, which can generate small-scale turbulent magnetic field in late-type stars, 
and have a weak dependency on rotation. Such a turbulent dynamo can naturally explain the weak dependency of activity 
on rotation among fast rotators in the saturation regime. 
Moreover, the generation of small-scale magnetic field supports some observational behaviour shown by very active stars, 
e.g., large spot coverage of stars in young open clusters may be due to many small and randomly located 
spots on the stellar photosphere \citep{jack2012,jack2013,fang2016}, 
and the possibility of multiple small flares such as nanoflares that might be responsible for 
the departures of relation between various chromospheric emissions among very active late-K and M dwarfs reported by 
previous studies and discussed above in current work. 
We thus conclude that a turbulent-like dynamo dominates in fast rotating cool stars in the saturated regime.

%%%%%%%%%%%%%%%%%%%%%%%
\section{Conclusion}
We have quantified CA (excess emissions in H$\alpha$, H$\beta$, and Ca~{\sc ii} K lines) in more than 700 stars from various open clusters including Pleiades, 
M34, Praesepe and Hyades, using LAMOST DR3 spectra. The CA was investigated as a function of stellar parameters: mass, age, and rotation. 
We also analysed the correlation between activity indicators of various atmosphere layers. 

CA was found to be overall increased as temperature decreases, showing a mass-dependency, which partially could be due to deeper convective zones in cooler stars. 
CA also shows a clear age-dependence, i.e., activity level of GK-type stars declines steadily from 100 Myrs to about 700 Myrs. 
Additionally, we found that the activity lifetime for M1-M2 type stars is around 600-700 Myr, which is consistent with previous studies. More interestingly, 
there appear two sequences among Pleiades GK-type stars in activity (spot coverage, chromospheric and coronal emission)-$T_{\text{eff}}$ diagram, 
paralleling with well known rotation sequences. We also found the existence of similar sequences in Praesepe and Hyades M-type members.

We found CA is correlated with rotation in the unsaturated regime, following a power law with power of $\beta$ between -2 and -2.5, 
which is marginally consistent with results from X-ray emissions, but with a larger turnoff point of Ro (e.g., $\text{Ro}_{\text{sat}}\sim0.2-0.3$).
A weak dependency of CA on rotation is noticed in the saturated regime of activity (Ro$\lesssim$0.1), 
e.g., a slight increase as Ro decreases (a power law with $\beta\approx0.2$), peaks around Ro$\sim$0.01, and decrease when Ro decreases for Ro$\lesssim$0.01. 
In addition, our analysis confirms the previous finding that slowly rotating, fully convective stars also follow the activity-rotation trend like the hotter stars. 

We found a good linear correlation between the luminosity emitted at H$\alpha$ and H$\beta$. 
The relations from excess emission of H$\alpha$ to that Ca~{\sc ii} K lines confirms that cool stars follow two different power law flux-flux relationships.
We also found weak correlation between chromospheric emission and photospheric activity indicators (cool spot filling factor and light variation amplitude), 
however, it shows dependency on spectral type and activity level, which provides clues on how spot configuration vary as a function of mass and activity. 
A good correlation between chromospheric and coronal emissions was also detected that confirms previous findings.

\section*{Acknowledgements}
We thank the referee for his/her observant comments and constructive suggestions that helps to improve the manuscript. 
This study is supported by the National Natural Science Foundation of China under grant No. 11390371, 11233004, U1431106, 11573035, 11625313, 11550110492, 
the National Key Basic Research Program of China (973 program) 2014CB845701. 
Y.B.K is thankful to the Chinese Academy of Sciences for supporting through CAS PIFI Fellowship. 
This work has made use of LAMOST data. The Guo Shou Jing Telescope (the Large sky Area 
Multi-Object fiber Spectroscopic Telescope, LAMOST) is a National Major Scientific Project built by the Chinese Academy of Sciences. Funding for the project 
has been provided by the National Development and Reform Commission. LAMOST is operated and managed by National Astronomical Observatories, Chinese Academy 
of Sciences. This research has made use of the VizieR catalogue access tool and the cross-match service provided by CDS, Strasbourg, France. This research 
has made use of the WEBDA database, operated at the Department of Theoretical Physics and Astrophysics of the Masaryk University.

%%%%%%%%%%%%%%%%%%%%%%%%%%%%%%%%%%%%%%%%%%%%%%%%%%



\bibliographystyle{mnras}

\begin{thebibliography}{99}

\bibitem[\protect\citeauthoryear{Ag{\"u}eros et al.}{2011}]{ague2011}
Ag{\"u}eros M. A., Covey K. R., Lemonias J. J., et al. 2011, \apj, 740, 110

\bibitem[\protect\citeauthoryear{An et al.}{2007}]{an++2007}
An D., Terndrup D. M., Pinsonneault M. H., Paulson D. B., Hanson R. B. \& Stauffer J. R. 2007, \apj, 655, 233

\bibitem[\protect\citeauthoryear{Baliunas et al.}{1995}]{bali1995}
Baliunas S. L., Donahue R. A., Soon W. H., et al., 1995, \apj, 438, 269

\bibitem[\protect\citeauthoryear{Baliunas et al.}{1998}]{bali1998}
Baliunas S. L., Donahue R. A., Soon W. \& Henry G. W. 1998, Cool Stars, Stellar Systems, and the Sun, 154, 153

\bibitem[\protect\citeauthoryear{Barnes}{2003a}]{barn2003a}
Barnes S. A. 2003a, \apj, 586, 464

\bibitem[\protect\citeauthoryear{Barnes}{2003b}]{barn2003b}
Barnes S. A. 2003b, \apjl, 586, L145

\bibitem[\protect\citeauthoryear{Bouy et al.}{2015}]{bouy2015}
Bouy H. et al., 2015, \aap, 577, 148

\bibitem[\protect\citeauthoryear{Brandt \& Huang}{2015}]{bran2015}
Brandt T. D. \& Huang C. X. 2015, \apj, 807, 24

\bibitem[\protect\citeauthoryear{Canterna, Crawford \& Perry}{1979}]{cant1979}
Canterna R., Crawford D. L., \& Perry C. L. 1979, \pasp, 91, 263

\bibitem[\protect\citeauthoryear{Carrera \& Pancino}{2011}]{carr2011}
Carrera R., \& Pancino E. 2011, \aap, 535, 30

\bibitem[\protect\citeauthoryear{Charbonneau}{2014}]{char2014}
Charbonneau P. 2014, \araa, 52, 251

\bibitem[\protect\citeauthoryear{Chen et al.}{2014}]{chen2014}
Chen Y., Girardi L., Bressan A., Marigo P., Barbieri M. \& Kong X. 2014, \mnras, 444, 2525

\bibitem[\protect\citeauthoryear{Copenhagen University et al.}{2011}]{cope2011}
Copenhagen University, Institute of Astronomy, Cambridge, UK, Real Instituto y Observatorio de la Armada en San Fernando 2011, yCat, 1327, 0

\bibitem[\protect\citeauthoryear{Covey et al.}{2016}]{cove2016}
Covey K. R., Agüeros M. A., Law N. M., et al. 2016, \apj, 822, 81

\bibitem[\protect\citeauthoryear{Cram \& Mullan}{1979}]{cram1979}
Cram L. E., \& Mullan D. J. 1979, \apj, 234, 579

\bibitem[\protect\citeauthoryear{Cui et al.}{2012}]{cui+2012}
Cui X.-Q et al., 2012, Res. Astron. Astrophys., 12, 1197

\bibitem[\protect\citeauthoryear{Cutri et al.}{2003}]{cutr2003}
Cutri R. M., et al. 2003, VizieR Online Data Catalog, 2246, 0

\bibitem[\protect\citeauthoryear{Delfosse et al.}{1998}]{delf1998}
Delfosse X., Forveille T., Perrier C. \& Mayor M. 1998, \aap, 1998, 331, 581

\bibitem[\protect\citeauthoryear{Delorme et al.}{2011}]{delo2011}
Delorme P., Collier Cameron A., Hebb L., Rostron J., Lister T. A., Norton A. J., Pollacco D. \& West R. G. 2011, \mnras, 413, 2218

\bibitem[\protect\citeauthoryear{Douglas et al.}{2014}]{doug2014}
Douglas S. T., Agüeros M. A., Covey K. R. 2014, \apj, 795, 161

\bibitem[\protect\citeauthoryear{Douglas et al.}{2016}]{doug2016}
Douglas S. T., Agüeros M. A., Covey K. R., Cargile P. A., Barclay T., Cody A., Howell S. B. \& Kopytova T. 2016, \apj, 822, 47

\bibitem[\protect\citeauthoryear{Durney, De Young \& Roxburgh}{1993}]{durn1993}
Durney B. R., De Young D. S. \& Roxburgh, I. W. 1993, Solar Physics., 145, 207

\bibitem[\protect\citeauthoryear{Fang et al.}{2016}]{fang2016}
Fang X.-S., Zhao G., Zhao J.-K., Chen Y.-Q., Bharat Kumar Y., 2016, \mnras, 463, 2494 

\bibitem[\protect\citeauthoryear{Fossati et al.}{2008}]{foss2008}
Fossati L., Bagnulo S., Landstreet J., Wade G., Kochukhov O., Monier R.,
Weiss W., Gebran M., 2008, \aap, 483, 891

\bibitem[\protect\citeauthoryear{Gallet \& Bouvier}{2013}]{gall2013}
Gallet F. \& Bouvier J. 2013, \aap, 556, A36

\bibitem[\protect\citeauthoryear{Geller, Latham \& Mathieu}{2015}]{gell2015}
Geller A. M., Latham D. W. \& Mathieu R. D. 2015, \aj, 150, 97

\bibitem[\protect\citeauthoryear{Gondoin}{2012}]{gond2012}
Gondoin P. 2012, \aap, 546, A117

\bibitem[\protect\citeauthoryear{Gondoin}{2013}]{gond2013}
Gondoin P. 2013, \aap, 556, A14 

\bibitem[\protect\citeauthoryear{Gray}{2003}]{gray2003}
Gray R. O., Corbally C. J., Garrison R. F., McFadden M. T. \& Robinson P. E. 2003, \aj, 126, 2048

\bibitem[\protect\citeauthoryear{Gray}{2006}]{gray2006}
Gray R. O., Corbally C. J., Garrison R. F., McFadden M. T., Bubar E. J., McGahee C. E., O'Donoghue A. A. \& Knox E. R. 2006, \aj, 132, 161

\bibitem[\protect\citeauthoryear{G{\"u}del}{2004}]{gude2004}
G{\"u}del M. 2004, \aapr, 12, 71

\bibitem[\protect\citeauthoryear{Hall \& Ramsey }{1992}]{hall1992}
Hall J. C. \& Ramsey L. W. 1992, \aj, 104, 1942

\bibitem[\protect\citeauthoryear{Hall}{2008}]{hall2008}
Hall J. C. 2008, Living Rev Solar Phys, 5, 2

\bibitem[\protect\citeauthoryear{Hartman et al.}{2010}]{hart2010}
Hartman J. D., Bakos G. {\'A}., Kov{\'a}cs G., Noyes R. W., 2010, \mnras, 408, 475

\bibitem[\protect\citeauthoryear{Hartman et al.}{2011}]{hart2011}
Hartman J. D., Bakos G. {\'A}., Noyes R. W., SipH ocz B., Kov{\'a}cs G., Mazeh T., Shporer A. \& P{\'a}l A. 2011, \aj, 141, 166

\bibitem[\protect\citeauthoryear{Hawley et al.}{1999}]{hawl1999}
Hawley S. L., Reid I. N., Gizis J. E., \& Byrne P. B. 1999, in Butler C. J., Doyle J. G., eds, ASP Conf. Ser. Vol. 158, 
Solar and Stellar Activity: Similarities and Differences. p. 63

\bibitem[\protect\citeauthoryear{Henry et al.}{1996}]{henr1996}
Henry T. J., Soderblom D. R., Donahue R. A. \& Baliunas S. L. 1996, \aj, 111, 439

\bibitem[\protect\citeauthoryear{Hodgkin et al.}{1995}]{hodg1995}
Hodgkin S. T., Jameson R. F., Steele I. A., 1995, \mnras, 274, 869

\bibitem[\protect\citeauthoryear{Huenemoerder \& Ramsey}{1987}]{huen1987}
Huenemoerder D. P. \& Ramsey L. W. 1987, \apj,  319, 392

\bibitem[\protect\citeauthoryear{Husser et al.}{2013}]{huss2013}
Husser T.-O., Wende-von Berg, S., Dreizler S., Homeier D., Reiners A., Barman T. \& Hauschildt P. H. 2013, \aap, 553, A6

\bibitem[\protect\citeauthoryear{Irwin et al.}{2006}]{irwi2006}
Irwin J., Aigrain S., Hodgkin S., Irwin M., Bouvier J., Clarke C., Hebb L. \& Moraux E. 2006, \mnras, 370, 954

\bibitem[\protect\citeauthoryear{Jackson \& Jeffries}{2010}]{jack2010}
Jackson R. J. \& Jeffries R. D., 2010, \mnras, 407, 465

\bibitem[\protect\citeauthoryear{Jackson \& Jeffries}{2012}]{jack2012}
Jackson R. J. \& Jeffries R. D. 2012, \mnras, 423, 2966

\bibitem[\protect\citeauthoryear{Jackson \& Jeffries}{2013}]{jack2013}
Jackson R. J. \& Jeffries R. D. 2012, \mnras, 431, 1883

\bibitem[\protect\citeauthoryear{Jacobson, Pilachowski \& Friel}{2011}]{jaco2011}
Jacobson H. R., Pilachowski C. A. \& Friel E. D. 2011, \aj, 142, 59

\bibitem[\protect\citeauthoryear{James et al.}{2010}]{jame2010}
James D. J., Barnes S. A., Meibom S., et al. 2010, \aap, 515, 100

\bibitem[\protect\citeauthoryear{Jones \& Prosser}{1996}]{jone1996}
Jones B. F. \& Prosser C. F. 1996, \aj, 111, 1193

\bibitem[\protect\citeauthoryear{Kov{\'a}cs et al.}{2014}]{kova2014}
Kov{\'a}cs G., Hartman J. D., Bakos G. {\'A}., et al. 2014, \mnras, 442, 2081

\bibitem[\protect\citeauthoryear{Lee et al.}{2008}]{lee+2008}
Lee Y. S., Beers T. C., \& Sivarani T., et al. 2008, AJ, 136, 2050

\bibitem[\protect\citeauthoryear{Liu et al.}{2015}]{liu+2015}
Liu X.-W., Zhao G. \& Hou J.-L., 2015, Res. Astron. Astrophys., 15, 1089

\bibitem[\protect\citeauthoryear{Lockwodd, Skiff \& Radick}{1997}]{lock1997}
Lockwood G. W., Skiff B. A., \& Radick R. R. 1997, \apj, 485, 789

\bibitem[\protect\citeauthoryear{Lockwodd et al.}{2007}]{lock2007}
Lockwood G. W., Skiff B. A., Henry G. W., Henry S., Radick R. R., Baliunas S. L., Donahue R. A. \& Soon W. 2007, \apjs, 171, 260

\bibitem[\protect\citeauthoryear{Luo et al.}{2015}]{luo+2015}
Luo A.-L. et al., 2015, Res. Astron. Astrophys., 15, 1095

\bibitem[\protect\citeauthoryear{Mamajek \& Hillenbrand}{2008}]{mama2008}
Mamajek E. E. \& Hillenbrand L. A. 2008, \apj, 687, 1264

\bibitem[\protect\citeauthoryear{Marsden, Carter \& Donati}{2009}]{mars2009}
Marsden S. C., Carter B. D., \& Donati J.-F 2009, \mnras, 399, 888

\bibitem[\protect\citeauthoryear{Mart{\'i}nez-Arn{\'a}iz et al.}{2011}]{mart2011}
Mart{\'i}nez-Arn{\'a}iz, R., L{\'o}pez-Santiago, J., Crespo-Chac{\'o}n, I. \& Montes, D. 2011, \mnras, 414, 2629

\bibitem[\protect\citeauthoryear{McQuillan, Mazeh \& Aigrain}{2014}]{mcqu2014}
McQuillan A., Mazeh T. \& Aigrain S. 2014, \apjs, 211, 24

\bibitem[\protect\citeauthoryear{Meibom et al.}{2011}]{meib2011}
Meibom S., Mathieu R. D., Stassun K. G., Liebesny P. \& Saar S. H. 2011, \apj, 733, 115

\bibitem[\protect\citeauthoryear{Melis et al.}{2014}]{meli2014}
Melis C., Reid M. J., Mioduszewski A. J., Stauffer J. R., Bower G. C., 2014, \sci, 345, 1029

\bibitem[\protect\citeauthoryear{Mermilliod, Mayor \& Udry}{2009}]{merm2009}
Mermilliod J.-C., Mayor M., \& Udry S. 2009, \aap, 498, 949

\bibitem[\protect\citeauthoryear{Montes et al.}{1995}]{mont1995}
Montes D., Fern{\'a}ndez-Figueroa M. J., de Castro E. \& Cornide M. 1995, \aap, 294, 165

\bibitem[\protect\citeauthoryear{Montes et al.}{1996}]{mont1996}
Montes D., Fern{\'a}ndez-Figueroa M. J., Cornide M. \& de Castro E. 1996, \aap, 312, 221

\bibitem[\protect\citeauthoryear{Montes et al.}{2004}]{mont2004}
Montes D., Crespo-Chac{\'o}n I., G{\'a}lvez M. C., Fern{\'a}ndez-Figueroa M. J., L{\'o}pez-Santiago J., 
de Castro E., Cornide M. \& Hern{\'a}n-Obispo M. 2004, Lecture Notes and Essays in Astrophysics, 1, 119

\bibitem[\protect\citeauthoryear{Newton et al.}{2017}]{newt2017}
Newton E. R., Irwin J., Charbonneau D., Berlind P., Calkins M. L. \& Mink J. 2017, \apj, 834, 85

\bibitem[\protect\citeauthoryear{Noyes et al.}{1984}]{noye1984}
Noyes R. W., Hartmann L. W., Baliunas S. L., Duncan D. K., Vaughan A. H. 1984, \apj, 279, 763

\bibitem[\protect\citeauthoryear{Önehag et al.}{2011}]{oneh2011}
Önehag A., Korn A., Gustafsson B., Stempels E., \& Vandenberg D. A. 2011, \aap, 528, A85

\bibitem[\protect\citeauthoryear{Perryman et al.}{1998}]{perr1998}
Perryman, M. A. C., Brown, A. G. A., Lebreton, Y., et al. 1998, \aap, 331, 81

\bibitem[\protect\citeauthoryear{Pizzolato et al.}{2003}]{pizz2003}
Pizzolato N., Maggio A., Micela G., Sciortino S. \& Ventura P. 2003, \aap,  397,

\bibitem[\protect\citeauthoryear{Radick et al.}{1987}]{radi1987}
Radick R. R., Thompson D. T., Lockwood G. W., Duncan D. K. \& Baggett W. E. 1987, \apj, 321, 459

\bibitem[\protect\citeauthoryear{Radick et al.}{1998}]{radi1998}
Radick R. R., Lockwood G. W., Skiff B. A. \& Baliunas S. L. 1998, \apjs, 118, 239

\bibitem[\protect\citeauthoryear{Rebull et al.}{2016}]{rebu2016}
Rebull L. M., Stauffer J. R., Bouvier J., et al. 2016, \aj, 152, 113

\bibitem[\protect\citeauthoryear{Reid, Hawley \& Mateo}{1995}]{reid1995}
Reid N., Hawley S. L. \& Mateo M. 1995, \mnras, 272, 828

\bibitem[\protect\citeauthoryear{Reiners, Joshi \& Goldman}{2012}]{rein2012}
Reiners A., Joshi N. \& Goldman B. 2012, \aj, 143, 93

\bibitem[\protect\citeauthoryear{Reiners, Schüssler \& Passegger}{2014}]{rein2014}
Reiners A., Schüssler M. \& Passegger V. M. 2014, \apj, 794, 144

\bibitem[\protect\citeauthoryear{Randich et al.}{1996}]{rand1996}
Randich S., Schmitt J. H. M. M., Prosser C. F. \& Stauffer J. R. 1996, \aap, 305, 785

\bibitem[\protect\citeauthoryear{Salaris, Weiss \& Percival}{2004}]{sala2004}
Salaris M., Weiss A., \& Percival S. M. 2004, \aap, 414, 163

\bibitem[\protect\citeauthoryear{Scholz \& Eislöffel}{2007}]{scho2007}
Scholz A. \& Eislöffel J. 2007, \mnras, 381, 1638

\bibitem[\protect\citeauthoryear{Scholz et al.}{2011}]{scho2011}
Scholz A., Irwin J., Bouvier J., SipH ocz B. M., Hodgkin S. \& Eislöffel J. 2011, \mnras, 413, 2595

\bibitem[\protect\citeauthoryear{Schrijver, Dobson \& Radick}{1992}]{schr1992}
Schrijver C. J., Dobson A. K., \& Radick R. R. 1992, \aap, 258, 432

\bibitem[\protect\citeauthoryear{Schuler et al.}{2003}]{schu2003}
Schuler S. C., King J. R., Fischer D. A., Soderblom D. R., \& Jones B. F. 2003, \aj, 125, 2085

\bibitem[\protect\citeauthoryear{Smolinski et al.}{2011}]{smol2011}
Smolinski J. P., Lee Y. S., Beers T. C., et al. 2011, \aj, 141, 89

\bibitem[\protect\citeauthoryear{Soderblom et al.}{1993}]{sode1993}
Soderblom D. R., Stauffer J. R., Hudon J. D., Jones B. F., 1993, \apjs, 85, 315

\bibitem[\protect\citeauthoryear{Soderblom, Jones \& Fischer}{2001}]{sode2001}
Soderblom, D. R., Jones, B. F. \& Fischer, D. 2001, \apj, 563, 334

\bibitem[\protect\citeauthoryear{Soderblom et al.}{2009}]{sode2009}
Soderblom D. R., Laskar T., Valenti J. A., Stauffer J. R., Rebull L. M., 2009, \aj, 138, 1292

\bibitem[\protect\citeauthoryear{Spiegel \& Zahn}{1992}]{spie1992}
Spiegel E. A. \& Zahn J.-P. 1992, \aap, 265, 106

\bibitem[\protect\citeauthoryear{Stauffer \& Hartmann}{1986}]{stau1986}
Stauffer J. R. \& Hartmann L. W. 1986, \apjs, 61, 531

\bibitem[\protect\citeauthoryear{Stauffer \& Hartmann}{1987}]{stau1987}
Stauffer J. R. \& Hartmann L. W. 1987, \apj, 318, 337

\bibitem[\protect\citeauthoryear{Stauffer et al.}{1991}]{stau1991}
Stauffer J. R., Giampapa M. S., Herbst W., Vincent J. M., Hartmann L. W. \& Stern R. A. 1991, \apj, 374, 142

\bibitem[\protect\citeauthoryear{Stauffer et al.}{1997}]{stau1997}
Stauffer J. R., Balachandran S. C., Krishnamurthi A., Pinsonneault M., Terndrup D. M. \& Stern R. A 1997, \apj, 475, 604

\bibitem[\protect\citeauthoryear{Stauffer, Schultz \& Kirkpatrick}{1998}]{stau1998}
Stauffer J. R., Schultz G. \& Kirkpatrick J. D. 1998, \apjl, 499, L199

\bibitem[\protect\citeauthoryear{Stauffer et al.}{2007}]{stau2007}
Stauffer J. R., Hartmann L. W., Fazio, G. G., et al. 2007, \apjs, 172, 663

\bibitem[\protect\citeauthoryear{Taylor \& Joner}{2005}]{tayl2005}
Taylor B. J., \& Joner M. D. 2005, \apjs, 159, 100

\bibitem[\protect\citeauthoryear{Taylor}{2006}]{tayl2006}
Taylor B. J. 2006, \aj, 132, 2453

\bibitem[\protect\citeauthoryear{van Leeuwen}{2009}]{vanl2009}
van Leeuwen, F. 2009, \aap, 497, 209

\bibitem[\protect\citeauthoryear{Vaughan \& Preston}{1980}]{vaug1980}
Vaughan A. H. \& Preston G. W. 1980, \pasp, 92, 385

\bibitem[\protect\citeauthoryear{Walkowicz, Hawley \& West}{2004}]{walk2004}
Walkowicz L. M., Hawley S. L., \& West A. A. 2004, \pasp, 116, 1105

\bibitem[\protect\citeauthoryear{West et al.}{2004}]{west2004}
West A. A., Hawley S. L., Walkowicz L. M., et al. 2004, \aj, 128, 426

\bibitem[\protect\citeauthoryear{West \& Hawley}{2008}]{west2008}
West A. A., \& Hawley S. L. 2008, \pasp, 120, 1161

\bibitem[\protect\citeauthoryear{West et al.}{2008}]{west+2008}
West A. A., Hawley S. L., Bochanski J. J., Covey K. R., Reid I. N., Dhital S., Hilton E. J. \& Masuda M. 2008, \aj, 135, 785

\bibitem[\protect\citeauthoryear{West et al.}{2015}]{west2015}
West A. A., Weisenburger K. L., Irwin J., Berta-Thompson Z. K., Charbonneau D., Dittmann J. \& Pineda J. S. 2015, \apj, 812, 3

\bibitem[\protect\citeauthoryear{Wilson}{1968}]{wils1968}
Wilson O. C. 1968, \apj, 153, 221

\bibitem[\protect\citeauthoryear{Wilson}{1978}]{wils1978}
Wilson O. C. 1978, \apj, 226, 379

\bibitem[\protect\citeauthoryear{Wright et al.}{2011}]{wrig2011}
Wright N. J., Drake J. J., Mamajek E. E. \& Henry G. W. 2011, \apj, 743, 48

\bibitem[\protect\citeauthoryear{Wright \& Drake}{2016}]{wrig2016} 
Wright N. J. \& Drake J. J. 2016, \nat, 535, 526

\bibitem[\protect\citeauthoryear{Wu et al.}{2011}]{wu++2011}
Wu Y., Luo A.-L., Li H.-N., et al. 2011, Res. Astron. Astrophys., 11, 924

\bibitem[\protect\citeauthoryear{Yang et al.}{2016}]{yang2016}
Yang X. L., Chen Y. Q. \& Zhao G. 2015, \aj, 150, 158

\bibitem[\protect\citeauthoryear{Zacharias et al.}{2012}]{zach2012}
Zacharias N., et al. 2012, VizieR Online Data Catalog, 1322, 0

\bibitem[\protect\citeauthoryear{Zhao et al.}{2006}]{zhao2006}
Zhao G., Chen Y.-Q., Shi J.-R., Liang Y.-C., Hou J.-L., Chen L., Zhang H.-W. \& Li A.-G., 2006, ChJAA, 6, 265

\bibitem[\protect\citeauthoryear{Zhao et al.}{2012}]{zhao2012}
Zhao G., Zhao Y.-H., Chu Y.-C., Jing Y.-P., Deng L.-C., 2012, Res. Astron. Astrophys., 12, 723

%%%%%%%%%%%%%%%%%%%%%%%%%%%%%%%%%%%%%%%%%%%
\end{thebibliography}


%%%%%%%%%%%%%%%%%%%%%%%%%%%%%%%%%%%%%%%%%%%%%%%%%%

%%%%%%%%%%%%%%%%% APPENDICES %%%%%%%%%%%%%%%%%%%%%
\appendix

\section{Chromospheric indicators: C\lowercase{a~{\sc ii}} K, H$\beta$}
\label{sec:otherlines}
\begin{figure*}
\centering
\includegraphics[width=\columnwidth]{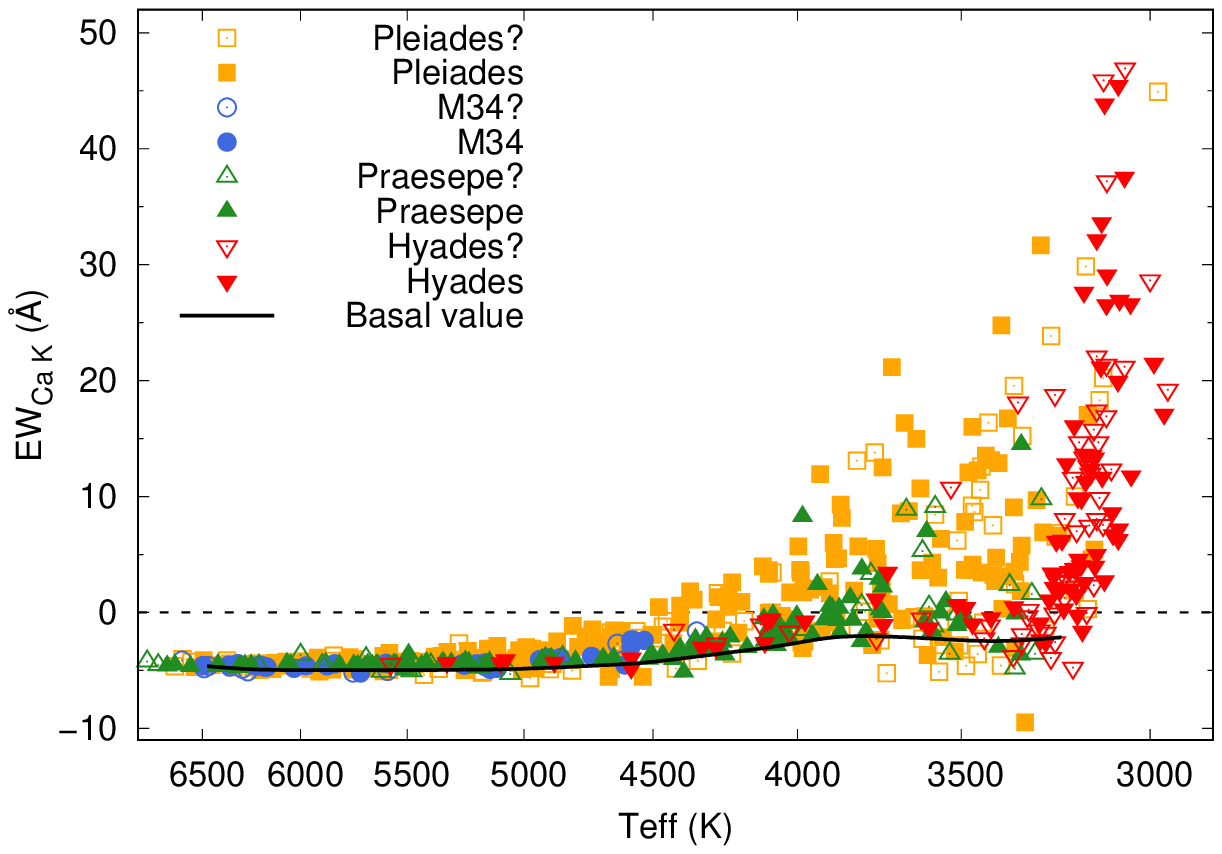}
\includegraphics[width=\columnwidth]{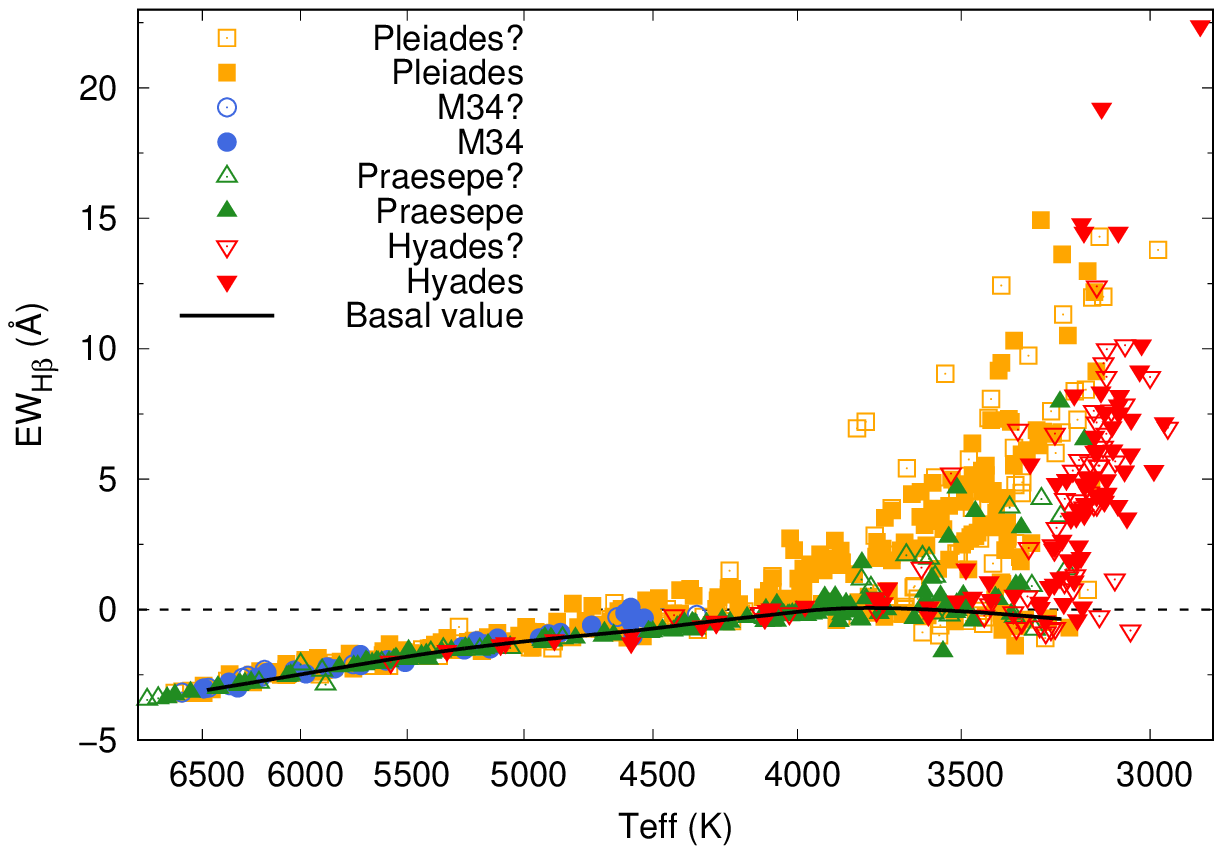}
\includegraphics[width=\columnwidth]{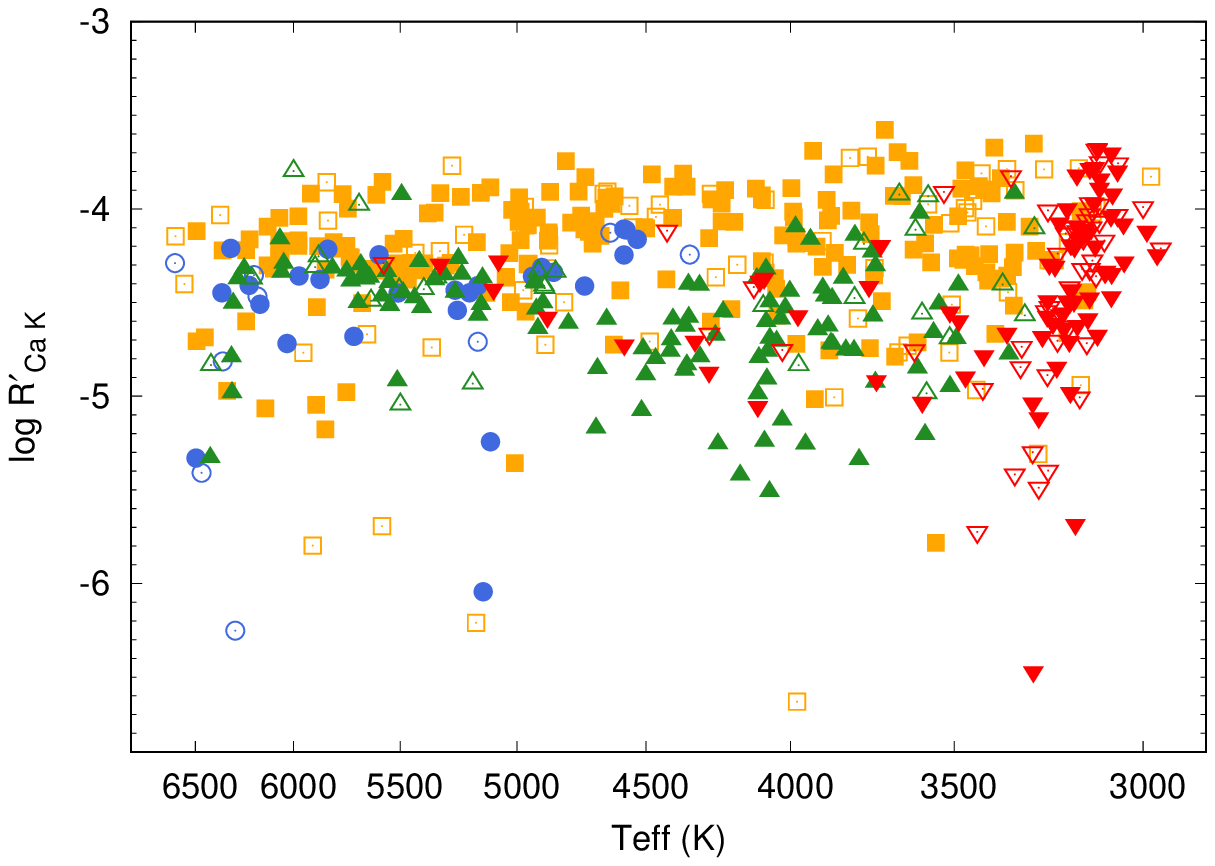}
\includegraphics[width=\columnwidth]{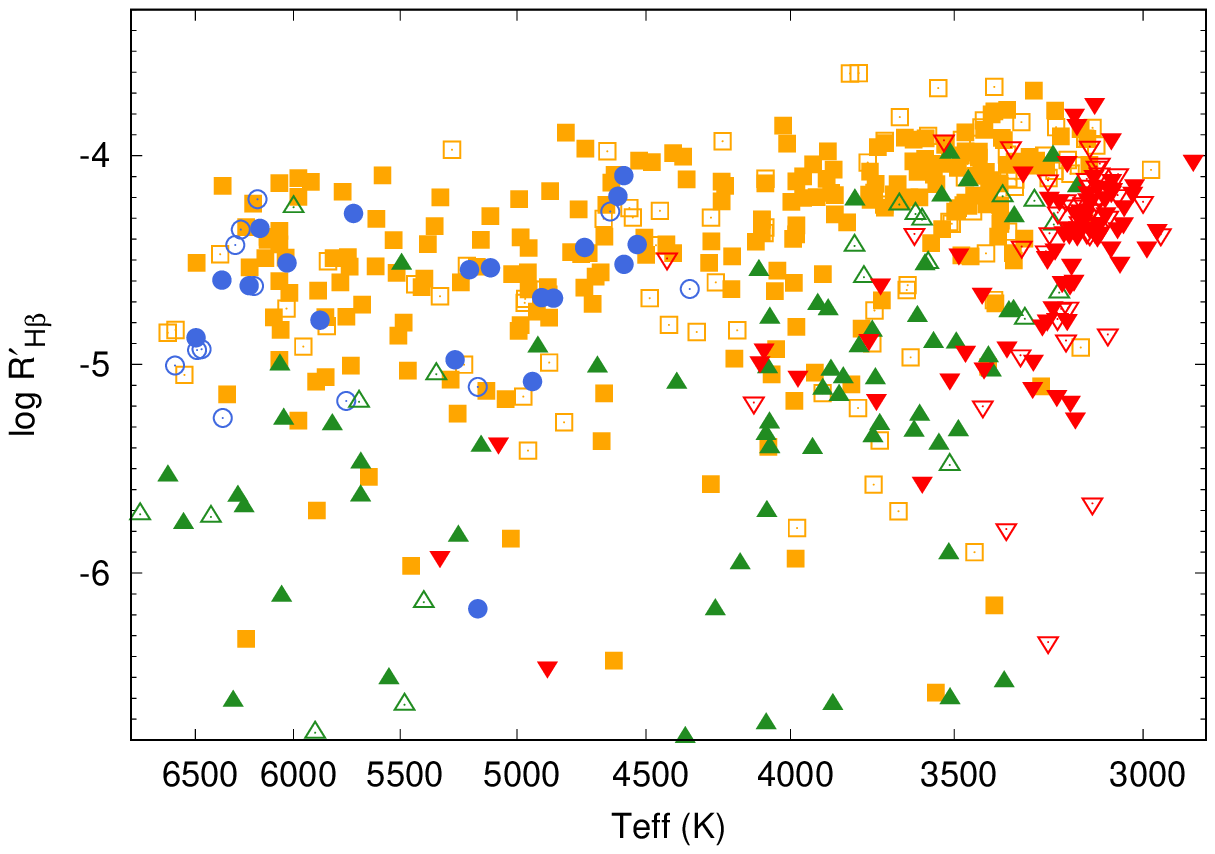}
\caption{Emissions of H$\beta$ and Ca {\sc ii} K lines for all sample stars.}
\label{fig:ewxx}
\end{figure*}

We measured equivalent widths of Ca~{\sc ii} K and H${\beta}$ lines, EW$_{\text{Ca K}}$ and EW$_{\text{H}\beta}$, 
and wavelength regime for these measurements are listed in Table~\ref{tab:ew_defin}. 
To get reliable measurements for these two blue lines, we merely measured EW$_{\text{Ca K}}$ for stars with $g$-band 
signal-to-noise ratio SNRg$>1.0$, and measured EW$_{\text{H}\beta}$ for those with SNRg$>3.0$. Such SNRg trim criteria cut off about 18 percent and 3 
percent of total sample stars, for Ca~{\sc ii} K and H${\beta}$, respectively. Fig.~\ref{fig:ewxx} shows the measurements of EW$_{\text{Ca K}}$ and 
EW$_{\text{H}\beta}$ as a function of effective temperature. Note that the corresponding basal values are also plotted, 
which were derived based on a large sample of inactive reference dwarf stars with solar metallicity, 
following the similar procedure that used to derive mean basal values for H$\alpha$ (see Paper I for more details). 
Similar to EW$^{'}_{\text{H}\alpha}$, we obtained the excess equivalent widths of the Ca~{\sc ii} K and H${\beta}$ lines, 
EW$^{'}_{\text{Ca K}}$ and EW$^{'}_{\text{H}\beta}$, respectively. Further, we derived the excess fractional luminosities, 
$R^{'}_{\text{Ca K}}$($\equiv L^{'}_{\text{Ca K}}/L_{\text{bol}}$) and $R^{'}_{\text{H}\beta}$ ($\equiv L^{'}_{\text{H}\beta}/L_{\text{bol}}$), 
by using corresponding $\chi$ ratios (see Appendix~\ref{sec:chi}). These measurements are shown in bottom panels of Fig.~\ref{fig:ewxx}.

Note that the derived mean value of inactive reference stars cut off at 6500 K. 
However, a very small fraction of sample stars are hotter than 6500 K (see Fig.~\ref{fig:sample_data}). 
We simply linearly extrapolated the reference value when necessary. 
In order to get reliable measurements for sample stars with $T_{\text{eff}}>6500$ K, 
we merely derived the excess equivalent widths for those stars with $T_{\text{eff}}$ less than 6900 K in this work.
%%%%%%%%%%%%%%%%%%%%%%%
%%%%%%%%%%%%%%%%%%%%%%%%%%%%%%%
\begin{table}
\centering
\caption{Equivalent width measurements of H${\alpha}$, H${\beta}$ and Ca~{\sc ii} K lines}
\label{tab:ew_defin}
\begin{tabular}{lcccccc}
   \hline
Line          & Line bandpass (\AA) &  Pseudo-continua (\AA) \\
   \hline
H$\alpha$     &  6557-6569          &   6547-6557,  6570-6580\\
H$\beta$      &  4855-4867          &   4842-4852,  4873-4883\\
Ca~{\sc ii} K &  3930-3937          &   3910-3915,  3950-3955 \\
   \hline
\end{tabular}
\end{table}
%%%

\section{Measurement Uncertainties}
\label{sec:errors}
\begin{figure}
\centering
\includegraphics[width=\columnwidth]{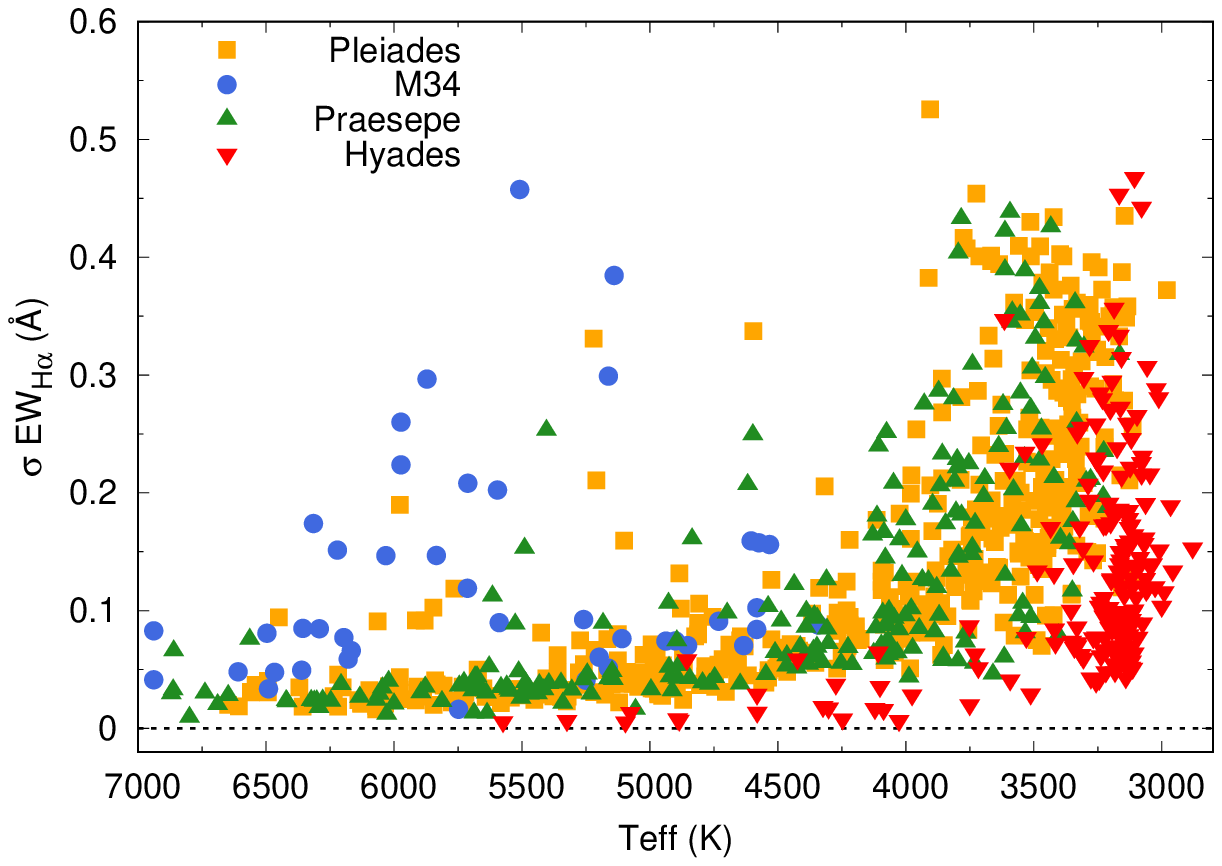}
\caption{Errors in measurements of EW$_{\text{H}\alpha}$. Note Hyades members have smaller errors because of the proximity.}
\label{fig:error}
\end{figure}

The uncertainties in the equivalent width measurements mainly result from the errors in the RV-corrected LAMOST spectra. 
The uncertainty in the LAMOST spectra at each wavelength (Poisson noise, and any noises during data reduction) is available and was 
directly propagated to the equivalent width measurements. Fig.~\ref{fig:error} shows the errors ($\sigma \text{EW}_{\text{H}\alpha}$) 
in EW$_{\text{H}\alpha}$ for stars in each open cluster, and the increment in errors with decreasing temperature is seen as expected. 
The typical error of EW$_{\text{H}\alpha}$ is less than 0.1~\AA~for GK-type sample 
stars, e.g., $\sim$0.05~\AA~for Pleiades members with temperatures around 5000 K; however, it become large for M-type stars, e.g., typically greater than 
0.1~\AA~and up to 0.5~\AA. Also, the lower SNR in the blue band of spectra increases the uncertainties in the measurements of Ca {\sc ii} K and H$\beta$ lines, 
especially for those M-type stars, e.g., the typical errors of EW$_{\text{H}\beta}$ are 0.05, 0.15, 1.0~\AA, for Pleiades members with temperatures around 
5500, 4500, 3500 K, respectively.

\begin{figure}
\centering
\includegraphics[width=\columnwidth]{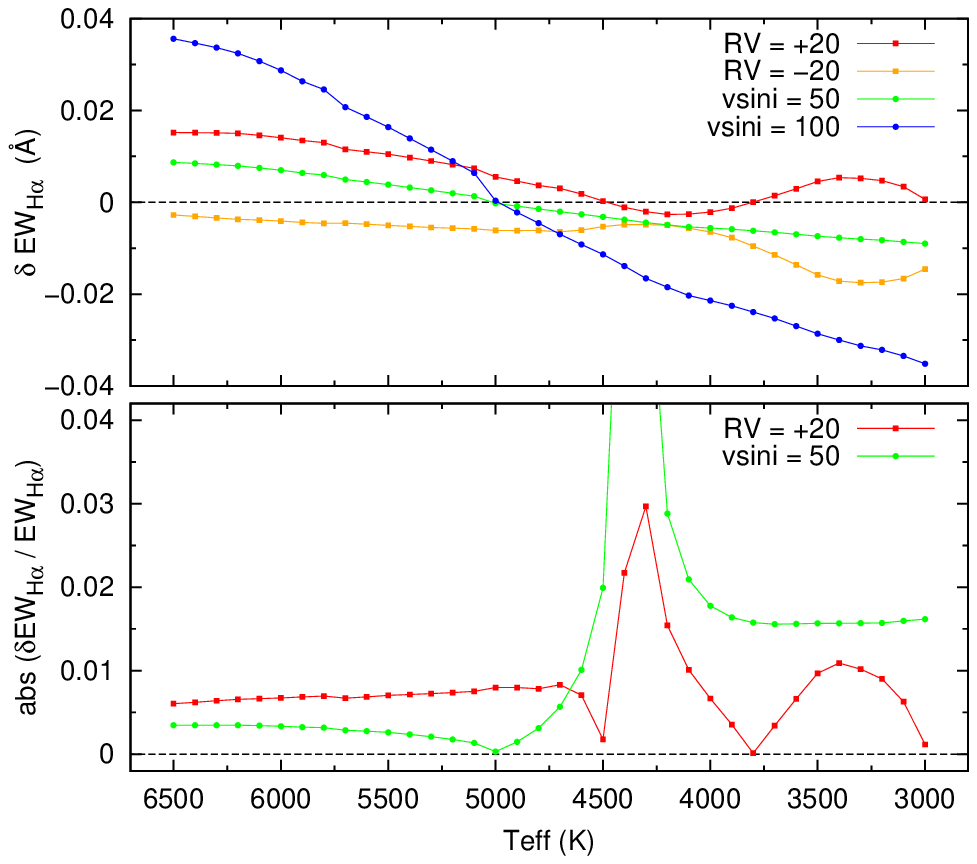}
\caption{Top: Estimated differential EW$_{\text{H}\alpha}$ values due to wavelength shifts, RV = $\pm20$ $km\,s^{-1}$, and rotational broadening,
$v\sin i$ = 50 and 100 $km\,s^{-1}$. Bottom: Relative differences compared to EW$_{\text{H}\alpha}$ scales, note that the large values around 4300 K result 
from the fact that EW$_{\text{H}\alpha}$ is near to zero for late K-type stars.}
\label{fig:rv_rot}
\end{figure}

The spectral resolution of LAMOST spectra (R$\sim$1800) limits the accuracy of RV measurements. To check how RV uncertainties affect EW$_{\text{H}\alpha}$ 
measurements, we have taken PHOENIX model spectra and degraded the resolution to 1800 by convolving a Gaussian profile, found that the differences of 
EW$_{\text{H}\alpha}$ due to RV shifts of $\pm$20 $km\,s^{-1}$ are less than 0.02~\AA~(a relative difference of less than 3 percent), as shown in 
Fig.~\ref{fig:rv_rot}. In addition, many cool Pleiades members are fast rotators \citep{rebu2016}, we thus evaluated 
the effects of rotational broadening on measurements of EW$_{\text{H}\alpha}$, and found the differences of EW$_{\text{H}\alpha}$ due 
to a $v\sin i$ of $\sim$100 $km\,s^{-1}$ are typically less than 0.04~\AA~(see Fig.~\ref{fig:rv_rot}), which is slightly larger than the 
errors due to RV uncertainties, but typically less than average errors of EW$_{\text{H}\alpha}$ measurements propagated from the uncertainties of the observed flux. 
So the uncertainties from RV-correction and rotational broadening are marginal compared to those due to uncertainty of observed spectra.

\begin{figure}
\centering
\includegraphics[width=\columnwidth]{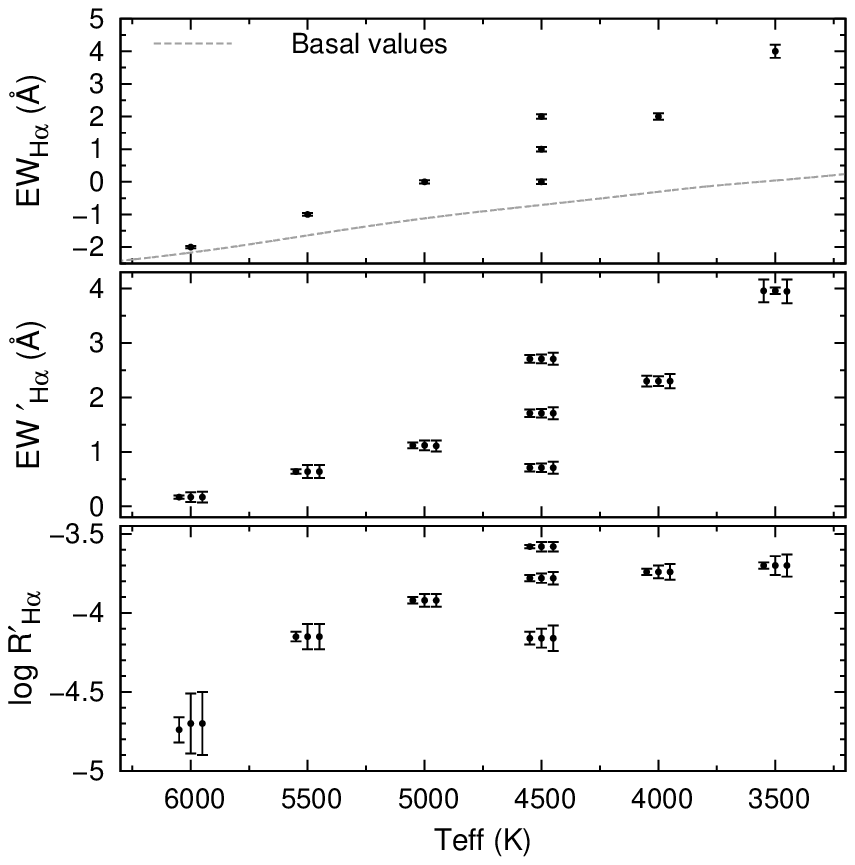}
\caption{Top: Examples of typical values of EW$_{\text{H}\alpha}$ with measured errorbars $\sigma(\text{EW}_{\text{H}\alpha})$. Middle and bottom: Resulted 
excess emissions with estimated errorbars from different uncertainty sources by using Monte Carlo simulations. The data points at the same temperature are 
separated by 50 K for display purpose: uncertainties from EW$_{\text{H}\alpha}$ (left), from $T_{\text{eff}}$ (centre), and from both of them (right). }
\label{fig:simulation}
\end{figure}

The uncertainties in EW$^{'}_{\text{H}\alpha}$ are mainly propagates from the errors of EW$_{\text{H}\alpha}$ measurement and corresponding basal value 
used at a given temperature. The basal value of equivalent width for a given line was derived based on a large sample of reference dwarfs with 
solar metallicity (e.g., for H$\alpha$ line, see Paper I). The scatter of EW$_{\text{H}\alpha}$ of reference stars in each 
50 K bin is typically smaller than the errors of measured EW$_{\text{H}\alpha}$, as demonstrated in the upper panels of Fig.~\ref{fig:ewha_teff}, 
where three times of the scatter ($\pm3\sigma$) are shown in grey lines. To estimate the uncertainties on 
the derived EW$^{'}_{\text{H}\alpha}$ resulting from the flux errors of spectra and uncertainties of 
temperature, we performed a Monte Carlo simulation 1000 times for six representative EW$_{\text{H}\alpha}$ values with $T_{\text{eff}}$ 
from 6000 to 3500 K, using the typical measurement error (see Fig.~\ref{fig:error}) as the standard deviation of EW$_{\text{H}\alpha}$, and setting 100 K as 
the standard deviation of temperature, as shown in Fig.~\ref{fig:simulation} and listed in Table~\ref{tab:mc}. It is clear that the EW$^{'}_{\text{H}\alpha}$ 
uncertainties are dominated by the uncertainties of temperatures for GK-type stars. It is not the same case with M-type stars since the 
reference EW$_{\text{H}\alpha}$ value is not so sensitive to temperature compared to hotter stars. 

The excess fractional luminosity $R^{'}_{\text{H}\alpha}$ depends on both derived EW$^{'}_{\text{H}\alpha}$ and estimated value of $\chi$ at a given temperature. 
To check the temperature selection affect on derived $R^{'}_{\text{H}\alpha}$, we performed a simple Monte Carlo simulations (see Table~\ref{tab:mc}), 
which show that temperature uncertainty dominates the derived $R^{'}_{\text{H}\alpha}$ uncertainty, 
as shown in Fig.~\ref{fig:simulation}. 

\begin{table}
\caption{Estimated uncertainties from Monte Carlo simulations}
\label{tab:mc}
\begin{tabular}{ccccccccc}
   \hline
Teff ($\sigma$) & EW$_{\text{H}\alpha} (\sigma)$ &$\sigma_{1}$&$\sigma_{2}$&$\sigma_{3}$&$\sigma_{4}$&$\sigma_{5}$\\
     (K)        &  (\AA)                         & (\AA)      &  (\AA)     &    (dex)   &   (dex)    &   (dex)      \\
   \hline
6000(100)        &  -2.0(0.03)               &    0.09    & 0.10       & 0.08       &   0.19     &  0.20      \\
5500(100)        &  -1.0(0.04)               &    0.12    & 0.12       & 0.03       &   0.08     &  0.08      \\
5000(100)        &  0.0(0.05)                &    0.09    & 0.10       & 0.02       &   0.04     &  0.04      \\
4000(100)        &  2.0(0.10)                &    0.09    & 0.13       & 0.02       &   0.04     &  0.05      \\
3500(100)        &  4.0(0.20)                &    0.06    & 0.22       & 0.02       &   0.06     &  0.07      \\
\hline
4500(100)        &  0.0(0.07)                &    0.08    & 0.11       & 0.04       &   0.06     &  0.08      \\
4500(100)        &  1.0(0.07)                &    0.08    & 0.11       & 0.02       &   0.03     &  0.04      \\
4500(100)        &  2.0(0.07)                &    0.08    & 0.11       & 0.01       &   0.03     &  0.03      \\
\hline
\multicolumn{7}{l}{$\sigma_{1}$: $\sigma (\text{EW}^{'}_{\text{H}\alpha}$) due to $\sigma(T_{\text{eff}})$;}\\ 
\multicolumn{7}{l}{$\sigma_{2}$: $\sigma (\text{EW}^{'}_{\text{H}\alpha}$) due to both $\sigma(\text{EW}_{\text{H}\alpha})$ and $\sigma(T_{\text{eff}})$;}\\
\multicolumn{7}{l}{$\sigma_{3}$: $\sigma (\log \text{R}^{'}_{\text{H}\alpha})$ due to $\sigma(\text{EW}_{\text{H}\alpha})$;}\\
\multicolumn{7}{l}{$\sigma_{4}$: $\sigma (\log \text{R}^{'}_{\text{H}\alpha})$ due to $\sigma(T_{\text{eff}})$;}\\
\multicolumn{7}{l}{$\sigma_{5}$: $\sigma (\log \text{R}^{'}_{\text{H}\alpha})$ due to both $\sigma(\text{EW}_{\text{H}\alpha})$ and $\sigma(T_{\text{eff}})$.}\
\ 
\end{tabular}
\end{table}

\section{The ratio of the continuum flux to the bolometric flux}
\label{sec:chi}
To quantify the strength of chromospheric activity as the ratio between H$\alpha$ emission luminosity and bolometric luminosity, \citet{walk2004} introduced a 
distance-independent value $\chi$, the ratio between the apparent continuum flux near H$\alpha$, $f(\lambda\text{6560})$, and the apparent bolometric flux, 
$f_{\text{bol}}$, namely, $\chi = f(\lambda\text{6560})/f_{\text{bol}}$. Then the measured equivalent width of H$\alpha$ emission line could be converted to 
the fractional luminosity by using the $\chi$ value, i.e., $L_{\text{H}\alpha}/L_{\text{bol}} = \chi \times \text{EW}_{\text{H}\alpha}$. 
$\chi$-method is widely used where flux-calibrated spectrum and distance information is 
not available for quantifying chromospheric activity level \citep[e.g.,][]{doug2014,west2015}.

\begin{figure}
\includegraphics[width=\columnwidth]{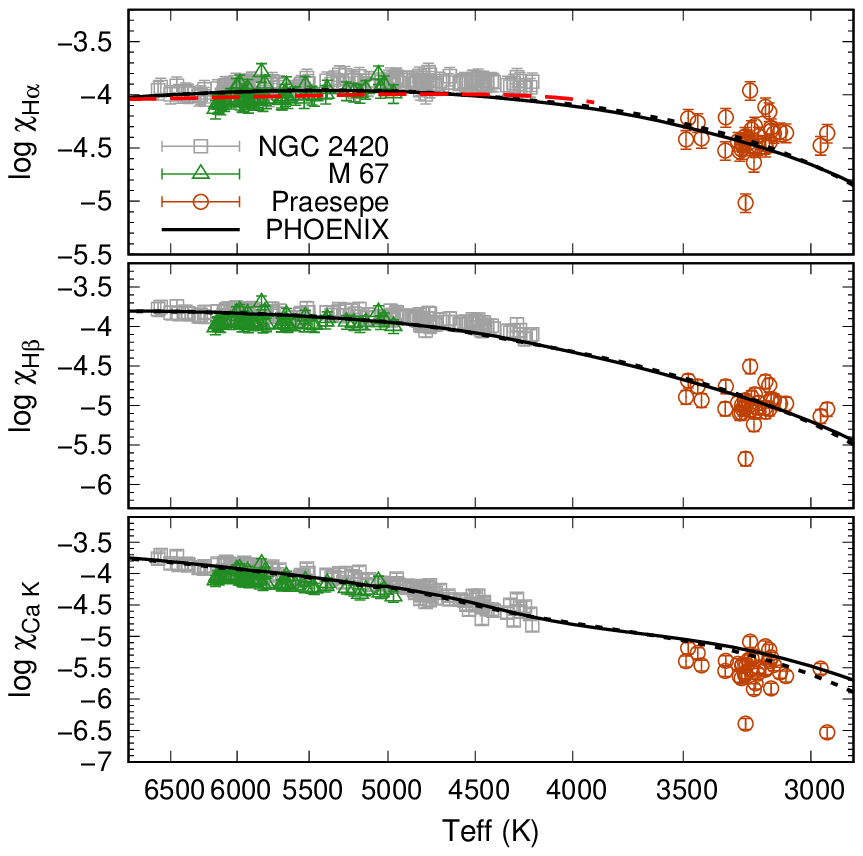}
\caption{$\chi$ values of H$\alpha$, H$\beta$, and Ca~{\sc ii} K lines, $\chi_{\text{H}\alpha}$, $\chi_{\text{H}\beta}$, and $\chi_{\text{Ca K}}$, 
respectively, derived based on SDSS spectra, and PHOENIX model spectra with $\log g$=4.5 (black solid lines) and $\log g$=5.0 (black dashed lines). 
The red long dashed line denotes the relation derived from empirically determined H$\alpha$ line surface flux provided by \citet{sode1993}.}
\label{fig:chi}
\end{figure}

$\chi$ values could be estimated based on observed spectra \citep[e.g.][]{west2008} or model spectra \citep[e.g.][]{rein2012}. In this work, we estimated the 
$\chi$ values as the ratio of the surface continuum flux, $F(\lambda_{c})$, to the stellar surface bolometric flux, $F_{\text{bol}}$, using the following 
formula, 
\begin{equation}
 \chi = \frac{F(\lambda_{c})}{F_{\text{bol}}} = \frac{F(\lambda_{c})}{\sigma T_{\text{eff}}^4},
 \label{chi}
\end{equation}
where $\sigma$ is Stefan-Boltzmann constant, $\sigma \approx 5.6704 \times 10^{-5} erg\, s^{-1}\, cm^{-2}\, K^{-4}$. We evaluated $F(\lambda_{c})$ for the 
lines of interest using PHOENIX ACES model spectra with solar metallicity \citep{huss2013}. We then obtained $\chi$ values for H${\alpha}$, Ca~{\sc ii} K and 
H${\beta}$ lines, as shown in Fig.~\ref{fig:chi}. 

To check the agreement of these $\chi$ values derived from models with those from the observations, 
we collected the flux calibrated SDSS spectra of stars in NGC~2420, M~67, and Praesepe. 
140 members of NGC~2420 with $\log g >4.0$ and 52 dwarfs in M~67 were selected from the catalogue of \citet{lee+2008}. 
We estimated their radii using the temperatures from \citet{lee+2008}, 
adopting the PARSEC model \citep{chen2014} with a metallicity of [Fe/H]$\sim-0.2$ \citep{jaco2011} and an age of $\sim2$ Gyr \citep{sala2004}, and the PARSEC model with a solar metallicity \citep{jaco2011} and an age of $\sim4$ Gyr \citep{oneh2011,gell2015}, for NGC 2420 and M67 respectively. 
We collected SDSS spectra with good quality ($SNR>10$) for 36 Praesepe stars of \citet{doug2014} based on the SDSS DR12 SSPP output 
catalogue \citep{lee+2008,smol2011}. We estimated their effective temperatures using the colours ($V-K_{s}$ and/or $r-K_{s}$), 
and derived corresponding radii adopting the PARSEC model with a metallicity of [Fe/H]$\sim+0.16$ \citep{carr2011} and an age of about 650 Myr. 
The distances adopted from WEBDA (http://www.univie.ac.at/webda/) for NGC 2420, M 67 and Praesepe are 3085, 908 and 187 $pc$, respectively. Based on the observed continuum flux, 
$f(\lambda_{c})$, and their corresponding radii ($R$) and distances ($D$), the $\chi$ ratios could be derived using the following relation,  
\begin{equation}
 \chi = \frac{F(\lambda_{c})}{F_{\text{bol}}} = \frac{(D/R)^{2}f(\lambda_{c})}{\sigma T_{\text{eff}}^4}, 
\end{equation}
as shown in Fig.~\ref{fig:chi}, wherein the error bar is equivalent to its uncertainty due to 10 percent error of distance. 
There exists a slight offset between members in NGC 2420 and M 67 at the same effective temperatures, 
which may be due to the uncertainties in distance, and/or the difference in basic physical parameters between these two clusters. 
Also shown in Fig.~\ref{fig:chi} by the red dashed line is a relation for H$\alpha$ line, 
which was derived from empirically determined surface flux of H$\alpha$ line provided by \citet{sode1993}. 
As shown by the figure, the $\chi$ values from model spectra are overall in good agreement with those from observations, 
there however exist some small discrepancies between them, e.g., $\sim$0.1 dex in H$\alpha$ line around 4500 K.

\section{Estimate of quiescent photosphere temperature and mass}
\label{sec:teffq} 
The cool spots on the stellar surface contribute light that would affect spectral features and broad-band colours, for which the derived effective temperature becomes 
the average surface temperature (flux-weighted mean temperature) in spotted stars. Therefore, no direct way to derive a quiescent temperature for 
active stars, particularly those Pleiades members with large spot coverages (see Paper I for more discussion on this topic). 

For Pleiades members, we followed the procedures in Paper I to estimate the quiescent temperature (see Paper I for more details). For M34 members, we collected 
V band magnitudes from literature \citep{jone1996,irwi2006,meib2011}, and got their $K_{s}$-band magnitudes from 2MASS archive 
\citep{cutr2003} via Vizier x-match service. We estimated their quiescent temperatures based on $V-K_{s}$ colours using PARSEC isochrones \citep{chen2014} 
with an age of 220 Myr and solar metallicity, wherein we adopted the reddening of E(B-V)=0.07 mag \citep{cant1979}.

\begin{figure}
\includegraphics[width=\columnwidth]{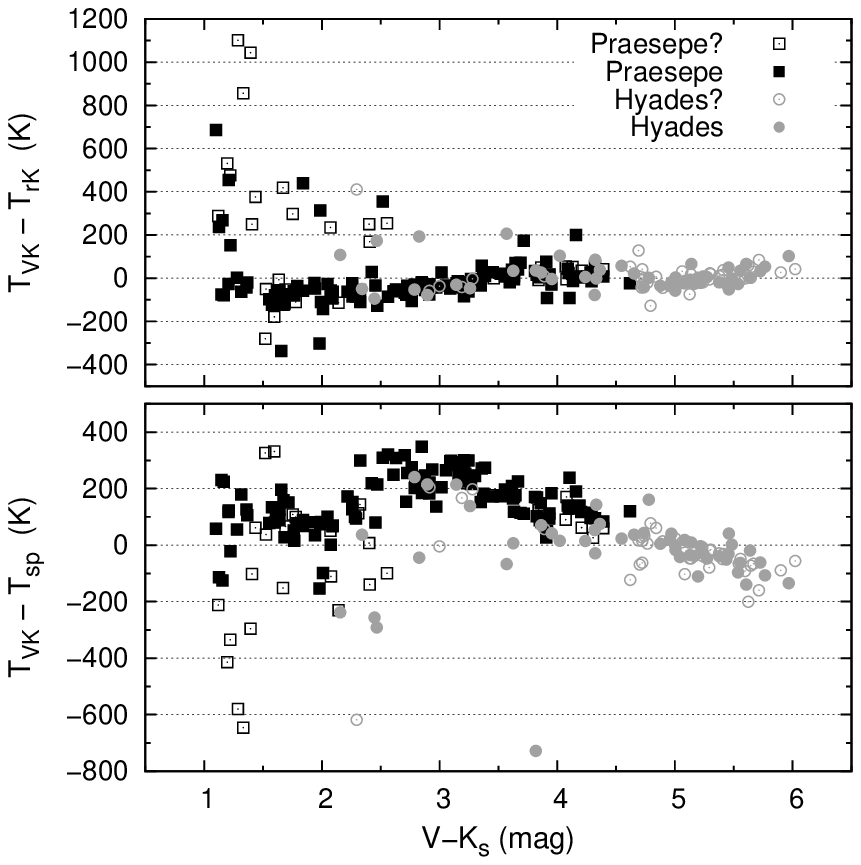}
\caption{The differences between $V-K_{s}$-based temperature ($T_{\text{VK}}$), $r-K_{s}$-based temperature ($T_{\text{rK}}$), and the temperature from 
spectral features ($T_{\text{sp}}$) among stars in Praesepe and Hyades. }
\label{fig:teffph_teffsp}
\end{figure}

For Praesepe and Hyades candidates, firstly we derived their effective temperatures based on $V-K_{s}$ colours, where V magnitudes taken from the fourth U.S. 
Naval Observatory CCD Astrograph Catalogue \citep[UCAC4;][]{zach2012}. A small fraction of sample stars have no V magnitudes, we then derived their temperatures 
based on $r-K_{s}$, where $r$ magnitudes from the Carlsberg Meridian Catalogue 15 \citep[CMC15;][]{cope2011}. The PARSEC models with an age of 650 Myr and a 
metallicity of [Fe/H]$\sim$0.1 were used to convert colours to temperatures, where the reddening E(B-V) = $0^{m}.027$ and E(B-V)=$0^{m}.001$~\citep{tayl2006} 
were adopted for Praesepe and Hyades, respectively. We then compared the colour-based temperatures with the spectra-based temperatures ($T_{\text{sp}}$, for 
FGK-type stars, we adopt the values provided by LASP, for M-type stars, we derived it via the molecular band CaH using the method in Paper I), and found that 
there exist complex differences between colour-based temperatures and spectra-based temperatures, as shown in Fig.~\ref{fig:teffph_teffsp}. We finally adopted 
the effective temperatures from spectral features as the quiescent temperatures for stars in these two clusters.

For the M-type slow rotators in Kepler field, firstly we estimated temperature for each target based on the molecular bands of CaH and TiO following the 
method in Paper I, and then took the average of our measured value and the value provided by \citet{mcqu2014}, as finally adopted quiescent temperature. 

To identify potential binaries and non-members for the member candidates in open clusters, we followed the procedures used in Paper I by using 
available CMDs, e.g., we identified these stars as photometrically binaries/non-members that have larger deviation compared with empirical single-star loci in 
V vs. $V-K_{s}$ diagram, namely, brighter stars with $-1.^{m}0 < \Delta V < -0^{m}.5$ were classified as probable binary members, while the fainter stars with 
$\Delta V >0^{m}.3$ or very much brighter stars with $\Delta V < -1^{m}.0$ were identified as probable non-member stars.

Once the effective temperatures are obtained for each star, we estimated the mass from temperature based on mass-$T_{\text{eff}}$ relation from PARSEC 
isochrones \citep{chen2014}. Note that, for these M-stars in Kepler field, we used the isochrones with a solar metallicity and an age of 2 Gyr. 
%%%%%%%%%%%%%%
\section{Rotation Periods Collection}
\label{sec:rotation}
For Pleiades, \citet{hart2010} detected rotation periods for nearly 400 members using the Hungarian-made Automated Telescope Network (HATNet) transit survey 
data. Recent studies based on dedicated photometric surveys, e.g., the Palomar Transient Factory (PTF) \citep{cove2016}, and the $K$2 \citep{rebu2016}, 
increased the measurements of rotation periods for more than 500 new members of this open cluster, vastly expanding the number of Pleiades members with 
rotation periods. For these members with multi-rotation measurements, we adopted values from the $K$2 \citep{rebu2016}, and for others, we adopted their 
rotation periods provided by \citet{cove2016} and \citet{hart2010}. Totally, 231 Pleiades members in our sample have available rotation periods.

For M34, we collected rotation periods from literature, \citet{irwi2006,jame2010,meib2011}. The latest measurements were adopted in this paper. 
Totally, we found 21 members have rotation periods. 

For stars in Praesepe, their rotation periods were collected from \citet{scho2007,delo2011,scho2011,ague2011,kova2014}. Totally, we got their rotation 
periods for 76 stars in Praesepe sample. For Hyades, we collected the rotation periods from \citet{radi1987,delo2011,hart2011,doug2016}. Totally, 24 stars in 
Hyades sample have rotation periods.

\section{On-line Tables}
We compiled the measurements for stars in Pleiades, M34, Praesepe and Hyades in on-line Table F1-F4, respectively, including effective 
(quiescent) temperatures, equivalent widths of H$\alpha$, H$\beta$ and Ca~{\sc ii} K lines, their excess equivalent widths, excess fractional luminosities, 
and spot filling factors.

%%%%%%%%%%%%%%%%%%%%%%%%%%%%%%%%%%%%%%%%%%%%%%%%%%
%%%%%%%%%%
% Don't change these lines
\bsp	% typesetting comment
\label{lastpage}
\end{document}